\documentclass{aa} 

\usepackage{natbib}
\bibpunct{(}{)}{;}{a}{}{,} % to follow the A&A style

\usepackage{stfloats}
\usepackage{multirow}
\usepackage{hyperref}
\usepackage{xcolor}
\hypersetup{
    colorlinks,
    linkcolor={red!50!black},
    citecolor={blue!50!black},
    urlcolor={blue!80!black}
}
\usepackage{graphicx}
\usepackage{pdflscape}
\usepackage{wasysym}
\usepackage{placeins}
\usepackage{txfonts}

\newcommand{\comment}[1]{}

\newcommand{\radm}{{\rm rad\,m^{-2}}}
\newcommand{\dm}{{\rm pc\,cm^{-3}}}
\newcommand{\logM}{\log (M_*/{\rm M}_{\odot})}

 \authorrunning{Kovacs et al. 2023}
 \titlerunning{DM and RM contribution of FRB host galaxies}
\begin{document}

   \title{The dispersion measure and rotation measure from fast radio burst host galaxies based on the IllustrisTNG50 simulation}

   \author{Timea Orsolya Kovacs \inst{1}, Sui Ann Mao \inst{1}, Aritra Basu\inst{2,1}, Yik Ki Ma\inst{1,3}, Ruediger Pakmor\inst{4}, Laura G. Spitler\inst{1} and Charles R. H. Walker\inst{1}}

    \institute{Max-Planck-Institut für Radioastronomie, Auf dem Hügel 69, D-53121 Bonn, Germany \and
    Thüringer Landessternwarte, Sternwarte 5, D-07778 Tautenburg, Germany \and
    Research School of Astronomy \& Astrophysics, Australian National University, Canberra, ACT 2611, Australia  \and
    Max-Planck-Institut für Astrophysik, Karl-Schwarzschild-Str. 1, D-85748 Garching, Germany }

   \date{Received 15 July 2023 / Accepted 18 July 2024}

% \abstract{}{}{}{}{} 
% 5 {} token are mandatory
 
  \abstract
   {Fast radio bursts (FRB) will become important cosmological tools in the near future, as the number of observed FRBs is increasing rapidly with more surveys being carried out. Soon a large sample of FRBs will have available dispersion measures (DM) and rotation measures (RM), which can be used to study the cosmic baryon density and the intergalactic magnetic field (IGMF). However, the observed DM and RM of FRBs consists of multiple contributions which must be quantified to estimate the DM and RM of the intergalactic medium (IGM).}
   {In this paper, we estimate one such contribution to DM and RM: that of FRB host galaxies. We show how this contribution changes with redshift, galaxy type, and the stellar mass of the galaxies. We also investigate its dependence on galaxy inclination, and on an FRB's offset from the center of the galaxy.}
  % methods heading (mandatory)To estimate the DM and RM contributions 
   {Using the IllustrisTNG50 simulations, we selected 16 500 galaxies at redshifts of 0$\leq z\leq$2, with stellar masses in the range $9 \leq \log (M_*/{\rm M}_{\odot}) \leq 12$. In each galaxy, we calculate the DM and RM contributions of 1000 sightlines, and from these, construct DM and RM probability density functions.}
   {We find that the rest frame DM distributions of all galaxies at a given redshift can be fitted by a lognormal function, and its median and width increases as a function of redshift.
    The rest frame RM distribution is symmetric, with median RM$_{\rm host, rf}$=0 rad m$^{-2}$, and it can be fitted by the combination of a Lorentzian and two Gaussians. The redshift evolution of the distribution width can be fitted by a curved power law.
   The parameters of these functions change for different subsets of galaxies with different stellar mass, inclination, and FRB offset. These changes are due to an increasing $n_e$ with redshift, SFR, and stellar mass, and we find a more ordered $B$ field at lower $z$ compared to higher $z$, suggested by more galaxies with $B$ field reversals and $B$ fields dominated by random $B$ field at higher $z$.}
   {We estimate the FRB host DM and RM contributions, which can be used in the future to isolate the IGM's contribution from the observed DM and RM of FRBs. We predict that to constrain an $\sigma_{\rm RM, IGM}$ of 2 $\radm$ to 95\% confidence level we need to observe $95\,000$ FRBs at $z=0.5$, but only 9\,500 FRBs at $z=2$.}


   \maketitle
%
%-------------------------------------------------------------------
\section{Introduction}
The seeds of intergalactic magnetic fields (IGMF) are an open question in astrophysics \citep{ 1996PhRvD..53..662B, 2005PhR...417....1B, 2012PhRvD..86j3005K, 2013PhRvD..87h3007K}. They could be primordial (magnetic fields that already existed soon after the Big Bang), astrophysical (magnetic fields that were caused by galaxy evolution, e.g. feedback processes, star formation, active galactic nuclei), or both. The true scenario could also have implications on the dynamics of inflation (due to inflationary magnetogenesis, e.g. \citealt{2013JCAP...10..004F,2021Galax...9..109V}) and the physics of the early Universe. Recent measurements exclude purely astrophysical and purely primordial origins and find an upper limit of $\sim4$\,nG on the co-moving magnetic field strength based on Cosmic Microwave Background (CMB) anisotropies \citep{2016A&A...594A..19P} and low-frequency radio observations with the Low-Frequency Array (LOFAR) of close pairs of extragalactic radio sources \citep{2020MNRAS.495.2607O}. \cite{2016MNRAS.462.3660H} found 0.1 nG magnetic field in voids based on the observed degree of isotropy of ultra high energy cosmic rays (UHECR, \citealt{2003PhRvD..68d3002S}). \cite{2023MNRAS.518.2273C} measured the magnetic field of filaments with LOFAR and inferred an IGMF of 0.04 $\textrm{--}$ 0.11\,nG on Mpc scales. Many of these methods measure the magnetic field only in filaments, thus we have limited constraints on the low-density Universe. Fast radio bursts (FRB) are a promising, unique tool for measuring the IGMF as they account for every ionized baryon along a line of sight (\citealt{2014ApJ...797...71Z,2016ApJ...824..105A}).  

FRBs are millisecond radio transients (for reviews, see \citealt{2019ARA&A..57..417C, 2019A&ARv..27....4P}); most are one-off sources and $\sim$ 3\% are confirmed repeaters \citep{2023ApJ...947...83C}. Their origin is still unknown, and there are many different FRB progenitor models, but the majority assume a connection to neutron stars \citep{2019PhR...821....1P}. Recently, more and more FRBs have been localized to distant galaxies (see e.g. \citealt{2017Natur.541...58C}, \citealt{2019Sci...365..565B}, and \citealt{2022evlb.confE..35M}) out to a redshift of $\sim$ 1 \citep{2023Sci...382..294R}.  Due to their extragalactic origin and high event rate (predicted to be 10$^3$\,$-$\,10$^4$ per day over the whole sky above a fluence of $\sim$2 Jy ms, \citealt{2018MNRAS.475.1427B}), FRBs can be used as cosmological probes.

FRBs experience propagation effects due to the intervening material they pass through, causing a dispersion in the observed pulse \citep{2019ARA&A..57..417C}, characterized by the dispersion measure (DM, see below for more details). Some FRBs are linearly polarized and thus also undergo Faraday rotation \citep{2015Natur.528..523M}, which causes the angle of polarization to change upon passing through a magneto-ionic medium. This is characterized by the rotation measure (RM, see below for more details). The observed DM of FRBs is 100 -- 3000 pc cm$^{-3}$ (\citealt{2016PASA...33...45P}, \citealt{2021ApJS..257...59C}), and the |RM| of FRBs ranges from $1.5$ to $\sim 10^5$ rad~m$^{-2}$ (\citealt{2018Natur.553..182M,2019A&ARv..27....4P}). The observed RM can even change sign with time \citep{2023Sci...380..599A}. FRBs can detect the baryonic content along the line-of-sight (LOS) into the distant Universe and are a powerful tool of cosmic magnetism (e.g. IGMF, \citealt{2016ApJ...824..105A}, \citealt{2014ApJ...797...71Z}), because they provide information on both DM and RM simultaneously. Furthermore, they provide a unique tool for the study of the missing baryons \citep{2014ApJ...780L..33M,2020Natur.581..391M}, cosmological parameters (\citealt{2015aska.confE..55M}, \citealt{2014ApJ...788..189G}, \citealt{2014PhRvD..89j7303Z}), Hubble constant ($H_{\rm 0}$) measurements \citep{2022MNRAS.516.4862J}, the intergalactic medium (IGM), and the interstellar medium (ISM) of our or other galaxies (e.g. \citealt{2023ApJ...954..179M}). 

The DM is the integral of the free electron density ($n_{e}$ [cm$^{-3}$]) along the LOS ($l$ [pc]) from a source to the position of an observer:
\begin{equation}
    {\rm DM} = \int_{\rm source}^{\rm observer} n_{e} {\rm d}l.
\end{equation}

Because DM is an integral along the entire LOS, there are multiple contributors to the observed DM of FRBs (DM$_{\rm obs}$): the immediate source environment (DM$_{\rm source}$), the host galaxy (DM$_{\rm host}$), the IGM (DM$_{\rm IGM}$), and the Milky Way (DM$_{\rm MW}$). Thus, the observed DM of a source at redshift $z_{\rm host}$ (i.e. $z_{\rm source}$ = $z_{\rm host}$) can be described in the following way:
    \begin{equation}
        {\rm DM}_{\rm obs} = \frac{\rm DM_{\rm source,rf}}{(1+z_{\rm host})} + \frac{\rm {\rm DM}_{\rm host,rf}}{(1+z_{\rm host})} + {\rm DM}_{\rm IGM} + {\rm DM}_{\rm MW},
    \end{equation}
where $\rm DM_{source,rf}$ and $\rm DM_{host,rf}$ are in the rest-frame, and can be converted to the observer's frame by the standard correction of 1/$(1+z_{\rm host})$, due to the cosmological time dilation
and the frequency shift. The magnitude of the different components' DM contributions are comparable based on models and simulations: 10$^{-6}$ -- $1000\,\dm$ from the local environment of an FRB (such as a supernova remnant, or surroundings of a magnetar \citealt{2018ApJ...861..150P,2022ApJ...933L...6L}), $\sim$200 to $\sim$1700 $\dm$ from the Milky Way and its halo (\citealt{2002astro.ph..7156C,2017ApJ...835...29Y,2015MNRAS.451.4277D,2020MNRAS.496L.106K}), $\sim$10 pc cm$^{-3}$ to a few thousand pc cm$^{-3}$ from the host galaxy (e.g. \citealt{2014PhRvD..89j7303Z,2020A&A...638A..37W,2020ApJ...900..170Z}), and a few thousand $\dm$ from the IGM \citep{2016ApJ...824..105A}. DM$_{\rm IGM}$ increases with redshift \citep{2020Natur.581..391M}, however, its exact value can be different if the LOS goes through a dense region (e.g. a cluster or a filament).

The RM of a source is defined similarly to DM, but the integral is weighted by the LOS component of the magnetic field ($B_\|$ [$\mu$G], parallel to the LOS):
\begin{equation}
    {\rm RM} = k \int_{\rm source}^{\rm observer} B_{\|} n_{e} {\rm d}l,
\end{equation}
where $k = 0.81$ $\mu$G$^{-1}$.

Similar to DM, the observed RM of a source at redshift $z_{\rm host}$ also consists of the components from the source ($\rm RM_{\rm source}$), host ($\rm RM_{\rm host}$), IGM ($\rm RM_{\rm IGM}$), and the Milky Way ($\rm RM_{\rm MW}$):
    \begin{equation}
        {\rm RM}_{\rm obs} = \frac{\rm RM_{\rm source,rf}}{(1+z_{\rm host})^2} + \frac{\rm RM_{\rm host,rf}}{(1+z_{\rm host})^2} + {\rm RM}_{\rm IGM} + {\rm RM}_{\rm MW}.
    \end{equation}
where $\rm RM_{source,rf}$ and $\rm RM_{host,rf}$ are in the rest-frame, and can be converted to the observer's frame by the standard correction of $1/(1+z_{\rm host})^2$. The contributions of the different components are comparable, such as 10$^{-2}$ and 10$^6$ rad m$^{-2}$ from the immediate surroundings (\citealt{2018ApJ...861..150P,2022ApJ...933L...6L}), 150 to 3000 $\radm$ from the MW (based on all-sky Galactic RM maps, e.g. \citealt{2012A&A...542A..93O, 2022A&A...657A..43H}), 10$^{-4}$ and 10$^{4}$ rad m$^{-2}$ from host galaxies based on previous simulations and models. The root mean square (rms) of RM$_{\rm IGM}$ is expected to increase towards higher redshift, but it is $\le$ 50 rad~m$^{-2}$ at all redshifts \citep{2016ApJ...824..105A,2019MNRAS.488.4220H}. If we can isolate the contribution of the IGM from the other contributions in the observed  RM, we can study the IGMF.

To be able to study the cosmic baryon density and the IGMF using the DM and RM measurements of FRBs, we need to constrain the other components, as they can all contribute significantly to the observed DM and RM mentioned above. In this paper, we focus on constraining the contribution of host galaxies. To do this, we calculated the DM and RM contribution of galaxies selected from the TNG50 simulation of the IllustrisTNG project (\citealt{2018MNRAS.473.4077P}, \citealt{2019ComAC...6....2N}), a state-of-the-art cosmological magneto-hydrodynamic simulation, which contains thousands of galaxies at every redshift (of different types), with relatively high resolution (70-140 pc), allowing us to provide good DM and RM estimates. We used 16.5 million sightlines to construct DM and RM Probability Density Functions (PDF) for different galaxy types to account for a large variety of possible FRB host galaxies.
The main goal of this paper is to model and constrain the contributions of host galaxies, with applicabilities of (1) the statistical removal of them for future IGM studies, and (2) informing the optimal criteria to sample FRBs for such studies. The secondary objective is to identify the astrophysical processes in the simulation that have led to the DM and RM trends.

This paper is organized as follows: in Section \ref{Methods}, we describe the simulation we used, our galaxy selection, and our DM and RM calculation methods. In Section \ref{Results}, we show how the rest frame DM and RM distributions depend on different galaxy properties, for example, redshift, stellar mass, and inclination. In Section \ref{Discussion}, we investigate how the observed DM and RM of hosts changes with redshift, its implications on measuring the RM of the IGM, and compare our results to previous works which used different simulations and models. In Section \ref{Conclusion}, we summarize our findings.

%--------------------------------------------------------------------
\section{Methods}
\begin{figure*}[!h]
    \centering

\includegraphics[width=18cm]{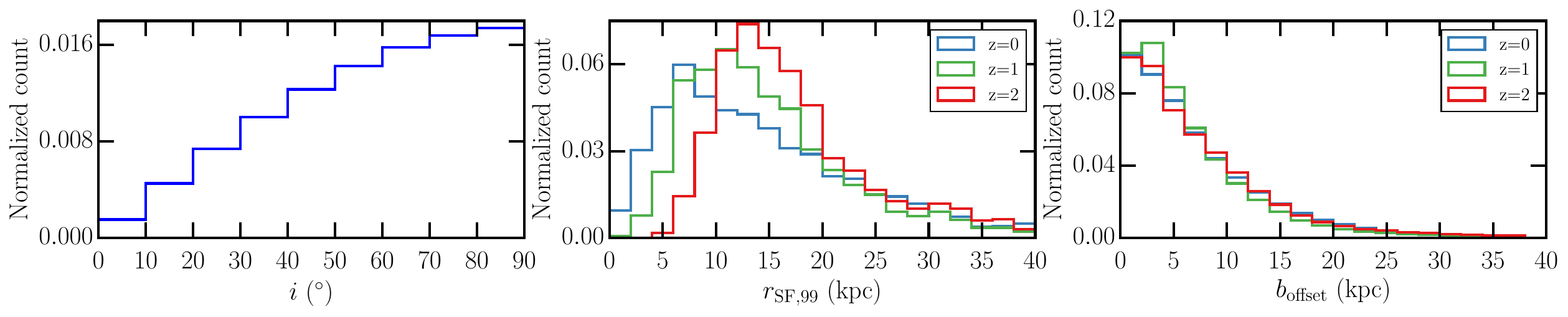}

    \caption{\textbf{Left:} The inclination ($i$) distribution of the galaxies, which is the same at every redshift. We define $i$ as the angle between the total angular momentum vector of the galaxy's stars and the LOS. \textbf{Middle:} The distribution of galaxy radii ($r_{\rm SF,99}$) at different redshifts. 
    \textbf{Right:} The distribution of the projected offsets ($b_{\rm offset}$) of the FRBs with respect to the center of their host galaxy at different redshifts.}
    \label{fig:incl}
\end{figure*}
\label{Methods}
In this section we summarize the TNG50 simulation of the IllustrisTNG project, we describe the selection process for our sample galaxies, and how we place FRBs within these galaxies. Finally, we show how we calculate the rest frame DM and RM host contributions of the galaxies.

\subsection{TNG50}
We have used IllustrisTNG, a state-of-the-art cosmological  magnetohydrodynamical (MHD) simulation  (\citealt{2018MNRAS.473.4077P}, \citealt{2019ComAC...6....2N}) to estimate the distribution of the DM and RM contribution of FRB host galaxies. 

\subsubsection{Description of the simulation}
The TNG project consists of three different simulation cubic volumes: TNG300, TNG100, and TNG50 with side length of approximately 300, 100, and 50 Mpc, respectively. We chose TNG50-1 because it has the highest physical resolution (\citealt{2019MNRAS.490.3196P}, \citealt{2019MNRAS.490.3234N}). 

In these simulations, the physical resolution depends on the gas density, there are a larger number of cells in high-density regions (e.g. galaxies). Since the DM and RM of FRBs are calculated as a LOS integral passing through the galaxy as a pencil beam, we want $n_e$ and $B_{||}$ that are realistic on the smallest scale possible for our study, as in reality both can change on a few 10s of pcs due to turbulence \citep{2008ApJ...680..362H}.

In the simulation volume of TNG50, the magnetic fields are modeled with the highest physical resolution in cosmological MHD simulations at the moment: the average cell size is 70 -- 140 pc in the star-forming regions of galaxies (which is 355 pc in TNG100, 715 pc in TNG300  \citealt{2019ComAC...6....2N}, and 370 pc in the Auriga simulations \citealt{2017MNRAS.467..179G}). The magnetic field information in TNG50 was saved in 20 snapshots by the TNG team, and the redshifts of the snapshots that we used in our study are listed in Table \ref{tab:gal_select}.

All the TNG simulations assume a cosmology consistent with the lambda cold dark matter ($\Lambda$CDM) model, and results from the Planck Collaboration (\citealt{2016A&A...594A..13P}, $\Omega_{\Lambda,0}$ = 0.6911, $\Omega_{m,0}$ = 0.3089, $\Omega_{b,0}$ = 0.0486, $\sigma_8$= 0.8159, $n_s$= 0.9667, and $h$ = 0.6774), with Newtonian self-gravity solved in an expanding Universe \citep{2019ComAC...6....2N}. The conversion of the simulation units to physical units was done following the recommended procedures for the TNG data \citep{2019ComAC...6....2N}.

\subsubsection{Supermassive black hole feedback in the simulation}
\label{AGN_feedback}

We note that the supermassive black hole (SMBH) feedback mechanism in the IllustrisTNG simulations (described in \citealt{2017MNRAS.465.3291W}, \citealt{2018MNRAS.479.4056W}) has a great effect on the properties of galaxies, which also appear in our results in Section \ref{Results}. Here we provide an overview of this feedback mechanism. The SMBH feedback changes mode for black hole masses above $\sim 10^8 M_{\odot}$, which usually corresponds to a galaxy stellar mass of around $\logM$ $\sim$ 10.5. Below this limit ($M_{\rm BH}$ $\lesssim$ 10$^8$), in the high accretion state, the feedback is purely thermal energy. For the low accretion state ($M_{\rm BH}$ $\gtrsim$ 10$^8$), more efficient kinetic energy is injected instead, quickly quenching star formation, and blowing strong outflows.

This results in a mostly star-forming population of galaxies at $\logM$ $\le$ 10.5, and a mostly quiescent population at $\logM$ $>$ 10.5; furthermore, this quenching can have an effect out to several 100s of kpc from the center of galaxies \citep{2020MNRAS.499..768Z}. X-ray observations show strong outflows from AGNs \citep{2010A&A...521A..57T}, which could cause the expulsion of gas from their host galaxies and the suppression of star formation \citep{2017FrASS...4...10C}, consistent with the quenching process in the simulation. 

\subsection{Galaxy selection}

\label{gal_selection}

Due to computational limitations, we only include galaxies (also called `subhalos'\footnote{The friends-of-friends algorithm organizes the dark matter particles into halos, and the subfind algorithm separates the halos into subhalos.}) with a stellar mass ($M_*$) in the range $\logM$ = 9 -- 12 (noting that out of 23 securely localized FRB-hosts, only 3 have  $9 > \logM$ \citealt{2023ApJ...954...80G}). Therefore, the study of dwarf galaxies, as well as satellite galaxies\footnote{Galaxies that are not the most massive subhalos of their parent halo.}, is outside the scope of this work. We also excluded subhalos flagged by the simulation as unsuitable for most analysis (as they are not of cosmological origin\footnote{Satellite galaxies that form within the virial radius of their parent halo, and their ratio of dark matter mass to total subhalo mass is less than 0.8 \citep{2019ComAC...6....2N}}), and galaxies labeled with star formation rate (SFR)~=~0 in the simulation output\footnote{Galaxies with SFR below the SFR resolution limit of 10$^{-5}$ M$_{*}$/yr}.

The number of galaxies in the above mass range ($\logM$ = 9 -- 12) experience a rapid fall off at $z>2$ in the simulation (see Table \ref{tab:gal_select}). This is as expected, because in the hierarchical structure formation of the $\Lambda$CDM model there are fewer massive galaxies at higher redshifts. We note that recent results from the \textit{James Webb} might challenge this view, as they show the presence of massive galaxies and a higher rate of galaxy disks at earlier cosmological times than expected from the $\Lambda$CDM model \citep{2022ApJ...938L...2F,2023Natur.616..266L}. Our results at $z<2$ should not be significantly affected, but at higher redshifts DM and RM might be higher in reality compared to our results from the simulation. Therefore, for our statistical analysis we focus on $z<2$, and only briefly investigate the overall redshift dependence of galaxies at $z>2$.

\begin{table}[]
    \centering
    \caption{The number of galaxies selected at each redshift.}

    \begin{tabular}{c c }
         \hline
         \hline
          $z$ & $N_{\rm halos}$  \\
         \hline
         
         0.0 & 1849 \\
         0.1 & 1834  \\
         0.2 & 1813 \\
         0.3 & 1752 \\
         0.4 & 1707 \\
         0.5 & 1695 \\
         0.7 & 1658 \\
         1.0 & 1607\\
         1.5 & 1389 \\
         2.0 & 1208 \\
         3.0 & 676  \\
         4.0 & 301  \\
         5.0 & 105  \\
        \hline
    \end{tabular}
    \label{tab:gal_select}
    \tablefoot{All galaxies have a stellar mass ($\logM$) between 9 and 12.}
\end{table}

In summary, we select galaxies from $z=0$ out to $z=2$, with $\logM$ in the range of 9 -- 12, resulting in a total of $16\,500$ galaxies (Table \ref{tab:gal_select}).

\subsection{Positions of FRBs}

\label{Methods:boffset}

We place FRBs in our selected galaxies assuming they are more likely to occur in denser regions. 
First, we define a radius for each galaxy that encloses 99\% of the star-forming gas of the subhalo ($r_{\rm SF,99}$). Fig. \ref{fig:incl} (middle) shows the distribution of $r_{\rm SF,99}$ at different redshifts, with slightly larger radii at higher redshifts, where the star-forming gas is not yet confined to the inner parts of the galaxies. We then randomly select 1000 gas cells inside this radius, which we adopt as the positions of our FRBs. As the physical resolution of the simulations is adaptive, this means we have more FRBs in denser regions of the galaxies, and most FRBs are in the disk. 

In Fig. \ref{fig:incl} (right) we show the distribution of the projected offset of FRBs from the center of their host galaxies ($b_{\rm offset}$). In order to clarify the definition of $b_{\rm offset}$ we show three example illustrations of $b_{\rm offset}$ in galaxies with different inclinations in Fig. \ref{fig:FRB_orientation}. We place more FRBs in the center of galaxies, and these distributions do not change significantly with redshift. We point out that out of a sample of 12 localized FRBs, the smallest $b_{\rm offset}$ is 0.6 $\pm$ 0.3 kpc and the largest  is 27.2 $\pm$ 22.6 kpc (\citealt{2020ApJ...903..152H}, where the errors are due to the uncertainties of the localization of the FRBs). This is consistent with our resulting range of $b_{\rm offset}$ <40 kpc. We note that different progenitors could have different spatial distribution in galaxies (see e.g. \citealt{2020A&A...638A..37W,2023MNRAS.518..539M}).

\subsection{Inclination of galaxies}

\label{Methods:inclination}

\begin{figure}
\centering
    \includegraphics[width=8.8cm]{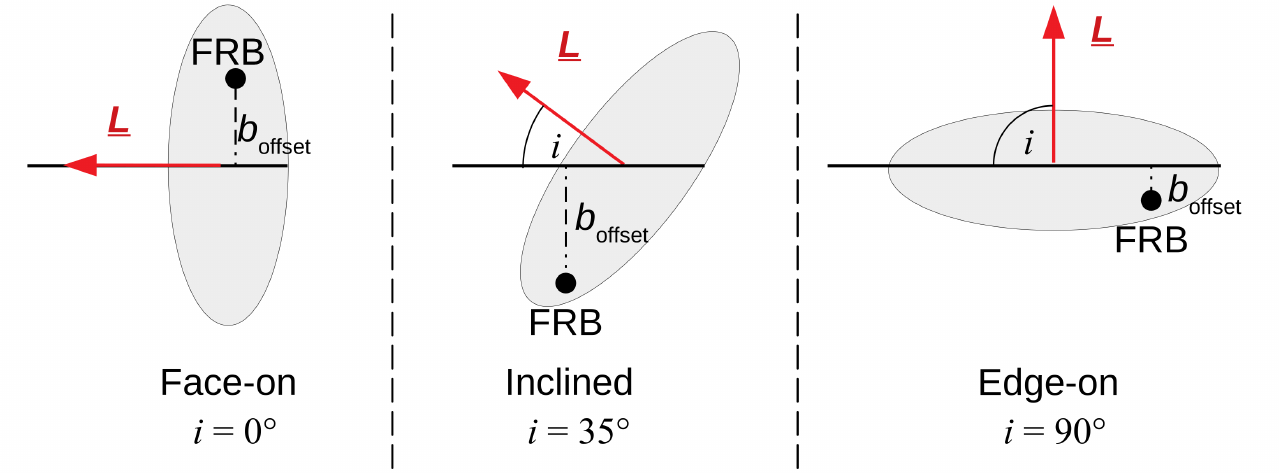}
    \caption{Illustration of the inclinations of host galaxies and the projected distance of FRBs from the center of the host galaxy ($b_{\rm offset}$), showed for face-on, inclined, and edge-on galaxies. The observer is on the left side. The total angular momentum vector of the host galaxy ($\vec{L}$) is perpendicular to the plane of its disk. The inclination ($i$) is the angle between the line of sight and $\vec{L}$. The filled circles indicate the positions of the FRBs.}
    \label{fig:FRB_orientation}
\end{figure}

We define the inclination ($i$) of the galaxies in the simulation as the angle between the total angular momentum vector ($\vec{L}$, which is perpendicular to the plane of the galaxy) and the line-of-sight. In Fig. \ref{fig:FRB_orientation} we show three example illustrations of different inclinations. 

We assume that galaxies are randomly oriented with respect to the line of sight. To ensure this, first we rotate each galaxy to face-on view, by rotating the total angular momentum vector of its stars to point to the observer. Then for each galaxy, we generate 1000 random rotation matrices using Fast Random Rotation Matrices \citep{ARVO1992117}, so that for each FRB position, the same galaxy has a unique orientation. After the rotations, the total angular momentum vectors are uniformly distributed on a sphere. We note that multiple orientations can result in the same inclination. An example of the distribution of $i$ is shown in Fig. \ref{fig:incl} (left). The magnetic field vectors and the position vectors of the cells of the galaxies are rotated by the same matrices.

\subsection{Calculating $n_e$}
 The ionization in the simulation that sets the electron density ($n_e$) is computed as the equilibrium state given radiative cooling and heating from the UV background and nearby AGN. However, $n_e$ must be calculated differently for star-forming and non star-forming cells (where SFR = 0 M$_{\odot}$/yr, i.e. below the SFR resolution of the simulation). For non star-forming cells, it can simply be computed from the electron abundance ($e_{\rm ab}$, the fractional electron number density with respect to the total hydrogen number density) and the hydrogen number density ($n_{\rm H}$):
\begin{equation}
    n_{e} [{\rm cm^{-3}}] = e_{\rm ab} \cdot n_{\rm H}.
\end{equation}
$e_{\rm ab}$ is available in the IllustrisTNG particle data. However, $n_{\rm H}$ has to be calculated using the total gas density of the cells ($\rho$):
\begin{equation}
    n_{\rm{H}} [{\rm cm^{-3}}] = \rho \cdot X_{\rm H} / m_{\rm p+},
\end{equation}
 where we assume the fraction of hydrogen to be $X_{\rm H}$ = 0.76 (based on the initial conditions of TNG, \citealt{2018MNRAS.473.4077P}), and $m_{\rm p+}$ is the mass of a proton.

For star-forming cells, we must consider the sub-grid model for star formation and the multiphase interstellar medium used by the simulation \citep{2003MNRAS.339..289S}. Following \cite{2003MNRAS.339..289S} (see also e.g. \citealt{2018MNRAS.481.4410P}, \citealt{2020ApJ...900..170Z}) we calculate  $n_e$ assuming each cell's interstellar medium consists of volume-filling hot ionized gas ($T \sim 10^7~{\rm K}$) and small neutral cold clouds ($T \sim 10^3~{\rm K}$), that our sightlines do not cross the cold clouds, and the temperature in the warm phase is hot enough to fully ionize hydrogen and helium.

Then by finding the mass fraction of the warm gas ($x_{\rm warm}$), and assuming it is 100\% ionized:
\begin{equation}
n_{ e} [{\rm cm^{-3}}] = x_{\rm warm} \cdot [X_{\rm H} + (Y_{\rm He} / 4 \cdot 2)]  \cdot \rho/ m_{\rm p+},
\label{Eq:n_e_SF}
\end{equation}
where $Y_{\rm He}$ (the fraction of helium) is 0.23. We note that the $n_e$ calculated using this subgrid model might be uncertain by a factor of 2--3, but a better estimate would require resolving the ISM. \cite{2020MNRAS.498.3193P} presented a new method of post-processing galaxies from MHD simulation to obtain better estimates on the small-scale ISM properties by including stellar clusters and their impact on the ISM. Comparing their radial and vertical profiles of $n_{\rm e}$ with a single galaxy from the Auriga simulations, they found that $n_e$ can vary by a factor of 2 lower or higher in different parts of the galaxy compared to when computed using Eq. \ref{Eq:n_e_SF}. We argue that because for most sightlines we integrate across a large part of the galaxies, these differences would average out, and our DMs and RMs are not affected significantly.

\subsection{Calculating DM and RM}
The DM is the line-of-sight integral of the electron density ($n_e$). For a given sightline in the simulation, we compute this by replacing the integral with a discrete sum between the FRB position in the inclined galaxy and $r_{\rm SF,99}$ of the galaxy. We use N steps of d$l$=20 pc step size, and the integration takes the closest cell to the position on the sightline in every step:
\begin{equation}
    {\rm DM} = \sum_{i=0}^{N} n_{e,i} {\rm d}l,
    \label{eq:DM}
\end{equation}
where $i$ is the index for the $N$ steps and $n_{e,i}$ is the electron density at each step. We also assume that the origin of the FRB is a point source, and its emission originates from scales smaller than 20 pc. We find that it is sufficient to use 1000 sightlines per galaxy, as two runs of the pipeline give the same overall distribution at a given redshift. We chose to rotate the galaxies randomly for each sightline, as a lower number of rotations can lead to biases, based on which galaxies get rotated to be face-on. We find that DM increases with increasing integral path length. When integrating to $r_{\rm SF,99}$ and 2 x $r_{\rm SF,99}$, 68 \% of our sightlines have a difference below 50 $\dm$ in DM and below 3 $\radm$ in RM. We note that the maximum difference can be as high as $\sim$ 8000 $\dm$ and $\sim$ 250\,000 $\radm$. However, we argue that as in Section \ref{Results} we use the median and the width of the distribution containing 68\% of sightlines, our results will not be significantly affected. For further discussion on the effect of different parameter choices (integral length and integral step size), see Appendix \ref{appendixA}. 

The RM is calculated similarly as DM:
\begin{equation}
    {\rm RM} = k \sum_{i=0}^{N} B_{||,i} n_{e,i} {\rm d}l,
    \label{eq:RM}
\end{equation}
where $B_{||,i}$ is the line of sight magnetic field at each step. The same parameter tests were performed as during DM calculations (see Appendix \ref{appendixA}).

\subsection{Pipeline summary}
In summary, we selected 16\,500 galaxies,  rotated each to a random inclination 1000 times and randomly chose 1000 FRB positions in each galaxy. Then we calculated the DM and RM contribution by the host galaxy for each FRB sightline.

For each sightline we save the galaxy ID, the position of the FRB, the $n_e$ of that cell, the calculated DM and RM, and the applied rotation matrix. For each galaxy we save the stellar mass, star formation rate, $r_{\rm SF,99}$, the properties of its magnetic field (average total magnetic field of the disk, and the average of its components -- azimuthal, radial and vertical), and the radial and vertical profile for the magnetic field and its components (for details see Appendix \ref{appendixB}). We also calculate these profiles for $n_e$ and $\rho$.

\section{Results}

\label{Results}

In this section, we calculate the DM$_{\rm host,rf}$ and RM$_{\rm host,rf}$ PDFs\footnote{All histograms presented in this work have been normalized to unit area.} for different subsets of galaxies, and investigate how the properties of these distributions change with redshift and other galaxy properties. Our goal is to provide an estimate for the host contribution of an observed FRB assuming we have information on its redshift, and/or its host galaxy properties. First we show the redshift dependence of the full sample, then the differences between star-forming and red galaxies (i.e. `quenched'), and how the distributions change with stellar mass. Assuming we also know the inclination of the observed host galaxy, or the $b_{\rm offset}$, we provide PDFs for different inclinations and offsets.
We summarize these trends in Section \ref{Trends_summary}.

\begin{figure}
    \centering
\includegraphics[width=8.8cm]{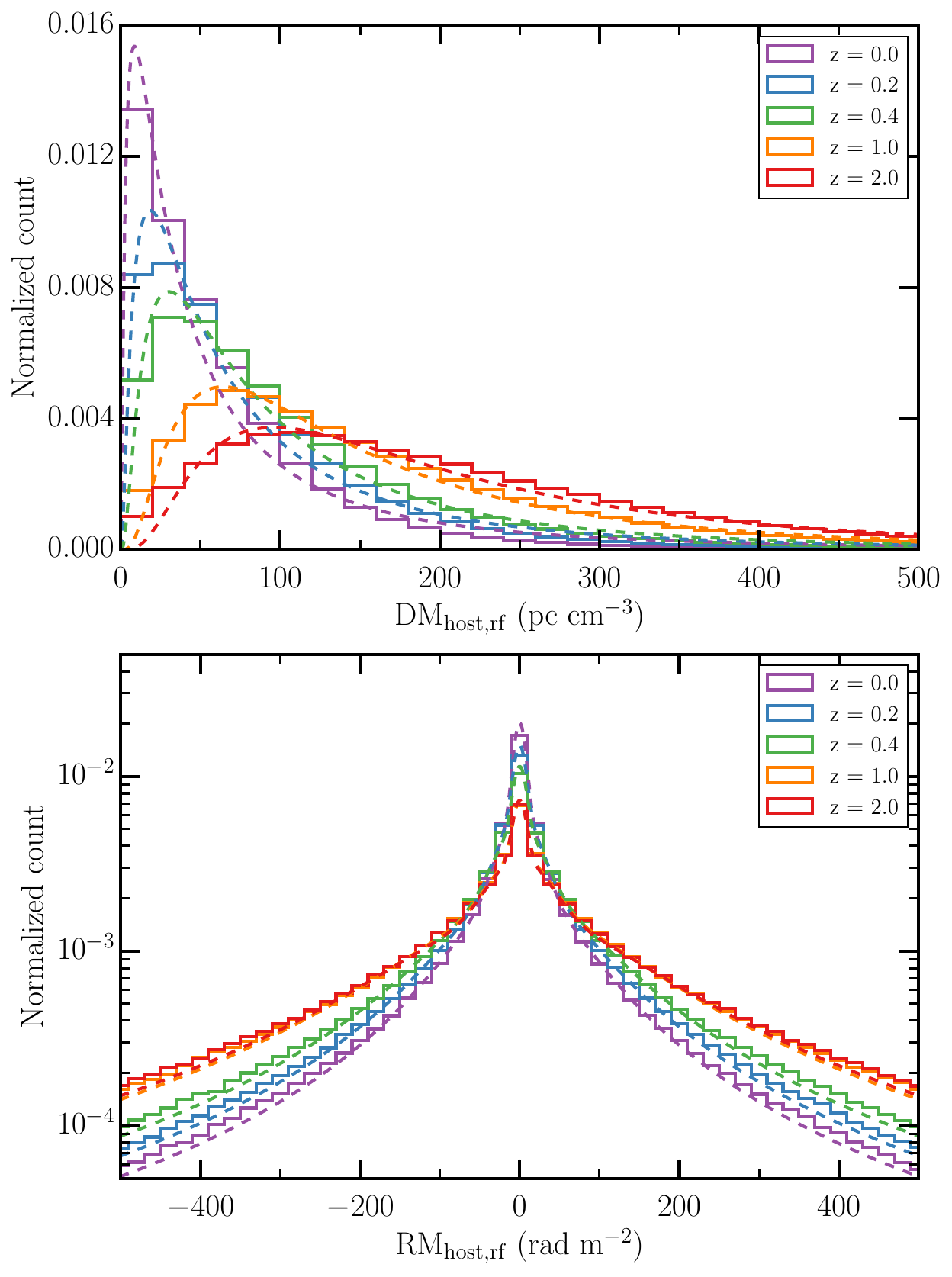}
    \caption{\textbf{Top:} The PDF of the rest frame DM contribution of host galaxies (of all galaxy types), using FRBs at different redshifts. The log normal fits are shown by the dashed lines, and the colors correspond to the different redshifts. 
    \textbf{Bottom:} The PDF of the rest frame RM contribution of host galaxies, using FRBs at different redshifts. 
    The PDFs are modeled by the sum of a Lorentzian and two Gaussian functions, and are shown by the dashed lines, and the colors correspond to the different redshifts. 
    }
    \label{fig:DM_dist}
\end{figure}

\subsection{Analysis of the sightlines}

We analyzed the calculated DM$_{\rm host,rf}$ and RM$_{\rm host,rf}$ in two ways, using their simple statistical properties and fitting PDFs to the distributions. We then fit how the statistical properties and the distribution parameters of DM$_{\rm host,rf}$ and RM$_{\rm host,rf}$ change across redshift, so that we can provide the estimated DM$_{\rm host,rf}$ and RM$_{\rm host,rf}$ distributions for observed FRBs at any redshift (for more details, see Appendix \ref{appendix:z_evol}). We list the parameters of these fits in Tables \ref{tab:DM_param} and \ref{tab:RM_param_fit}. 

First, we investigated the statistical properties of the lines of sight to provide straightforward estimates that enable simple comparisons to observed DMs and RMs. We calculated the median and the 1$\sigma$ width of the distributions, which we define as the difference between the 84th and 16th percentile ($w_{\rm DM, rest}$ and $w_{\rm RM, rest}$, containing 68\% of the data). The median RM = 0.00 $\pm$ 0.03 $\radm$ at every redshift.

\begin{figure}
    \centering
\includegraphics[width=8.8cm]{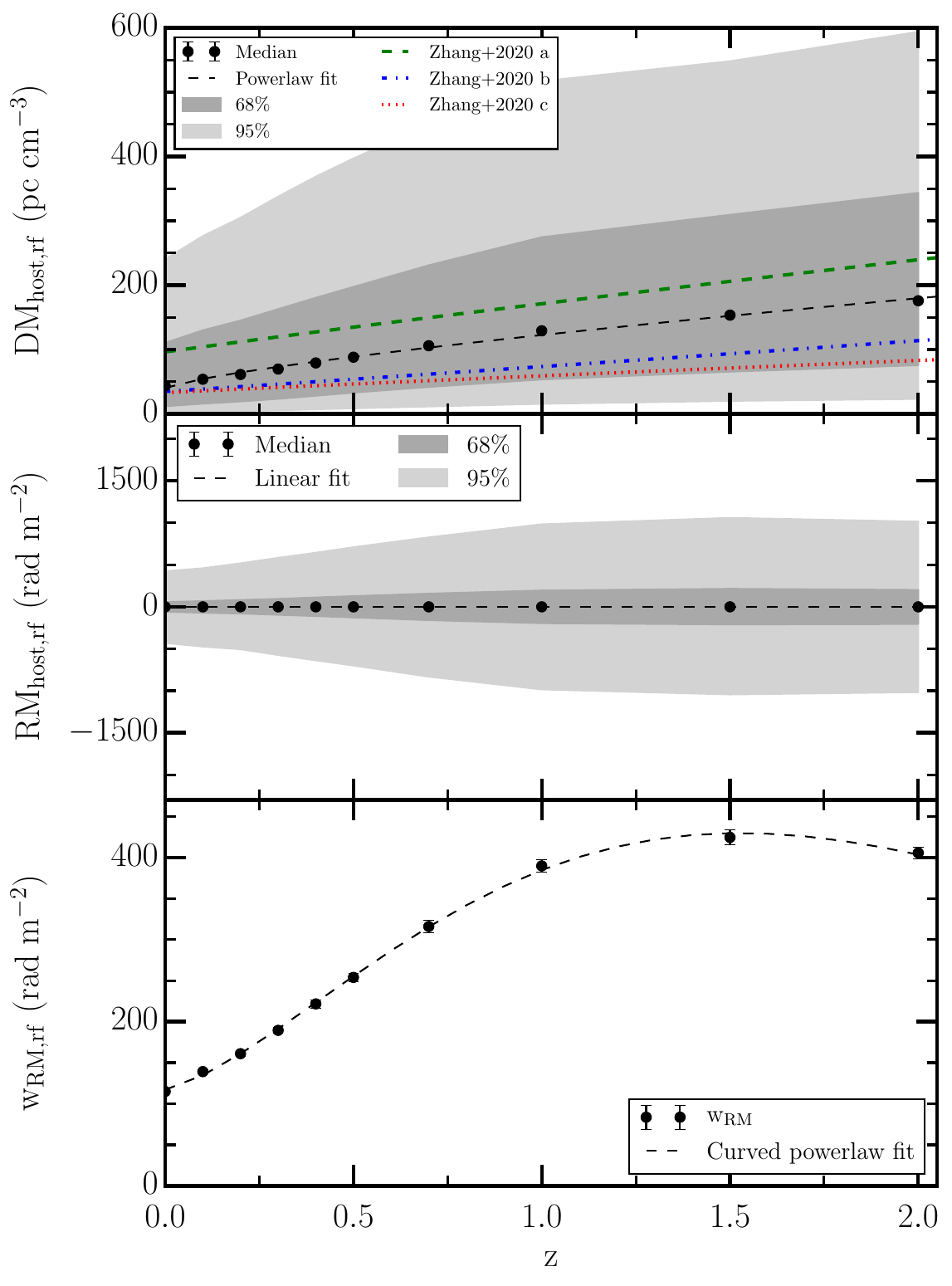}

    \caption{\textbf{Top:} The median of DM$_{\rm host,rf}$ 
    as a function of redshifts for all galaxy types. The shaded regions correspond to the 1$\sigma$ and 2$\sigma$ width of the distributions. 
    The power law fit is plotted as a dashed black line.
    For comparison, we also plotted the results from \cite{2020ApJ...900..170Z}, in the case of host galaxies like the host of FRB180916 (a), galaxies similar to the hosts of non-repeating FRBs (b) - hosts known at the time of \cite{2020ApJ...900..170Z}'s paper, and galaxies similar to the host of the repeating FRB121102 (c). For further details on the comparison, see Section \ref{Dis:DM_comp}. \textbf{Middle:} The median of RM$_{\rm host,rf}$ as a function of redshift.
    The shaded regions correspond to the 1$\sigma$ and 2$\sigma$ width of the distributions.
    \textbf{Bottom:} The 1$\sigma$ width of the distributions of RM$_{\rm host,rf}$ (w$_{\rm RM,rf}$) as a function of redshift, and the curved power law fit (black dashed line). The errors are from the bootstrap method.}
    \label{fig:DM_z_ev}
\end{figure}

Next, we fit mathematical PDFs to provide a way for the reader to have a more accurate model of DM$_{\rm host,rf}$ and RM$_{\rm host,rf}$. We calculate the PDFs individually for each redshift, and for each PDF we take into account the DM and RM of every FRB at the given redshift. The PDFs for redshifts 0.0, 0.2, 0.4, 1.0, and 2.0 are shown in Fig. \ref{fig:DM_dist}. We only show the PDFs for half of the snapshots in our analysis to make the plot clearer, but we will show in Section \ref{Res:DMvsz} that the parameters of the distributions change smoothly with redshift (see and Fig. \ref{fig:DM_z_ev}).

We found that the PDF of DM$_{\rm host,rf}$ can be described by a log normal distribution:
\begin{equation}
    f_{\rm DM, host}(x) = \frac{1}{\sqrt{2\pi}x\sigma}  \exp(-(\ln(x)-\mu)^2/(2\sigma^2)),
    \label{Eq:DM_PDF}
\end{equation}
parametrized with $\mu$ and $\sigma$, which are the mean and standard deviation of the DM's natural logarithm, and not the DM itself. It is worth noting that the tail of the distribution is slightly overestimated by the log normal fit. Other distributions that can be fitted to the data include the gamma function, however this overestimates the 16th percentile by a factor of 4. The difference between the median of the actual PDF and the median from the log normal fit (exp($\mu$)) is below 10 pc cm$^{-3}$ at all redshifts.

The RM distribution is symmetric (shown in Fig. \ref{fig:DM_dist} bottom), and can be fitted by the sum of one Lorentzian and two Gaussian functions, similar to the results of \cite{2018MNRAS.477.2528B} (for more details see Section \ref{Comparison:rm}):
\begin{multline}
    f_{\rm RM,host} (x) = a_1\cdot \left[\frac{\gamma^2}{\pi \gamma (\gamma^2 + x^2)}\right] \\
+ a_2 \exp{\left[-\frac{1}{2} \left(\frac{x} {\sigma_1}\right)^2\right]} 
+ a_3 \exp{\left[-\frac{1}{2}\left(\frac{x}{\sigma_2}\right)^2\right] },
\label{Eq:RM_PDF}
\end{multline}
where $a_1$, $a_2$ and $a_3$ are normalization fractions, $\gamma$ is the parameter of the Lorentzian function, and $\sigma_1$ and $\sigma_2$ are parameters of the Gaussian functions. We note that the exact definition of the function is different from that of \cite{2018MNRAS.477.2528B}, for example we did not define a variable for the mean, but assumed it to be 0 $\radm$. We also experimented with a purely Lorentzian fit, and different combinations of Gaussian and Lorentzian fits, but the combination from \cite{2018MNRAS.477.2528B} captured the long tails and peak of the distribution best. The $w_{\rm RM, rf}$ of the fitted PDFs differs from that of the real PDFs by less than 20 rad~m$^{-2}$.

We note in case we would have used a longer integral path length (e.g. extending to 2 x $r_{\rm SF,99}$), our estimated median DM might be higher by $\sim$ 20 \% and $w_{\rm DM, rf}$ by 10\%. Our RM distribution would be less affected, with $w_{\rm RM, rf}$ only increasing by a few \%. The smaller difference in RM is most likely caused by frequent changes in the magnetic field direction in the galaxy's halo or in the circumgalactic medium.

\subsection{Redshift evolution of DM$_{\rm host,rf}$ and RM$_{\rm host,rf}$}

\label{Res:DMvsz}

In this section, we investigate how the rest frame distributions of DM$_{\rm host,rf}$ and RM$_{\rm host,rf}$ change across redshift, considering all galaxies (regardless of their galaxy type or other properties).

We find that the median of DM$_{\rm host,rf}$ and the $w_{\rm DM, rest}$ both increase from $z=0$ to $z=2$ following a power law, by a factor of $\sim$4 and $\sim$2.5, respectively. We find that the RM distribution also becomes broader with redshift, with $w_{\rm RM, rest}$ increasing by a factor $\sim$ 3.5 up to $z = 1$. At $1<z<1.5$, $w_{\rm RM, rest}$ flattens, and starts to decrease at $z=1.5$. The $w_{\rm RM, rest}$ can be fitted by a curved power law as a function of redshift. For more details on the functional forms, see Appendix \ref{appendix:z_evol}. As we might observe FRBs at higher redshifts in the future with the SKA \citep{2020MNRAS.497.4107H}, we checked if our derived forms for DM$_{\rm host,rf}$ and RM$_{\rm host,rf}$ evolution at $2<z\leq5$ hold for higher redshifts, and found some deviations, thus our results presented in this paper should be used only up to $z=2$\footnote{More information can be found in Chapter 5 of \cite{phdthesis}.}. As the snapshots in the simulation at $z>2$ contain significantly fewer galaxies than at $z \leq 2$ (see Table \ref{tab:gal_select}), we do not include them in the paper. 

We investigate the reason behind these redshift evolutions. In the case of DM, the increase can be caused by either an increase in $n_e$ or in the integral path length ($L$). Additionally, RM also depends on the strength and direction of the line-of-sight magnetic field. To be able to interpret the role of $n_e$ and $B$ in the resulting DM and RM, we calculated the electron density, gas density, and magnetic field radial and vertical profiles, and other properties of the magnetic field. To calculate the magnetic field properties, we followed the methods in \cite{2017MNRAS.469.3185P} and \cite{2018MNRAS.481.4410P}, who studied Milky-Way-like galaxies from the Auriga simulations. The range of magnetic field strengths are in agreement with what we have found in the examined galaxies (see Appendix \ref{appendixB}).

\subsubsection{DM - electron density changes with redshift}
\label{DM_vs_sf}

 \begin{figure}
    \centering
    \includegraphics[width=8cm]{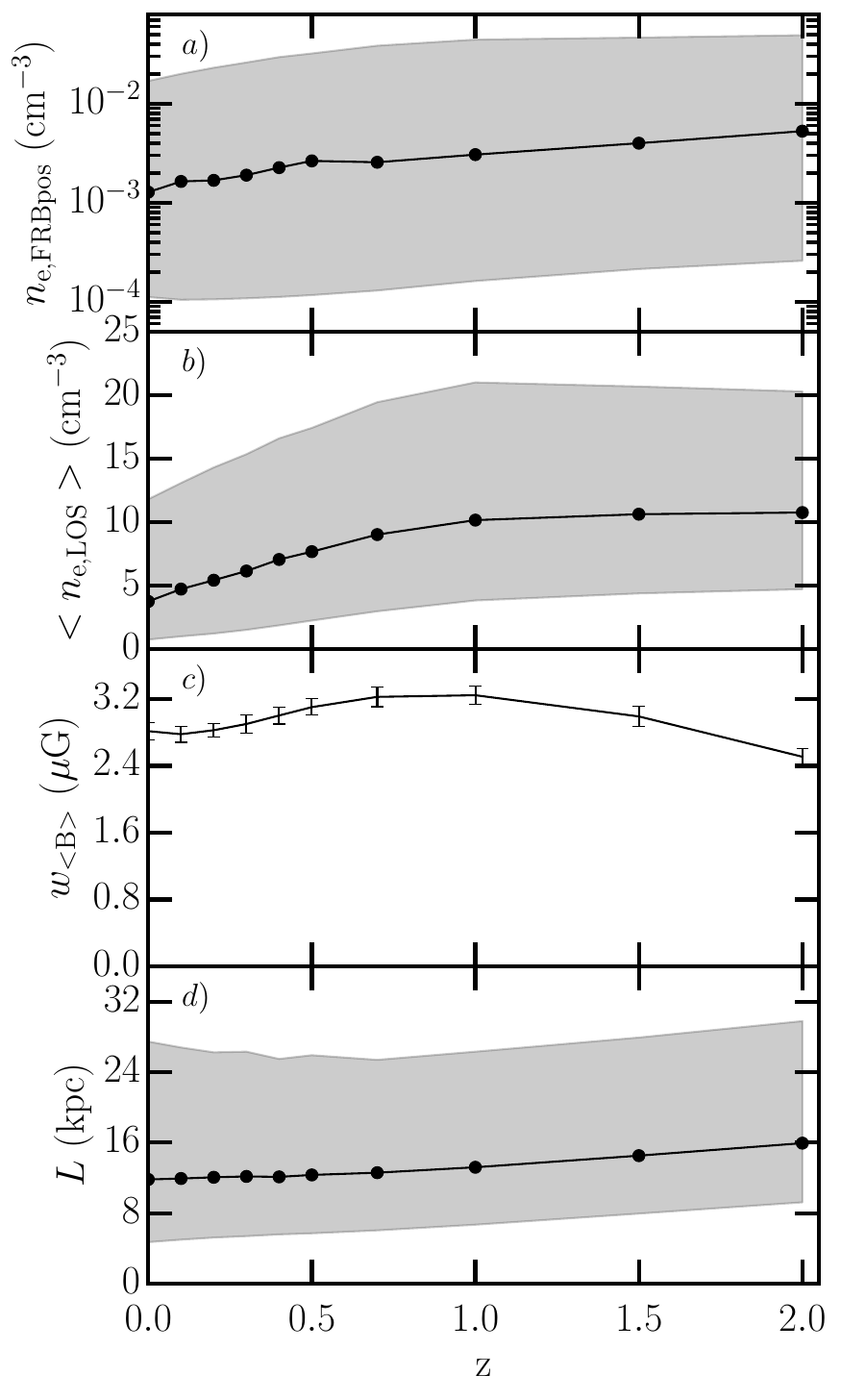}

    \caption{We show how the median of different properties and their ranges of values between the 16th and 84th percentile change with redshift, in order to help to understand what drives the DM and RM trends. \textbf{a:} The $n_e$ of the starting position of the FRBs ($n_{\rm e,FRBpos}$) as a function of redshift.
    \textbf{b:} The average $n_e$ along the line of sight ($\langle n_{\rm e,LOS} \rangle$) as a function of redshift.
    \textbf{c:} The distribution width of the average $B_{\rm ||}$ along the line of sight ($\langle B_{\rm ||, LOS}\rangle$) as a function of redshift.}

    \label{fig:property_trends}
\end{figure}

Based on observations (e.g. \citealt{2017MNRAS.465.3220K}), galaxies at higher redshift have higher star formation rates, subsequently higher electron densities, such as a factor of 5 greater at $z=1.5$ than at $z=0$. In the simulation, we find that the median SFR of the selected galaxies is also larger by a factor of $\sim$9 at $z=2$ compared to $z=0$, similarly to observed galaxies, and the number of quenched galaxies increases from a few \% to $\sim$ 20 \%. The 1$\sigma$ width of the SFR distribution also increases by a factor of $\sim$10 from $z=0$ to $z=2$. Furthermore, we find that from $z=0$ to $z=2$, the median $n_e$ of the locations of FRBs ($n_{\rm e,FRBpos}$) increases by a factor of $\sim$4, and the range of values increases by a factor of $\sim$3 (shown in panel a of Fig. \ref{fig:property_trends}). As the 1000 positions for each galaxy were randomly selected, their properties can tell us about the general environment in these galaxies. In summary, we find that $n_{\rm e,FRBpos}$ significantly increases with redshift, which can cause a significant increase in DM.
 
Another cause of larger DMs at higher redshifts may be a longer $L$ through the host galaxy. As we define the end point of the integral as the radius of the galaxies ($r_{\rm SF,99}$), this increase can be due to an increase in the radius. We find that indeed, the median of $r_{\rm SF,99}$ and the path length increases by a factor of 1.3 from $z=0$ to $z=2$ (shown in panel d of Fig. \ref{fig:property_trends}), noting that at $z<1$ it only changes by a few \%.

We can calculate the average electron density along an FRB sightline as we know the path length ($L$), if we rearrange Eq. (\ref{eq:DM}): $\langle n_{\rm e,LOS}\rangle$ = $\frac{\rm DM}{L}$ cm$^{-3}$. We find that this increases with redshift up to $z=2$ by a factor of 2.9, but between $z=1$ and $z=2$, it only increases by a few \% (shown in panel b of Fig. \ref{fig:property_trends}). We previously saw that at $z>1$ the path length increases, which could explain why DM keeps increasing despite $\langle n_{\rm e,LOS}\rangle$ only increasing slightly. This could mean that at higher redshifts the galaxies have star-forming gas out to a larger distance compared to lower redshifts, but $n_e$ is low in the outskirts of galaxies, causing us to measure a lower $\langle n_{\rm e,LOS}\rangle$ than expected from DM. This idea is supported by the fact that the difference between the radius containing 95\% and 99\% of the star-forming gas in galaxies are larger at higher redshifts. At $z=2$, the difference is 40\% of $r_{\rm SF,99}$ on average, while it is only 20\% at $z=0$.

Thus, we conclude that the increasing trend of median DM with increasing redshift can be mainly explained by higher electron densities caused by higher SFR, but an increase in path lengths also contributes to a lesser degree. Note that, a factor of two longer path length would result in only 20\% increase in median DMs at each redshift bin (see Appendix \ref{appendixA}).

\subsubsection{RM - magnetic field changes with redshift}

\begin{figure}
    \centering
    \includegraphics[width=8.8cm]{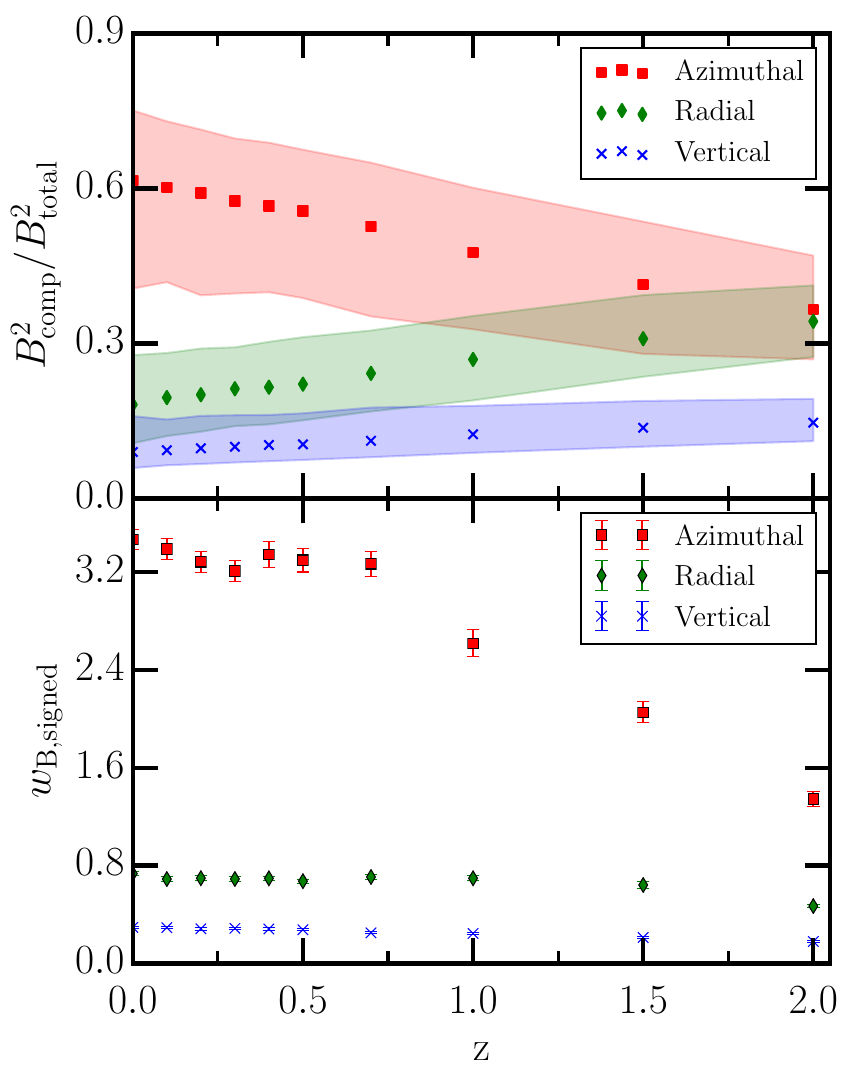}
    \caption{\textbf{Top:} The median of the different magnetic field components' ratios to the total magnetic field ($B_{\rm component}^2/B_{\rm total}^2$, where the components are azimuthal, radial, and vertical) as a function of redshift.
    \textbf{Bottom:} The distribution width of the average of the signed magnetic field strength maps as a function of redshift. 
    }
    \label{fig:B_comp}
\end{figure}

While the $w_{\rm RM,rf}$ increase at $z<1$ can be partly attributed to increasing electron densities, the decrease at $z>1$ is not seen in DM (which includes changes in $n_e$ and $L$), thus it has to be caused by the magnetic field; either from changes in its strength or structure. We note that a longer $L$ has little impact on our results (see Appendix \ref{appendixA}), only causing a $w_{\rm RM,rf}$ larger by $\sim$2\%.

We calculate the average total magnetic field strength in the disk of the galaxies by averaging each galaxy's total magnetic field strength map (see Appendix \ref{appendixB}). We find a small increase ($\sim$ 20\%) from $z=0$ to $z=1$ and a small decrease ($\sim$ 20\%) from $z=1$ to $z=2$ in the total magnetic field strength. Although this is only a small factor, we also need to consider the factor of 4 increase we see in DM. Until $z=1$, both $n_e$ and the average B field strength increase, so $w_{\rm RM,rf}$ also increases. For $z>1$ only $n_e$ increases and the average B field strength decreases, which can be partly the reason why the $w_{\rm RM,rf}$ stops its steep increase. However, it can not fully explain the decreasing RM, thus we also need to explore the magnetic field structure.

If both DM and RM are measured for the same FRB sightline: $\langle B_{\rm ||, LOS}\rangle=~1.23~\frac{\rm RM}{\rm DM} \mu$G, assuming a constant magnetic field with no field reversals, resulting in a lower limit of the actual magnetic field strength. The distribution width of $\langle B_{\rm ||, LOS}\rangle$ increases until $z=1$ by a factor of 1.15, and then decreases by 20\% from $z=1$ to $z=2$, becoming smaller than at $z=0$ (shown in panel c of Fig. \ref{fig:property_trends}), in contrast to the average total magnetic field strength which decreases to the same value it was at $z=0$. Since we estimated  $\langle B_{\rm ||, LOS}\rangle$ with the assumption of no field reversals, one possible explanation for the decrease in $\langle B_{\rm ||, LOS}\rangle$ is the presence of more reversals or random field at $z>1$ compared to $z<1$. As the changes in field direction along the line of sight can cancel each other, more field reversals and random fields can also cause lower $w_{\rm RM,rf}$. On the other hand, the presence of a strong large-scale ordered field can cause high |RM| values, which would increase $w_{\rm RM,rf}$. 

To investigate whether the number of galaxies with ordered field changes across redshift, we separate the magnetic field strength into azimuthal, radial and vertical components. We generated magnetic field strength maps of the field components, with the sign of the field preserved (thin projections of the magnetic field components centered on the mid-plane for each magnetic field component). We investigate the presence of 1) large-scale $B$ fields, and 2) $B$ field reversals or random fields in the galaxies. By calculating the average magnetic field strength of the face-on maps with preserved field signs (from here on called signed magnetic field strength), we can separate the two cases. If its absolute value is a few $\mu$G the galaxy has a large-scale field without reversal, and if this value is close to 0 $\mu$G the field is random or has a field reversal. The distribution of these signed magnetic field strengths is symmetric about 0 $\mu$G. In the case of the radial and vertical field, the width of this distribution is low ($<1\mu$G) and does not change with redshift. This is due to their lower field strength compared to the azimuthal component, and that they are more likely to have random field. In the bottom plot of Fig.\ref{fig:B_comp} we show how the width of the signed azimuthal magnetic field strength distribution ($w_{\rm B,signed}$) changes with redshift. At $z\geq0.7$ it quickly decreases. At $z=0$ we find 26\% (53\%) of galaxies have a mean azimuthal field larger than 2 $\mu$G (1 $\mu$G), and at $z=2$ we find only 6\% (22\%) of galaxies have a mean azimuthal field larger than 2 $\mu$G (1 $\mu$G). We find that the number of galaxies with reversals in the azimuthal field or with azimuthal fields dominated by random fields increases towards higher redshift. 

In summary, with increasing redshift, we find a slight, 20\% decrease in total field strength, and we find that galaxies have a less ordered magnetic field, where random fields and reversals dominate, causing lower $w_{\rm RM,rf}$.

\subsection{Dependence of DM$_{\rm host,rf}$ and RM$_{\rm host,rf}$ on different galaxy properties}

\begin{figure*}[!h]
    \centering
    \includegraphics[width=16cm]{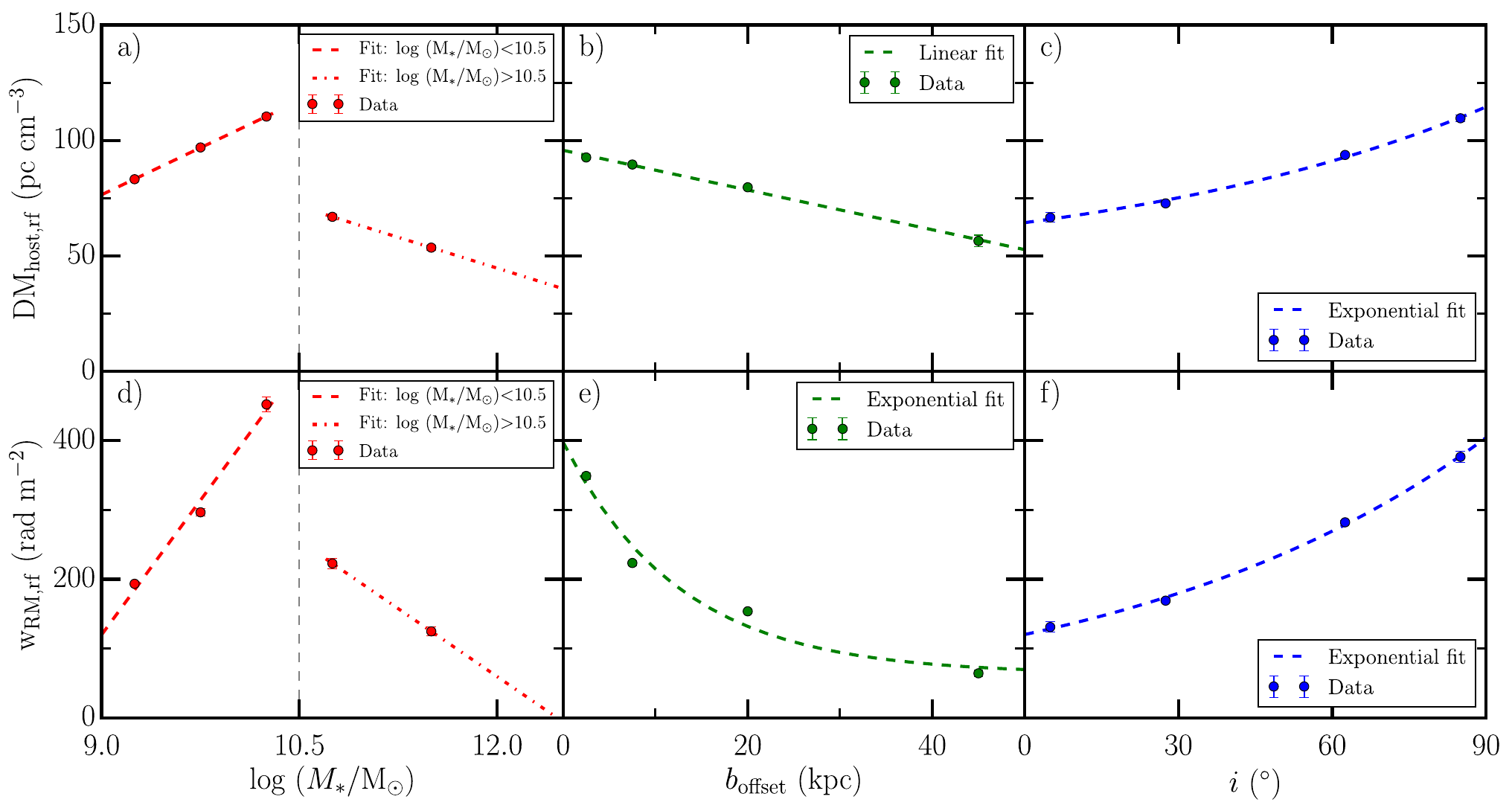}
    \caption{The top row shows how the median DM$_{\rm host,rf}$ changes as a function of stellar mass (left), $b_{\rm offset}$ (middle) and galaxy inclination (right). The bottom row shows the same for the RM$_{\rm host,rf}$ distribution width ($w_{\rm RM, rf}$). This figure shows these trends for galaxies at $z=0.5$. We note that the sudden drop at $\logM=10.5$ is due to the change in supermassive black hole feedback mechanism in the TNG50 simulation (see Section \ref{AGN_feedback}).}

    \label{fig:fits_all}
\end{figure*}

\label{Res:prop}

We investigate if the DM$_{\rm host,rf}$ and RM$_{\rm host,rf}$ are different in galaxies with different properties, such as star-forming activity, stellar mass, the position of the FRB -- specifically the projected offset from the center of galaxies ($b_{\rm offset}$, see Section \ref{Methods:boffset}), and the inclination of the host galaxy. These are all properties that should be possible to derive for localized FRBs, and observed FRB host galaxies could be separated based on them in the future. Star forming and red (i.e. quenched) galaxies can be separated by their color–magnitude diagram \citep{2004ApJ...608..752B}, their inclination can be measured by the ratio of their major and minor axis (assuming circular disk galaxies).

To separate the star-forming and quenched galaxies in our TNG50 galaxy sample, we adapt a method from \cite{2019MNRAS.485.4817D}, who derived the Main Sequence (MS) of galaxies in TNG100. The MS is a tight relation between the stellar mass and SFR of star-forming galaxies based on observations (e.g. \citealt{2007ApJ...660L..43N}), and this relation also scales with redshift. The stellar mass of galaxies is readily available in the TNG data release. The $b_{\rm offset}$ can be calculated from the position data, and we define the inclination angle as the angle between the total angular momentum vector of the galaxy's stars and the line of sight (see Section \ref{Methods:inclination}).

We separated star-forming and red galaxies, divided the galaxies into five stellar mass,  four projected offset, and four inclination bins. The shapes of the distributions of the subsets of sightlines are the same as the ones for the full sample.
In Figs. \ref{fig:DM_subsets} and \ref{fig:RM_subsets} we show the median DM and $w_{\rm RM,rf}$ as a function of redshift from each subset group, all of which can be fitted by a power law and a curved power law, respectively. The fitted parameters for the different groups are listed in Table \ref{tab:DM_param}.

While the shape of the PDFs remains the same, the parameters of the PDFs change with different properties, which we describe below, and summarize in Fig. \ref{fig:fits_all} (shown for galaxies at $z=0.5$).

\subsubsection{Star-forming and quenched galaxies}

We find that star-forming galaxies have a wider DM and RM distribution, and a higher median DM and $w_{\rm DM,rf}$ by a factor of $\sim$ 3 and $\sim 6$, respectively, compared to quenched galaxies (see e.g. Figs. \ref{fig:DM_subsets}a and \ref{fig:RM_subsets}a). The same trend can be seen at all redshifts. We only have a few red galaxies in our sample, as most of the sampled galaxies are star-forming (97\% in total, with 95\% at $z$=0 and 99\% at $z$=2). As a result, at $z\ge$ 1.5 we have fewer than ten quenched galaxies in a redshift bin, thus the trends for quenched galaxies can not be seen clearly.
    
To test if the different distribution shape is caused by the small sample size of the quenched galaxies, we randomly selected the same number of galaxies from the star-forming sample as there are in the quenched sample. The resulting DM and RM PDFs closely resemble the PDFs of all star-forming galaxies, suggesting the quenched galaxies have a different PDF unrelated to the sample size. On the contrary, the difference in PDFs might be due to the fact that the structure and properties (e.g. SFR, stellar mass) of star-forming spiral and red elliptical galaxies are different (see e.g. \citealt{2014ARA&A..52..291C} for a review), thus their electron densities can also differ. Furthermore, the magnetic field strength and structure of different galaxies can vary: ellipticals might have weaker large-scale fields because they lack a differentially rotating disk \citep{2012SSRv..166..215B}.

Based on optical spectroscopic observations, $n_e$ is correlated with the SFR of galaxies \citep{2015MNRAS.451.1284S,2017MNRAS.465.3220K} A high SFR density increases the number of young massive stars, which in turn, by stellar winds and shocks will have a larger energy input to HII regions, and the diffuse ionized gas, increasing the electron density of galaxies. However, the red galaxies have already started or gone through quenching, their cold star-forming gas is depleted, and the electron densities decreased, which causes the lower DM.

We do find a difference in $n_e$ in the simulation: the average $n_{\rm e,FRBpos}$ is 4 times larger in star-forming galaxies than in quenched galaxies. The $\langle n_{\rm e,LOS}\rangle$ is 5 times smaller for quenched galaxies.  However, we also find that the path length in quenched galaxies is usually two times larger on average. But this is not surprising, as the quenched galaxies have large stellar masses ($\sim$ 10 times larger than average SF galaxies) and galaxy sizes ($\sim$ 3 times larger), similar to large ellipticals. Only a few quenched galaxies ($\sim$ 15\%) have radii below 5 kpc, and we found that 95\% of these galaxies have log (M$_*$/M$_{\odot}$) $<$ 10.5.
So in summary, while the path lengths are larger, the electron densities are lower in red galaxies, which causes their smaller DM contributions.

We find that the width of the distribution of $\langle B_{\rm ||,LOS}\rangle$ along the line of sight in star-forming galaxies is 1.5 times larger than that of red galaxies. This suggests a difference in the magnetic field properties of quenched and star-forming galaxies, in addition to the differences in $n_e$. 
It has been shown that early-type galaxies have an irregular gas distribution in the TNG simulations (due to the interaction of AGN feedback with the surrounding gas), and the magnetic field traces the gas density closely and is therefore also irregular \citep{2018MNRAS.480.5113M}. We see similar trends in our sample: quenched galaxies are often not well described by a double exponential radial magnetic field strength profile, with a maximum $B$ field strength at larger radii instead of at the center of the galaxy. Their central magnetic field strengths are lower than those for star-forming galaxies. They also exhibit "wiggles" (magnetic field strength fluctuating with radius) in the radial profiles. Similar features can be seen in the gas density profiles, but the $n_e$ profiles can still be described by a double exponential function. This irregularity in the magnetic field can cause lower $w_{\rm RM,rf}$.  

\subsubsection{Stellar mass}  
In Fig. \ref{fig:fits_all} a) and d) we show that the median of DM and $w_{\rm RM,rf}$ increases linearly with stellar mass, up to very high mass galaxies ($\logM$ > 10.6 at $z<1$, $\logM$ > 11 at $z>1$). For more massive galaxies, the DM and RM PDF becomes narrower, the median DM drops to half of its value and $w_{\rm RM,rf}$ narrows. 
At every redshift, the $w_{\rm RM,rf}$ decreases linearly after the drop, while the median DM does not always decrease linearly. The difference between the fits of the redshift dependence of the different stellar mass bins are more visible in the case of RM. At low redshifts ($z$<0.3), the galaxies with lower stellar mass have a higher median DM and $w_{\rm RM,rf}$ than galaxies with larger masses.

The increase can be explained by the MS relation between the stellar mass and SFR, as SFR increases with stellar mass, thus the electron densities do too. The sudden drop occurs in massive galaxies with $\logM$ > 10.5, which are analogous to large ellipticals with low electron densities and star formation rates. In the simulation, this is likely caused by the change in the SMBH feedback mechanism in galaxies with a stellar mass of around $\logM$ $\sim$ 10.5 (described in \citealt{2017MNRAS.465.3291W}, \citealt{2018MNRAS.479.4056W}, for more details see Section \ref{AGN_feedback}). This subsequently lowers their gas densities, electron densities and SFR. The $\langle n_{\rm e,LOS} \rangle$ in galaxies with lower stellar masses is lower by a factor of 2. 
Although the path length increases as a function of stellar mass, the $n_e$ decrease is more prominent, producing low DMs in very massive galaxies. 

The quenching process also has an effect on the magnetic field of the galaxies, as they start to have a different radial $B$ field profile, with lower magnetic field strength at the center and a more irregular magnetic field. This, in combination with lower electron densities, causes a lower $w_{\rm RM,rf}$.

We investigated if the massive galaxies in our sample are truly undergoing quenching. We find that 70\% to 90\% (corresponding to different redshifts) of galaxies we classified as quenched have a stellar mass above $\logM$ $\sim$ 10.5. However, only <30\% of $\logM > 10.5$ galaxies are classified as quenched. This is because we only consider galaxies red if they are 1 dex below the MS. So while for galaxies with smaller stellar masses we find an average SFR close to the SFR of the main sequence, for massive galaxies we find SFRs that lie below the MS. These galaxies might have already started the quenching process. The fraction of quenched galaxies decreases at higher redshift, and more massive galaxies are classified as star-forming. We also note that $<$4\% of star-forming galaxies are massive and very active in star formation: with $\logM > 11$ and a mean SFR $\sim$ 10 M$_{\odot}$/yr. As we showed in Section \ref{DM_vs_sf}, $n_e$ is correlated with SFR, and a lower SFR can explain lower DMs.

\subsubsection{Projected offset of FRBs} 
 We find that the median DM$_{\rm host,rf}$ and $w_{\rm RM, rf}$ decrease with larger FRB offsets. The magnitude of the change from $b_{\rm offset} = 0$ kpc to $b_{\rm offset} = 50$ kpc is more significant in the case of $w_{\rm RM, rf}$ (decrease by a factor of $\sim$8), than the median DM$_{\rm host,rf}$ (factor of $\sim$2). In Fig. \ref{fig:fits_all} b), we show that the median of DM decreases linearly, and in Fig. \ref{fig:fits_all} e) we see how $w_{\rm RM, rf}$ decreases exponentially. 
 
This is not surprising, as we expect both $n_e$ and $B$ field to change as a function of distance from the center of the galaxy (see e.g. \citealt{2007A&A...470..539B}). As a result, the DM and RM distributions can also be affected.

\begin{figure*}
    \sidecaption
    \includegraphics[width=12cm]{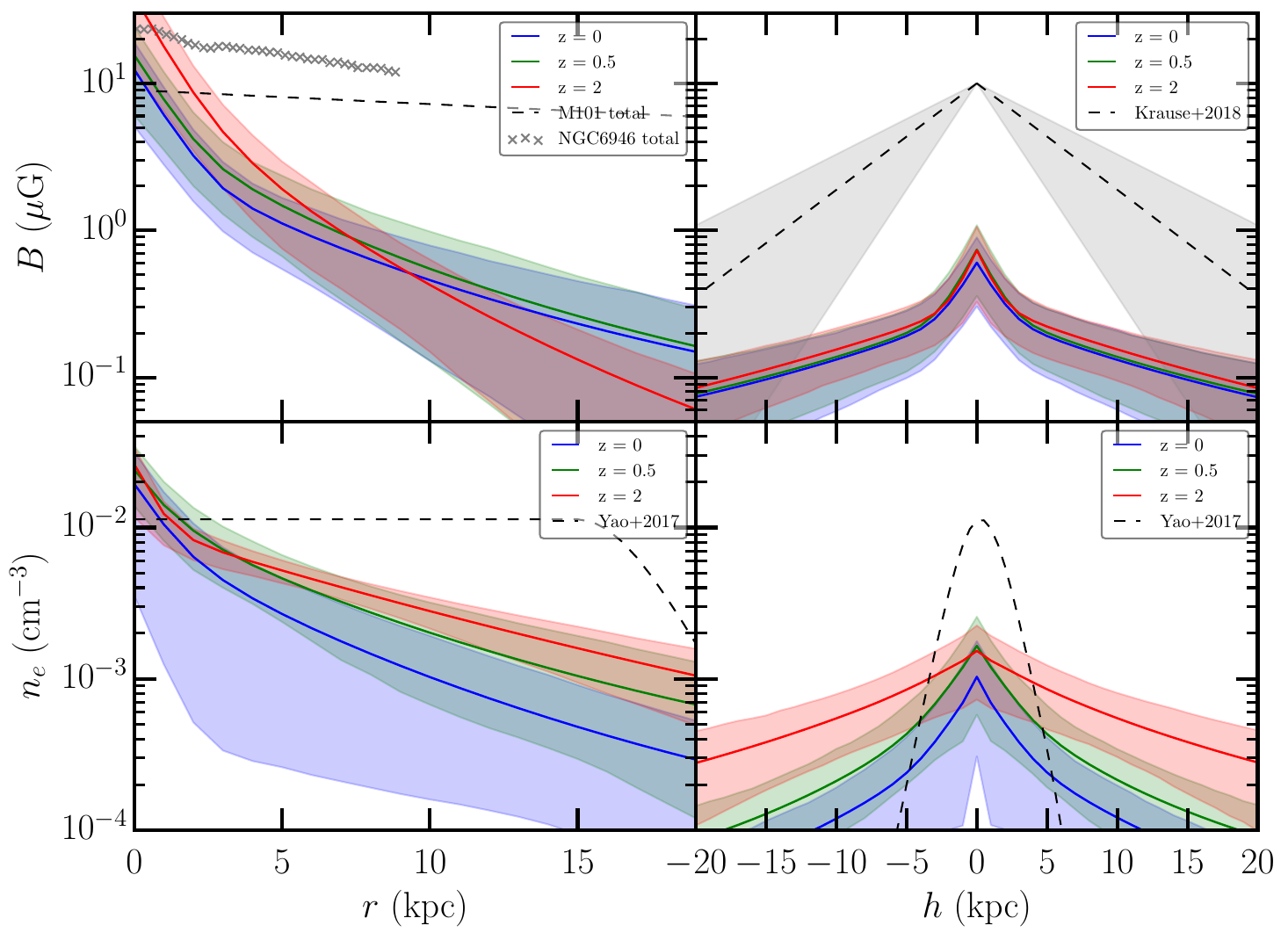}
    \caption{\textbf{Top row}: The averaged radial and vertical profile of B of galaxies at various redshifts, and their 1$\sigma$ width. In the top left panel, we show the total magnetic strength profile of NGC6946 \citep{2007A&A...470..539B} and M101  \citep{2016A&A...588A.114B}. In the top right panel, we show the average vertical $B$ profiles of CHANG-ES galaxies \citep{2018A&A...611A..72K} with a scale-height of 6 $\pm$ 3 kpc. \textbf{Bottom row}: The averaged radial and vertical $n_{e}$ profile  of galaxies at various redshifts, and their 1$\sigma$ width. We overplotted the $n_{e}$ profiles of the thick disk from \cite{2017ApJ...835...29Y}.}
    \label{fig:B_profile}
\end{figure*}

We find that in the simulation the electron density is higher and the magnetic field is the strongest at the center of galaxies (except in quenched galaxies), and it decreases by a double exponential as a function of radius and distance from the midplane (see the profiles in Fig. \ref{fig:B_profile}, and Appendix \ref{appendixB}). RM decreases more rapidly than DM, because both $n_e$ and $B$ exponentially decrease with radius. We note that the magnetic field strength profiles are lower in magnitude than those from the observations of NGC6946 \citep{2007A&A...470..539B}, M101 \citep{2016A&A...588A.114B} and the CHANG-ES galaxies \citep{2018A&A...611A..72K}. This can be due to the limited spatial resolution of IllustrisTNG, causing a missing turbulent field (similar to the Auriga simulations \citealt{2017MNRAS.469.3185P}).

We also find that the path lengths increase with $b_{\rm offset}$, along with galaxy radius, because galaxies with larger radii have a higher chance of having an FRB with larger offset.

\subsubsection{Inclination}

We divided the sightlines into four inclination bins at each redshift. We find that the median DM increases exponentially with inclination, shown in Fig. \ref{fig:fits_all} c): face-on galaxies ($i$ < 10$^\circ$) have a smaller median DM than galaxies with larger inclinations (10$^\circ$< $i$ < 80$^\circ$), while edge-on galaxies ($i$ > 80$^\circ$) have the highest DM range (larger by a factor of $\sim$ 2 compared to face-on galaxies). The same trends are found at all other redshifts. In Fig. \ref{fig:fits_all} f) we also see $w_{\rm RM,rf}$ increasing exponentially with galaxy inclination: edge-on host galaxies have an $w_{\rm RM,rf}$ a factor of $\sim$ 4 larger compared to face-on galaxies, which is more significant compared to DM. The increase with inclination can be seen at every redshift, but at $z \ge 1.5$, the increase is only linear in contrast to the exponential increase at $z$~<~1.5. These differences between the trends suggest that the magnetic field also has an effect.

As we have 1000 sightlines for each galaxy, with different inclinations for each of those 1000 sightlines, we expect no variation of galaxy properties (e.g. stellar mass, SFR, radial and vertical profiles of $n_e$ and $B$) with inclination. And, as we select the FRB positions randomly, $n_{\rm e,FRBpos}$ also only fluctuates by 4\% for the different inclination bins. Thus, we conclude that the trend with inclination is a geometric effect, and is independent of the physical properties of individual galaxies. The DM of edge-on galaxies is larger because those sightlines have a longer path through the dense ISM inside the galaxy's disk, compared to the case of face-on galaxies, where the sightline only passes through a short path in the disk.

We find the $\langle n_{\rm e,LOS}\rangle$ is larger by a factor of $\sim$ 1.5-2 for edge-on compared to face-on galaxies, which is due to larger $n_e$ in galaxy disks. The median path length decreases slightly with inclination, but the mean path length stays the same. This could be explained by how in edge-on view, both small and long path lengths are possible depending on the line of sight. As our FRB sources are mostly found in a disk, an FRB from a source at the edge of the disk can go through the whole disk and the halo of the galaxy, or only the halo, depending on whether the source is on the near-side or the far-side of the disk with respect to the observer. In face-on view, the path lengths are closer to each other, as all sightlines go through the halo.
    
From the differences between the DM and RM trends, we suspect the geometry of the magnetic field also plays a role. We found that the azimuthal field dominates over the two other magnetic field components at most redshifts. In the case of an edge-on galaxy, the azimuthal field can be parallel to the line of sight, causing a higher possible RM (if there is no field reversal). In the case of a face-on galaxy, the vertical field (which is weaker than the azimuthal field) will contribute the most to the line-of-sight magnetic field, resulting in lower RMs. We note this contribution can be close to 0 in the case of symmetric vertical fields. 

We find that at $z\geq1$ the field was less ordered compared to lower redshifts and the azimuthal field does not dominate over the other components, resulting in a weaker inclination dependence. In the top panel of Fig. \ref{fig:B_comp} we show that the relative strength of the azimuthal field increases with time, becoming 5-6 times larger than the other two components. In contrast, at $z=2$, the azimuthal and radial components of the magnetic field have similar relative strengths, and are only twice as strong as the vertical field strength. The increase in the azimuthal field strength is caused by the ordering of the field by the differential rotation of the galaxy disks \citep{2009A&A...494...21A,2013lsmf.book..215B}. The total magnetic field is already amplified at higher redshifts to saturation, but the ordering of the field only happens at redshifts of $z$ = 1 -- 2 (similar to \citealt{2017MNRAS.469.3185P}). However, the mean-field dynamo would cause an increase in all magnetic field strength components (see e.g. \citealt{shukurov_subramanian_2021}, Section 13.5), in contradiction to the results from TNG50, suggesting the simulation might not contain the mean-field dynamo, as it does not reach the required spatial resolution in spite of currently being the cosmological MHD simulation with the highest spatial resolution. Nevertheless, purely differential rotation would not cause a decrease in the radial and vertical magnetic field strength with time either, and these would stay constant. The decrease we see in these components could be explained by other processes, such as outflows \citep{2015ApJ...808...28C}, accretion \citep{2000A&A...358.1142M}, dissipation or reconnection.

\subsection{Summary of trends}
\label{Trends_summary}

We quantified the redshift evolution of DM$_{\rm host, rf}$ and RM$_{\rm host, rf}$ (see section \ref{Res:DMvsz}). We found that the median and 1$\sigma$ distribution width (w$_{\rm DM,rf}$) of DM$_{\rm host,rf}$ increases towards higher redshift as a function of a power law, however this increase becomes insignificant at $z>4$. The change of w$_{\rm RM,rf}$ as a function of redshift can be described as a curved power law at $z\leq$2: at $z<1.5$ it increases towards higher redshift, and at $z>1.5$ it starts to decrease. At $z>2$ it falls even further. In section \ref{Res:prop}, we see the same trends with redshift if we separate the galaxies by star formation, stellar mass, $i$, and $b_{\rm offset}$. We showed that quenched galaxies have a lower median DM$_{\rm host,rf}$ and  w$_{\rm RM,rf}$ than star-forming galaxies, and median DM$_{\rm host,rf}$ and w$_{\rm RM,rf}$ increases with stellar mass until $\logM$=10.5, where they suddenly drop and start to decrease. We have also found that the median DM$_{\rm host,rf}$ decreases linearly, and w$_{\rm RM,rf}$ decreases exponentially with $b_{\rm offset}$. Lastly, we showed that the median DM$_{\rm host,rf}$ and w$_{\rm RM,rf}$ exponentially increase with $i$.

\section{Discussion}

\label{Discussion}

In this section, we investigate the implications of our results on the future studies of the IGM and estimate the number of FRBs needed to measure its magnetic field. We put our results in context with previous works, which used different models and simulations.

\subsection{Implication on future studies of the magnetic field of the IGM}
\label{IGM_implication}

\cite{2014PhRvD..89j7303Z} has pointed out that if DM$_{\rm host,rf}$ does not increase as fast as the $z$+1 factor from the redshift dilution decreases it, FRBs can be easily used as cosmological probes without much contamination from the host galaxy DMs. Thus, in this section, we first show how the statistical properties of the observed DM and RM distributions change with redshift, see if we can use them as cosmological probes, and how many FRBs we would need to measure the magnetic field of the IGM using FRBs with known redshift. Finally, we show how we could provide a constraint on DM$_{\rm host,rf}$ and RM$_{\rm host,rf}$ if the FRB is localized and more information is known about its host galaxy.

\subsubsection{DM$_{\rm host}$ and RM$_{\rm host}$ in the observer's frame}

 Below, we investigate how the DM and RM contributions of the host galaxies change with redshift in the observer's frame (DM$_{\rm host,obs}$ and RM$_{\rm host,obs}$). 
\begin{figure}
    \centering
\includegraphics[width=8.8cm]{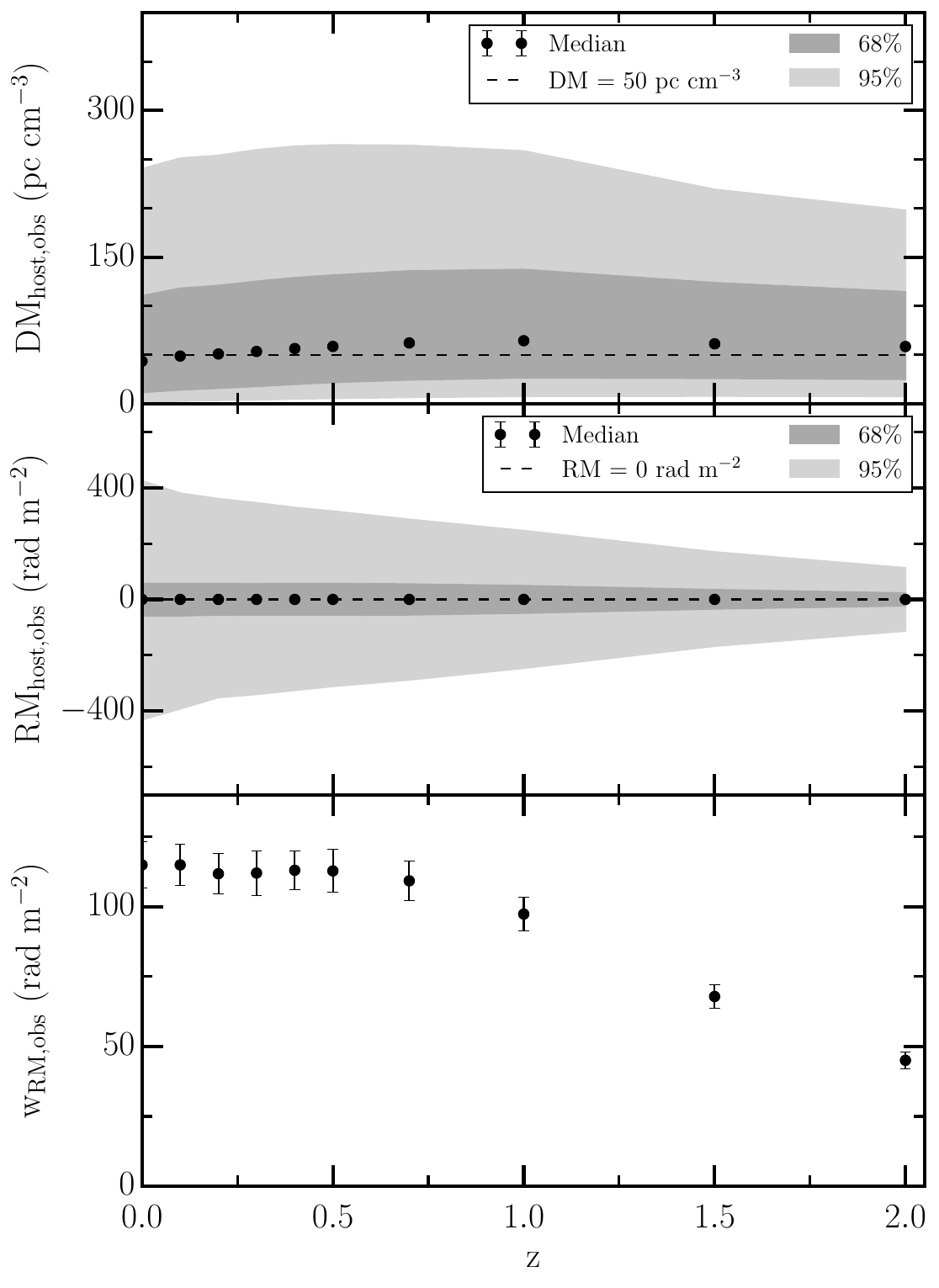}

    \caption{Same as Fig. \ref{fig:DM_z_ev}, but in the observer's frame. \textbf{Top}: The possible observed DM contribution of host galaxies based on 68\% and 95\% of our sightlines. The black points are the medians of DM at each redshift.  
    \textbf{Middle:} The same as top, but for the possible observed RM contribution of host galaxies. 
    \textbf{Bottom:} The width of the observed RM distribution as a function of redshift.}
    \label{fig:observed}
\end{figure}
We calculated DM$_{\rm host,obs}$ and RM$_{\rm host,obs}$ by correcting DM$_{\rm host,rf}$ and RM$_{\rm host,rf}$ calculated from the simulation with the redshift of the galaxies: 
\begin{equation}
    {\rm DM}_{\rm host, obs} = \frac{{\rm DM}_{\rm host, rf}}{(z_{\rm host}+1)} \quad \rm   
\end{equation}    
and
\begin{equation}
    {\rm RM}_{\rm host, obs} = \frac{{\rm RM}_{\rm host, rf}}{(z_{\rm host}+1)^2}.
    \end{equation}

 In Fig. \ref{fig:observed}, we show the 68\% and 95\% of DM and RM for each redshift. Considering all sightlines at all redshifts, 95\% of DM$_{\rm host,obs}$ is between 4 and 250 pc cm$^{-3}$, and 95\% of |RM$_{\rm host,obs}$| is less than 310 rad m$^{-2}$. These ranges decrease at higher redshift: only slightly in the case of DM$_{\rm host,obs}$, but significantly in the case of RM$_{\rm host,obs}$. 

The differences in the DM$_{\rm host, obs}$ at different redshifts is smaller than what we have seen in the rest frame DMs. The median of DM$_{\rm host,obs}$ is between 46 and 69 pc cm$^{-3}$, it first increases slightly, then decreases after $z = 1$.  Thus, the median DM$_{\rm host, obs}$ is similar at every redshift. The $w_{\rm DM, obs}$ also does not change significantly (1$\sigma$ width: 92$-$115~$\dm$, and 2$\sigma$ width: 202$-$278 pc cm$^{-3}$). This means if we have no information about the host galaxy, we can subtract the same host DM contribution from each FRB's total observed DM, independently of the host's redshift. This is in contrast to the results of \cite{2020AcA....70...87J}, who found even the observed DM increases with redshift (probably due to their longer integral path lengths). If we had considered a longer integral path length (e.g. extending to 2x$r_{\rm sf,99}$) the DM medians would be 20\% higher at every redshift.

We show in the middle panel of Fig. \ref{fig:observed} that the median of RM$_{\rm host, obs}$ is 0 rad m$^{-2}$ at every redshift. The 2 $\sigma$ width of the RM$_{\rm host, obs}$ distribution linearly decreases with redshift: it drops from 800 rad m$^{-2}$ at $z=0$ to 200 rad m$^{-2}$ at $z=2$. In the bottom panel of \ref{fig:observed} we show the changes of $w_{\rm RM, obs}$: at $z\leq$ 0.5 it does not change significantly ($w_{\rm RM, obs}$ = 111 $-$ 119 rad m$^{-2}$), but it starts to decrease after $z=0.7$.

We find that the increase in DM$_{\rm host,rf}$ and RM$_{\rm host,rf}$ with increasing redshift is weaker, than the decrease due to redshift dilution. As both DM$_{\rm IGM}$ and RM$_{\rm IGM}$ should increase towards higher redshift, if we observe FRBs at high redshift, we can get better constraints on cosmological parameters and the IGMF.

\subsubsection{Number of FRBs needed for IGM studies}
\label{Section:FRB_num}

FRBs can be used as a promising probe of the IGMF \citep{ 2014ApJ...797...71Z, 2016ApJ...824..105A}. However, since the RMs measured towards a sample of FRBs contains contribution from the FRB host galaxy and the IGMF, we investigate the minimum number of polarized FRBs required to statistically infer the IGMF. We note that in the past most surveys have not recorded the polarization data of FRBs, however, the current and future surveys are going to also observe polarization, which will result in an increase in the number of FRBs with measured RMs.

We perform a two-sample Kolmogorov-Smirnov (KS) test on the distributions of RM, with and without the contribution from the IGMF, in order to estimate the number of polarized FRBs needed to constrain the IGMF.  We assume that the RM contributed by the Milky Way has been robustly subtracted, the RM contributed by the FRB host galaxies at a particular redshift follows the distribution shown in the bottom panel of Fig.~\ref{fig:DM_dist}, and the RM from the local environment is negligible. We model the statistical distribution of RM arising in the IGMF ($\rm RM_{IGM}$) as a Gaussian distribution with mean zero, and standard deviation $\sigma_{\rm RM, IGM}$.

For the purpose of this work, we consider $\sigma_{\rm RM, IGM}=2$, 10, 20, and $40\,\radm$, following the findings of \citealt{2016ApJ...824..105A}. They found that $\sigma_{\rm RM, IGM}$ increases from $z=2$ to $z=7$, from 16 to 45 $\radm$ considering all IGM (including hot gas in clusters), and from 1.3 to 9 $\radm$ for only filaments. In order to estimate the statistical difference in the distributions of RM for $N_{\rm FRB}$ FRBs, with and without the contribution of IGMF, we randomly draw $N_{\rm FRB}$ values of $\rm RM_{host,obs}+RM_{IGM}$ and $\rm RM_{host,obs}$, and determine the $p$-value by applying a KS test. 
We performed 1000 Monte-Carlo simulations for a given $N_{\rm FRB}$, where $N_{\rm FRB}$ ranges between 10 and $10^6$. For a given $\sigma_{\rm RM, IGM}$, the $N_{\rm FRB}$ for which at least 95\% of the Monte-Carlo samples has $p < 0.05$, is considered as the minimum number of FRBs needed to discern the contribution of IGMF at 95\% confidence level. We did these tests for sightlines at $z=0.5$ and $z=2$. 

\begin{figure}
    \centering
\includegraphics[width=8.8cm]{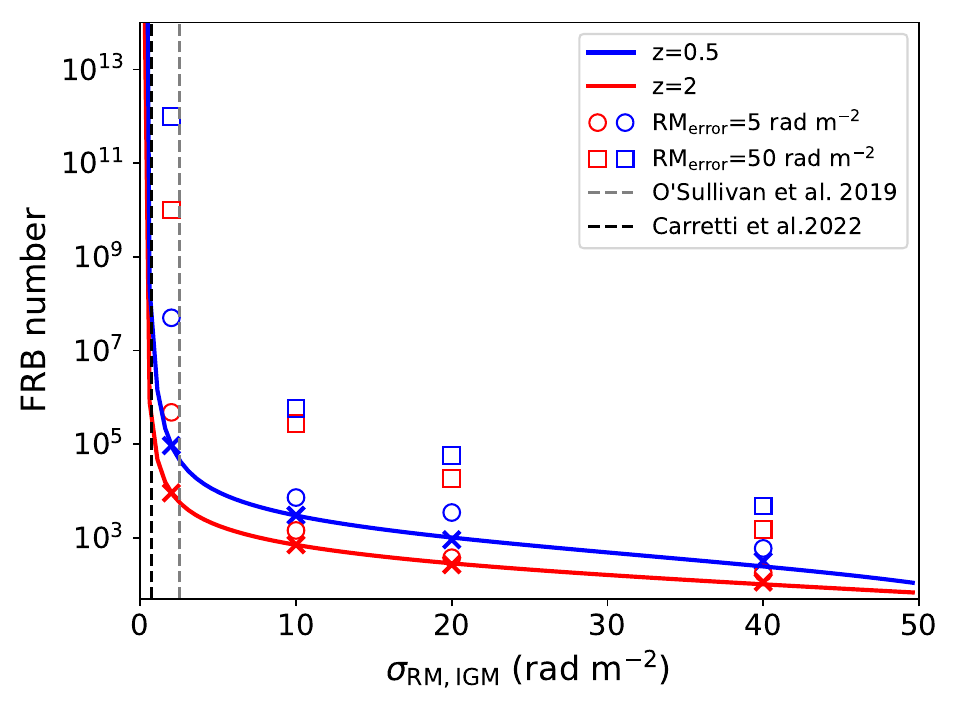}
    \caption{The number of FRBs needed to detect a given $\sigma_{\rm RM, IGM}$ at two redshifts ($z=0.5$ and $z=2$). We consider the IGMF possible to detect if the KS test returns a p-value below 0.05 for 95\% of Monte Carlo tests (for details, see Section \ref{Section:FRB_num}). The data points show the FRB number needed for significant detection for a given $\sigma_{\rm RM, IGM}$, and the solid line is the reciprocal exponential fit. We also show the number of FRBs needed considering an RM error of 5 and 50 $\radm$ as empty circles and squares, respectively. We indicate the observations of \cite{2019A&A...622A..16O} and \cite{2022MNRAS.512..945C}. }
    \label{fig:FRB_num}
\end{figure}

In Fig.~\ref{fig:FRB_num} we show how many FRBs we need to detect an IGMF with a given $\sigma_{\rm RM, IGM}$ at these redshifts. We also overplotted LOFAR measurements of the magnetic field in filaments (derived using radio lobes \citealt{2019A&A...622A..16O}, and using a catalog of sources with extragalactic RM \citealt{2022MNRAS.512..945C}). To detect a $\sigma_{\rm RM, IGM}$ of $40\,\radm$, we need at least 350 FRBs at $z=0.5$ or 150 FRBs at $z=2$. For a small $\sigma_{\rm RM, IGM}$ of 2 $\radm$ we need at least 95\,000 FRBs at $z=0.5$ or 9\,500 FRBs at $z=2$. We need fewer high redshift FRBs than low redshift FRBs to detect the same $\sigma_{\rm RM, IGM}$ due to the width of the distribution of the observed RM$_{\rm host}$ decreasing with redshift. We find that $N_{\rm FRB}$ as a function of $\sigma_{\rm RM, IGM}$ can be fitted by a function in the following form: 
\begin{equation}
    N_{\rm FRB} = a_{\rm IGM} \exp(-b_{\rm IGM}/\sigma_{\rm RM, IGM})+c_{\rm IGM},
    \label{eq:IGM_FRBnum}
\end{equation}
where the $a_{\rm IGM}$, $b_{\rm IGM}$ and $c_{\rm IGM}$ parameters are listed in Table \ref{tab:IGM_params} for $z=0.5$ and $z=2$.

\begin{table}[]
    \centering
    \caption{The parameters of the fit (Eq. \ref{eq:IGM_FRBnum}) for $N_{\rm FRB}$ as a function of $\sigma_{\rm RM, IGM}$ at $z=0.5$ and $z=2$.}
    \begin{tabular}{l c c c}
        \hline
        \hline

        $z$ & $a_{\rm IGM}$ & $b_{\rm IGM}$ & $c_{\rm IGM}$\\
        \hline
        0.5 & 3706.0 $\pm$ 413.6 &$-$7.0 $\pm$ 0.2 & $-$4118.0 $\pm$ 547.7 \\
        2.0 &1759.0 $\pm$ 202.9 &$-$4.0 $\pm$ 0.2 &$-$1825.0 $\pm$ 227.4 \\
        \hline

    \end{tabular}
    \label{tab:IGM_params}
\end{table}

We also investigate how adding measurement uncertainty to the observed RM would affect the number of FRBs needed. We added a Gaussian noise with $\sigma$ = 1, 5, 10, 20, and 50 $\radm$ to our RMs. We found that the number of FRBs needed to detect the IGMF exponentially increases with the error of RM.  An error of 1 to 10 $\radm$ would increase the number of FRBs by 1.5 to 5 times (depending on $\sigma_{\rm IGM}$).
In the case of $\sigma_{\rm IGM}$> 10 $\radm$ higher errors also increase the required number of FRBs by a similar magnitude.
For $\sigma_{\rm IGM}$ = 2 $\radm$, an RM$_{\rm error}$ = 20 $\radm$ and RM$_{\rm error}$ = 50 $\radm$ increases the required number of FRBs by 20 and 300 times, respectively. This additional uncertainty can also be considered the contribution of the immediate environment, if it is on an order of $1-50$ $\radm$. Thus, the increase in the required number of FRBs is true as well as in the case of RM from the source environment. We note that using a longer integral path length would also slightly increase the number of FRBs needed to detect the IGM by 10\% (see Appendix \ref{appendixA}).

We conclude that in the case of a $\sigma_{\rm RM, IGM}$ = 40 $\radm$ we will be able to constrain the IGMF with a few hundred - thousand FRBs (even if we take into account the measurement error of RM), which will be realistically achieved in the next years. In the case of $\sigma_{\rm RM, IGM}$ = 2 $\radm$, we will need significantly more FRBs. For this calculation, we have only used RM$_{\rm host}$, for which a different integral path length only leads to a change by a few $\radm$ in 68\% of the sightlines, thus  would most likely not affect our results.

\subsubsection{Localized FRBs}
 We compared our results to the observations of one example FRB (FRB190608, for details see Appendix \ref{FRBexample}). We find that by placing the FRB at a more specific location (minor axis) we get a similar DM range, but a more precise RM compared to the results from our pipeline, suggesting RM is more sensitive to the exact place in the galaxy. We suggest selecting sightlines based on their inclination if one wants to use our calculated RMs to constrain the contribution of a host galaxy. However, we note that in all cases, the range of possible values is wide for a single host galaxy, thus to derive the host contributions, a large sample of FRBs is needed. This calculation also provides a sanity check that our calculated DMs and RMs are reasonable, as we find that our DM and RM from the pipeline are in broad agreement with the measurements.

\subsection{Comparison to results from different simulations and models}

\label{Comparison}

There have been some previous works in the past years that have estimated DM$_{\rm host}$, and a few works that calculated RM$_{\rm host}$, using different simulations and models. To put our results from the high spatial resolution IllustrisTNG simulation in context, here we compare them to some of the previous works, to see if they show a difference.

\subsubsection{DM in previous works}
\label{Dis:DM_comp}

Here we discuss the different $n_e$ models, a semi analytic galaxy model, and other versions of the IllustrisTNG simulation that have been previously used in the literature.

\begin{enumerate}
    \item Electron density models:
    A few of the previous studies have used the NE2001 model for $n_e$ \citep{2002astro.ph..7156C} to compute the DM$_{\rm host}$ for MW-type galaxies (see e.g. \citealt{2015RAA....15.1629X,2019MNRAS.488.4220H}).
    In these studies, the DM$_{\rm host}$ is found to be a factor of two higher compared to our work for galaxies with MW mass.
    Furthermore, similar to our work, the DM$_{\rm host}$ is found to increase exponentially with increasing inclination (see Fig. \ref{fig:fits_all}), but with a steeper decline \citep{2015RAA....15.1629X}.
    Such differences arise because, in the NE2001 model, $n_e$ has a thick disk profile where $n_e \gtrsim 10^{-2}\,{\rm cm}^{-3}$ is found out to a radius of $\sim$ 15 kpc. In contrast, in TNG50, $n_e$ falls to $\sim10^{-2}$ cm$^{-3}$ within 12\,kpc on an average. Also, $n_e$ in TNG50 has a larger value vertically. As a result, in this work, face-on view provides a larger DM, and edge-on view provides a smaller DM compared to \cite{2015RAA....15.1629X}.
    Using a simple $n_e$ model that follows the stellar distribution in spiral galaxies, \cite{2020A&A...638A..37W} found a similar log normal DM distribution as seen in Fig. \ref{fig:DM_dist}. Overall, we find that the statistical distribution of DM$_{\rm host}$ to be broadly consistent with previous studies that have used different $n_e$ models.
    
\item Semi-analytic models:

We also compared our results to that obtained using a semi-analytic galaxy formation model (SAM), GALFORM \citep{2016MNRAS.462.3854L}, by \cite{2020MNRAS.498.4811H}. We find excellent agreement in the range of DMs found using SAM to that from TNG50. For example, \cite{2020MNRAS.498.4811H} report the likelihood of DMs for a sample, derived based on the FRB detection capabilities of ASKAP, CHIME and Parkes radio telescope with redshift, to lie in the range 0.1 to 450 pc cm$^{-3}$. By considering 3$\sigma$ (99.7\%) of our data points from all the redshift bins used by us, we find DM to lie in the range 10$^{-4}$ to 490 rad m$^{-2}$, similar to \cite{2020MNRAS.498.4811H}. This similarity is not surprising because, SAMs and hydrodynamic simulations have been shown to generally agree with each other, except for dissimilarity in the evolution of gas properties (see, e.g., \citealt{2018MNRAS.474..492M} for a comparison between EAGLE and GALFORM).

\item Other simulation volumes of IllustrisTNG: 

Previously, IllustrisTNG simulations performed with lower resolutions have also been used to study DM of the host or intervening galaxies, i.e., TNG100 \citep{2020ApJ...900..170Z,2020AcA....70...87J}, TNG300 \citep{2024A&A...683A..71W}. The different TNG resolutions have systematic effect on the inferred $n_e$ (see \citealt{2018MNRAS.473.4077P}) arising from systematics in black hole masses and gas fractions, because the same halo masses have larger stellar masses at higher resolution.

Overall, we find agreement in the trends observed for DM$_{\rm host}$ variation with stellar mass up to $\logM$ < 10.5 and decrease in $b_{\rm offset}$ (Fig. \ref{fig:fits_all}) with that from IllustrisTNG100-3 and 100-1 \citep{2020AcA....70...87J}. However, the absolute values of DM$_{\rm host}$ are significantly higher by 2--3 times at $z\le 2$ and up to 6 times at $z=2$ compared to our work. In Fig. \ref{fig:DM_z_ev}, we compare the results from \cite{2020ApJ...900..170Z}, who used data from IllustrisTNG100-1.\footnote{Note that, \citep{2020ApJ...900..170Z} three different galaxy selection criteria--- (i) 200 galaxies similar to the host of FRB180916, (ii) 200 galaxies similar to the hosts of non-repeating FRBs known at the time of \citep{2020ApJ...900..170Z}’s paper and (iii) 1000 galaxies similar to the host of the repeating FRB121102, up to z=1.5. Cases (i) and (ii) are similar to our work, while case (iii) includes dwarf galaxies, $\logM<8$, that were excluded in our study.} Although there were slight systematic offsets in the median DM$_{\rm host}$ with respect to our results, they were consistent with the 1$\sigma$ dispersion found in our study. It is important to note that, compared to our work, the integral path lengths were longer by up to 10 times in \cite{2020ApJ...900..170Z}, and by up to 30 times in \cite{2020AcA....70...87J}, that leads to the higher values of DM in their studies. Based on tests in Appendix \ref{appendixA}, we would have also derived 20\% larger DMs if we have considered an integral path length of twice as long, suggesting that integrating even further out would have resulted in similar magnitudes of DM as reported in the works above.

Using TNG300, \cite{2024A&A...683A..71W} found the rest frame DM of intervening galaxies increases with increasing redshift, which is suppressed when converted to the observer's frame, similar to the trend seen for our DM$_{\rm host}$.

\end{enumerate}

\subsubsection{RM in previous works}

\label{Comparison:rm}

As there are only two previous studies about the RM contribution of FRB host galaxies from models, we also compare our results to a work about intervening galaxies in front of background quasars.

\begin{enumerate}
    \item Large-scale magnetic field models:
    We compare our results to \cite{2018MNRAS.477.2528B}, who calculated the RM of disk galaxies in front of background sources (thus their sightlines go through the entire galaxy), using a large-scale axisymmetric spiral magnetic field and a radially decreasing electron density model. We used the same function as them to model the RM distribution of FRBs (see Section \ref{Results}): the sum of a Lorentzian and two Gaussian functions. \cite{2018MNRAS.477.2528B} found that the width of the Lorentzian increases with increasing magnetic field strength in the center of galaxies ($B_{\rm center}$). We find that the Lorentzian width of our full sample at $z=0$, with median $B_{\rm center}$ $\sim$ 10 $\mu$G, is consistent with the one predicted by \cite{2018MNRAS.477.2528B}\footnote{Once we correct for the slightly different definition of the Lorentzian function in the two works}.
    Depending on the galaxy selection, our Lorentzian width for a given $B_{\rm center}$ can be 50\% higher, but \cite{2018MNRAS.477.2528B} notes they predict $B_{\rm center}$ with 50\% uncertainty. Additionally, small differences could also be due to the assumption of a 500 pc thick disk by \cite{2018MNRAS.477.2528B}, while TNG50 has $n_e$ and $B$ out to larger vertical heights. Overall, our results are broadly consistent.

\cite{2019MNRAS.488.4220H} calculated the RM contribution of a possible MW-like spiral host galaxy (using NE2001 and a model of the MW galactic-scale magnetic field -- \citealt{2012ApJ...757...14J}). Considering the full sample of our galaxies, we find a mean |RM| of 100 rad m$^{-2}$, almost twice as large as their result (65 rad m$^{-2}$). However, if we only select MW mass galaxies, our results are consistent (63 rad m$^{-2}$). This is surprising as we have found different DM distributions compared to the NE2001 model. We note that as \cite{2019MNRAS.488.4220H} did not consider that $n_e$ or $B$ change with redshift, they predicted that the observed RM would fall to |RM| $\sim$ 7 rad m$^{-2}$ at $z$=2, in contrast to our result of $\sim 50~\radm$.

    \item Semi-analytic model
    \cite{2020MNRAS.498.4811H} calculated RM using galaxy models from \cite{2019MNRAS.483.2424R}, who studied the evolution of magnetic fields of few million galaxies using the combination of the semi-analytic galaxy formation model GALFORM \cite{2016MNRAS.462.3854L} and the MAGNETIZER code (\citealt{2020ascl.soft08011R}, which solves the non-linear turbulent mean-field dynamo equation).
    
    \cite{2020MNRAS.498.4811H} report the likelihood (the same way as for DM) of the observed host |RM|s for the full sample to lie in the range 0.01 to 320 rad m$^{-2}$. Considering 2 $\sigma$ (95\% of our data), we get a range of 0.2 to 490 rad m$^{-2}$, which shows slightly larger possible |RM|s than those from \cite{2020MNRAS.498.4811H} while still consistent with their results. However, if we consider 3 $\sigma$ (99.7\%), our range includes significantly higher |RM| values (10$^{-8}$ to 2650 rad m$^{-2}$). This could be because $\sim$ 60\% of galaxies in their sample have very weak magnetic field (<0.05 $\mu$G), whereas only $\sim$15\% of our galaxies have weak magnetic fields.

\end{enumerate}

Overall, we found that our results are consistent with previous results, even though they were derived with different methods. This can be due to the fact that all of these models and simulations mainly resolve the large-scale magnetic field of galaxies. We might find differences if we can resolve the small-scale field, for example once a simulation with even higher spatial resolution than IllustrisTNG50 becomes available. We note our path lengths (median $\sim$ 14 kpc) are comparable to the ones used in the  works above. Additionally, according to our tests (see Appendix \ref{appendixA}), a different integral path length is unlikely to affect our results.

\section{Conclusions}
\label{Conclusion}
We have calculated the DM and RM contribution of FRB host galaxies (DM$_{\rm host, rf}$ and RM$_{\rm host, rf}$) using the state-of-the-art MHD simulation TNG50 of IllustrisTNG project for a large sample of galaxies ($\sim$16\,500). We investigated how the median DM$_{\rm host, rf}$ and the width of the RM$_{\rm host, rf}$ distribution ($w_{\rm RM, rf}$) change with redshift, stellar mass, inclination, and FRB projected offsets from the center of galaxies ($b_{\rm offset}$).

Our main findings are:
\begin{itemize}
    \item The distributions of DM$_{\rm host, rf}$ can be fitted by a log normal function, and the shape of the distributions of RM$_{\rm host, rf}$ can be fitted by a combination of one Lorentzian and two Gaussian functions. The shape of the distributions does not change with the host galaxy's redshift, stellar mass, star formation, inclination, or $b_{\rm offset}$. However, we find that the parameters of these distributions change with the properties listed above (see Tables \ref{tab:DM_param} and \ref{tab:RM_param_fit}).
    \item We find that the median of DM$_{\rm host, rf}$ increases as a function of redshift. This can be explained by galaxies at higher redshifts having higher SFRs and therefore higher electron densities. This can be seen as an increase in the electron density profiles of galaxies in the simulation at higher redshifts. 
    \item The median of RM$_{\rm host, rf}$ is always 0 rad m$^{-2}$, and $w_{\rm RM, rf}$ increases with redshift up to $z=1.5$. After that, it quickly decreases. This is caused by a change in the magnetic field properties: the average of the total magnetic field strength in the disk increases by a factor of 1.2 up to z = 1, and at z=2 it drops to the same value as at $z=0$. At $z=2$ the azimuthal and radial components have similar relative strengths, but at $z=0$ the relative strength of the azimuthal field is 5-6 times larger than the other two components. We also find the presence of more azimuthal reversals or random fields at $z > 1$  compared to $z < 1$, showing the $B$ field becomes ordered as redshift decreases.
    \item The median of DM$_{\rm host, rf}$ increases with stellar mass, up to $\logM$>10.5), beyond which it drops. The same trend can be seen for $w_{\rm RM, rf}$. This is caused by the increase in $n_e$ due to increasing SFR with stellar mass, and the quenching process beginning in galaxies with $\logM \sim 10.5$, which also causes $B$ fields with irregularities and lower central $B$ field.
    \item We show that the median DM$_{\rm host, obs}$ of our entire sample of galaxies from the simulation does not change significantly with redshift, remaining between 46 and 69 $\dm$, in spite of DM$_{\rm host, rf}$ increasing with redshift. This can be useful in cases where we do not know the redshift of the FRB's galaxy, as we can assume the same range of DM$_{\rm host, obs}$ at all redshifts. 
    \item We find that $w_{\rm RM, rf}$ is decreasing with redshift, which means we can constrain the host's contribution more precisely at high redshift. We find that we would need more than 95\,000 polarized FRBs at $z=0.5$ to measure an $\sigma_{\rm RM,IGM}$ $\sim$2 rad m$^{-2}$ with a confidence level of 95\%, but we would only need 9\,500 FRBs at $z=2$ for the same precision. As more surveys are carried out recording polarization data of FRBs, the number of FRBs with measured RM is expected to increase significantly.
    \item The fitted DM and RM PDFs can be used in the frameworks of \cite{2020A&A...638A..37W} and \cite{2020MNRAS.498.4811H} to estimate the redshift of FRBs and to constrain the IGMF, providing additional choices for the host galaxy DM and RM PDFs.
    \item We apply our method to estimate the host DM and RM contribution for the well localized FRB190806. We found our DM and RM estimates are consistent with observations, thus our database of sightlines can be used to constrain the DM and RM contributions of host galaxies.

\end{itemize}
We note that the TNG50 model probably does not model the action of the mean-field dynamo, due to the limited spatial resolution currently available in cosmological simulations. A future (improved) MHD simulation may be able to include the mean-field dynamo, which might slightly affect our RM results.

We provide an estimate of the host galaxies' DM and RM contribution, which will allow future studies to separate the DM and RM of the IGM from the observed DM and RM of the FRBs. These results will help studies of the magnetic field of the IGM and the cosmic baryon density. The list of 16.5 million DM and RM values, together with the galaxy IDs in TNG50, positions of the FRBs and galaxy inclinations can be found on zenodo\footnote{\url{https://zenodo.org/records/12797564}}, and example scripts to use the PDFs in this work will be published on github\footnote{\url{https://github.com/tkovacs04/FRB-RMDM}}. This will allow future works to use their own subset of FRBs, with different assumptions on host galaxy properties (stellar mass and star formation rate) and FRB redshift distributions to fit their own distributions. 

\begin{acknowledgements}
 We thank Rainer Beck for his valuable suggestions and comments, and also thank Fabian Walter for his suggestions and help. The IllustrisTNG simulations were undertaken with compute time awarded by the Gauss Centre for Supercomputing (GCS) under GCS Large-Scale Projects GCS-ILLU and GCS-DWAR on the GCS share of the supercomputer Hazel Hen at the High Performance Computing Center Stuttgart (HLRS), as well as on the machines of the Max Planck Computing and Data Facility (MPCDF) in Garching, Germany.
 
 Computations in this paper were performed on the HPC system Raven at the MPCDF. This research has made use of NASA’s Astrophysics Data System Bibliographic Services. This publication is adapted from part of the PhD thesis of the lead author (\citealt{phdthesis}, Chapter 5).

\end{acknowledgements}

\bibliographystyle{aa}
\bibliography{bibliography}

\begin{thebibliography}{104}
\expandafter\ifx\csname natexlab\endcsname\relax\def\natexlab#1{#1}\fi

\bibitem[{{Akahori} {et~al.}(2016){Akahori}, {Ryu}, \& {Gaensler}}]{2016ApJ...824..105A}
{Akahori}, T., {Ryu}, D., \& {Gaensler}, B.~M. 2016, \apj, 824, 105

\bibitem[{{Anna-Thomas} {et~al.}(2023){Anna-Thomas}, {Connor}, {Dai}, {Feng}, {Burke-Spolaor}, {Beniamini}, {Yang}, {Zhang}, {Aggarwal}, {Law}, {Li}, {Niu}, {Chatterjee}, {Cruces}, {Duan}, {Filipovic}, {Hobbs}, {Lynch}, {Miao}, {Niu}, {Ocker}, {Tsai}, {Wang}, {Xue}, {Yao}, {Yu}, {Zhang}, {Zhang}, {Zhu}, \& {Zhu}}]{2023Sci...380..599A}
{Anna-Thomas}, R., {Connor}, L., {Dai}, S., {et~al.} 2023, Science, 380, 599

\bibitem[{{Arshakian} {et~al.}(2009){Arshakian}, {Beck}, {Krause}, \& {Sokoloff}}]{2009A&A...494...21A}
{Arshakian}, T.~G., {Beck}, R., {Krause}, M., \& {Sokoloff}, D. 2009, \aap, 494, 21

\bibitem[{Arvo(1992)}]{ARVO1992117}
Arvo, J. 1992, in {Graphics Gems III (IBM Version)}, ed. D.~KIRK (San Francisco: Morgan Kaufmann), 117--120

\bibitem[{{Bannister} {et~al.}(2019){Bannister}, {Deller}, {Phillips}, {Macquart}, {Prochaska}, {Tejos}, {Ryder}, {Sadler}, {Shannon}, {Simha}, {Day}, {McQuinn}, {North-Hickey}, {Bhandari}, {Arcus}, {Bennert}, {Burchett}, {Bouwhuis}, {Dodson}, {Ekers}, {Farah}, {Flynn}, {James}, {Kerr}, {Lenc}, {Mahony}, {O'Meara}, {Os{\l}owski}, {Qiu}, {Treu}, {U}, {Bateman}, {Bock}, {Bolton}, {Brown}, {Bunton}, {Chippendale}, {Cooray}, {Cornwell}, {Gupta}, {Hayman}, {Kesteven}, {Koribalski}, {MacLeod}, {McClure-Griffiths}, {Neuhold}, {Norris}, {Pilawa}, {Qiao}, {Reynolds}, {Roxby}, {Shimwell}, {Voronkov}, \& {Wilson}}]{2019Sci...365..565B}
{Bannister}, K.~W., {Deller}, A.~T., {Phillips}, C., {et~al.} 2019, Science, 365, 565

\bibitem[{{Basu} {et~al.}(2018){Basu}, {Mao}, {Fletcher}, {Kanekar}, {Shukurov}, {Schnitzeler}, {Vacca}, \& {Junklewitz}}]{2018MNRAS.477.2528B}
{Basu}, A., {Mao}, S.~A., {Fletcher}, A., {et~al.} 2018, \mnras, 477, 2528

\bibitem[{{Baym} {et~al.}(1996){Baym}, {B{\"o}deker}, \& {McLerran}}]{1996PhRvD..53..662B}
{Baym}, G., {B{\"o}deker}, D., \& {McLerran}, L. 1996, \prd, 53, 662

\bibitem[{{Beck}(2007)}]{2007A&A...470..539B}
{Beck}, R. 2007, \aap, 470, 539

\bibitem[{{Beck}(2012)}]{2012SSRv..166..215B}
{Beck}, R. 2012, \ssr, 166, 215

\bibitem[{{Beck}(2013)}]{2013lsmf.book..215B}
{Beck}, R. 2013, in Large-Scale Magnetic Fields in the Universe, ed. R.~{Beck}, A.~{Balogh}, A.~{Bykov}, R.~A. {Treumann}, \& L.~{Widrow}, Vol.~39, 215--230

\bibitem[{{Bell} {et~al.}(2004){Bell}, {Wolf}, {Meisenheimer}, {Rix}, {Borch}, {Dye}, {Kleinheinrich}, {Wisotzki}, \& {McIntosh}}]{2004ApJ...608..752B}
{Bell}, E.~F., {Wolf}, C., {Meisenheimer}, K., {et~al.} 2004, \apj, 608, 752

\bibitem[{{Berkhuijsen} {et~al.}(2016){Berkhuijsen}, {Urbanik}, {Beck}, \& {Han}}]{2016A&A...588A.114B}
{Berkhuijsen}, E.~M., {Urbanik}, M., {Beck}, R., \& {Han}, J.~L. 2016, \aap, 588, A114

\bibitem[{{Bhandari} {et~al.}(2018){Bhandari}, {Keane}, {Barr}, {Jameson}, {Petroff}, {Johnston}, {Bailes}, {Bhat}, {Burgay}, {Burke-Spolaor}, {Caleb}, {Eatough}, {Flynn}, {Green}, {Jankowski}, {Kramer}, {Krishnan}, {Morello}, {Possenti}, {Stappers}, {Tiburzi}, {van Straten}, {Andreoni}, {Butterley}, {Chandra}, {Cooke}, {Corongiu}, {Coward}, {Dhillon}, {Dodson}, {Hardy}, {Howell}, {Jaroenjittichai}, {Klotz}, {Littlefair}, {Marsh}, {Mickaliger}, {Muxlow}, {Perrodin}, {Pritchard}, {Sawangwit}, {Terai}, {Tominaga}, {Torne}, {Totani}, {Trois}, {Turpin}, {Niino}, {Wilson}, {Albert}, {Andr{\'e}}, {Anghinolfi}, {Anton}, {Ardid}, {Aubert}, {Avgitas}, {Baret}, {Barrios-Mart\'{\i}}, {Basa}, {Belhorma}, {Bertin}, {Biagi}, {Bormuth}, {Bourret}, {Bouwhuis}, {Br{\^a}nza{\c s}}, {Bruijn}, {Brunner}, {Busto}, {Capone}, {Caramete}, {Carr}, {Celli}, {Moursli}, {Chiarusi}, {Circella}, {Coelho}, {Coleiro}, {Coniglione}, {Costantini}, {Coyle}, {Creusot}, {D\'{\i}az}, {Deschamps}, {De Bonis}, {Distefano}, {Palma}, {Domi},
  {Donzaud}, {Dornic}, {Drouhin}, {Eberl}, {Bojaddaini}, {Khayati}, {Els{\"a}sser}, {Enzenh{\"o}fer}, {Ettahiri}, {Fassi}, {Felis}, {Fusco}, {Gay}, {Giordano}, {Glotin}, {Gregoire}, {Gracia-Ruiz}, {Graf}, {Hallmann}, {van Haren}, {Heijboer}, {Hello}, {Hern{\'a}ndez-Rey}, {H{\"o}{\ss}l}, {Hofest{\"a}dt}, {Hugon}, {Illuminati}, {James}, {de Jong}, {Jongen}, {Kadler}, {Kalekin}, {Katz}, {Kie{\ss}ling}, {Kouchner}, {Kreter}, {Kreykenbohm}, {Kulikovskiy}, {Lachaud}, {Lahmann}, {Lef{\`e}vre}, {Leonora}, {Loucatos}, {Marcelin}, {Margiotta}, {Marinelli}, {Mart\'{\i}nez-Mora}, {Mele}, {Melis}, {Michael}, {Migliozzi}, {Moussa}, {Navas}, {Nezri}, {Organokov}, {P{\v a}v{\v a}la{\c s}}, {Pellegrino}, {Perrina}, {Piattelli}, {Popa}, {Pradier}, {Quinn}, {Racca}, {Riccobene}, {S{\'a}nchez-Losa}, {Salda{\~n}a}, {Salvadori}, {Samtleben}, {Sanguineti}, {Sapienza}, {Sch{\"u}ssler}, {Sieger}, {Spurio}, {Stolarczyk}, {Taiuti}, {Tayalati}, {Trovato}, {Turpin}, {T{\"o}nnis}, {Vallage}, {Van Elewyck}, {Versari}, {Vivolo}, {Vizzocca},
  {Wilms}, {Zornoza}, \& {Z{\'u}{\~n}iga}}]{2018MNRAS.475.1427B}
{Bhandari}, S., {Keane}, E.~F., {Barr}, E.~D., {et~al.} 2018, \mnras, 475, 1427

\bibitem[{{Brandenburg} \& {Subramanian}(2005)}]{2005PhR...417....1B}
{Brandenburg}, A. \& {Subramanian}, K. 2005, \physrep, 417, 1

\bibitem[{{Carretti} {et~al.}(2023){Carretti}, {O'Sullivan}, {Vacca}, {Vazza}, {Gheller}, {Vernstrom}, \& {Bonafede}}]{2023MNRAS.518.2273C}
{Carretti}, E., {O'Sullivan}, S.~P., {Vacca}, V., {et~al.} 2023, \mnras, 518, 2273

\bibitem[{{Carretti} {et~al.}(2022){Carretti}, {Vacca}, {O'Sullivan}, {Heald}, {Horellou}, {R{\"o}ttgering}, {Scaife}, {Shimwell}, {Shulevski}, {Stuardi}, \& {Vernstrom}}]{2022MNRAS.512..945C}
{Carretti}, E., {Vacca}, V., {O'Sullivan}, S.~P., {et~al.} 2022, \mnras, 512, 945

\bibitem[{{Chamandy} \& {Taylor}(2015)}]{2015ApJ...808...28C}
{Chamandy}, L. \& {Taylor}, A.~R. 2015, \apj, 808, 28

\bibitem[{{Chatterjee} {et~al.}(2017){Chatterjee}, {Law}, {Wharton}, {Burke-Spolaor}, {Hessels}, {Bower}, {Cordes}, {Tendulkar}, {Bassa}, {Demorest}, {Butler}, {Seymour}, {Scholz}, {Abruzzo}, {Bogdanov}, {Kaspi}, {Keimpema}, {Lazio}, {Marcote}, {McLaughlin}, {Paragi}, {Ransom}, {Rupen}, {Spitler}, \& {van Langevelde}}]{2017Natur.541...58C}
{Chatterjee}, S., {Law}, C.~J., {Wharton}, R.~S., {et~al.} 2017, \nat, 541, 58

\bibitem[{{CHIME/FRB Collaboration} {et~al.}(2021){CHIME/FRB Collaboration}, {Amiri}, {Andersen}, {Bandura}, {Berger}, {Bhardwaj}, {Boyce}, {Boyle}, {Brar}, {Breitman}, {Cassanelli}, {Chawla}, {Chen}, {Cliche}, {Cook}, {Cubranic}, {Curtin}, {Deng}, {Dobbs}, {Dong}, {Eadie}, {Fandino}, {Fonseca}, {Gaensler}, {Giri}, {Good}, {Halpern}, {Hill}, {Hinshaw}, {Josephy}, {Kaczmarek}, {Kader}, {Kania}, {Kaspi}, {Landecker}, {Lang}, {Leung}, {Li}, {Lin}, {Masui}, {McKinven}, {Mena-Parra}, {Merryfield}, {Meyers}, {Michilli}, {Milutinovic}, {Mirhosseini}, {M{\"u}nchmeyer}, {Naidu}, {Newburgh}, {Ng}, {Patel}, {Pen}, {Petroff}, {Pinsonneault-Marotte}, {Pleunis}, {Rafiei-Ravandi}, {Rahman}, {Ransom}, {Renard}, {Sanghavi}, {Scholz}, {Shaw}, {Shin}, {Siegel}, {Sikora}, {Singh}, {Smith}, {Stairs}, {Tan}, {Tendulkar}, {Vanderlinde}, {Wang}, {Wulf}, \& {Zwaniga}}]{2021ApJS..257...59C}
{CHIME/FRB Collaboration}, {Amiri}, M., {Andersen}, B.~C., {et~al.} 2021, \apjs, 257, 59

\bibitem[{{Chime/Frb Collaboration} {et~al.}(2023){Chime/Frb Collaboration}, {Andersen}, {Bandura}, {Bhardwaj}, {Boyle}, {Brar}, {Cassanelli}, {Chatterjee}, {Chawla}, {Cook}, {Curtin}, {Dobbs}, {Dong}, {Faber}, {Fandino}, {Fonseca}, {Gaensler}, {Giri}, {Herrera-Martin}, {Hill}, {Ibik}, {Josephy}, {Kaczmarek}, {Kader}, {Kaspi}, {Landecker}, {Lanman}, {Lazda}, {Leung}, {Lin}, {Masui}, {McKinven}, {Mena-Parra}, {Meyers}, {Michilli}, {Ng}, {Pandhi}, {Pearlman}, {Pen}, {Petroff}, {Pleunis}, {Rafiei-Ravandi}, {Rahman}, {Ransom}, {Renard}, {Sand}, {Sanghavi}, {Scholz}, {Shah}, {Shin}, {Siegel}, {Smith}, {Stairs}, {Su}, {Tendulkar}, {Vanderlinde}, {Wang}, {Wulf}, \& {Zwaniga}}]{2023ApJ...947...83C}
{Chime/Frb Collaboration}, {Andersen}, B.~C., {Bandura}, K., {et~al.} 2023, \apj, 947, 83

\bibitem[{{Chittidi} {et~al.}(2021){Chittidi}, {Simha}, {Mannings}, {Prochaska}, {Ryder}, {Rafelski}, {Neeleman}, {Macquart}, {Tejos}, {Jorgenson}, {Day}, {Marnoch}, {Bhandari}, {Deller}, {Qiu}, {Bannister}, {Shannon}, \& {Heintz}}]{2021ApJ...922..173C}
{Chittidi}, J.~S., {Simha}, S., {Mannings}, A., {et~al.} 2021, \apj, 922, 173

\bibitem[{{Combes}(2017)}]{2017FrASS...4...10C}
{Combes}, F. 2017, Frontiers in Astronomy and Space Sciences, 4, 10

\bibitem[{{Conselice}(2014)}]{2014ARA&A..52..291C}
{Conselice}, C.~J. 2014, \araa, 52, 291

\bibitem[{{Cordes} \& {Chatterjee}(2019)}]{2019ARA&A..57..417C}
{Cordes}, J.~M. \& {Chatterjee}, S. 2019, \araa, 57, 417

\bibitem[{{Cordes} \& {Lazio}(2002)}]{2002astro.ph..7156C}
{Cordes}, J.~M. \& {Lazio}, T.~J.~W. 2002, arXiv e-prints, astro

\bibitem[{{Day} {et~al.}(2020){Day}, {Deller}, {Shannon}, {Qiu(邱昊)}, {Bannister}, {Bhandari}, {Ekers}, {Flynn}, {James}, {Macquart}, {Mahony}, {Phillips}, \& {Xavier Prochaska}}]{2020MNRAS.497.3335D}
{Day}, C.~K., {Deller}, A.~T., {Shannon}, R.~M., {et~al.} 2020, \mnras, 497, 3335

\bibitem[{{Dolag} {et~al.}(2015){Dolag}, {Gaensler}, {Beck}, \& {Beck}}]{2015MNRAS.451.4277D}
{Dolag}, K., {Gaensler}, B.~M., {Beck}, A.~M., \& {Beck}, M.~C. 2015, \mnras, 451, 4277

\bibitem[{{Donnari} {et~al.}(2019){Donnari}, {Pillepich}, {Nelson}, {Vogelsberger}, {Genel}, {Weinberger}, {Marinacci}, {Springel}, \& {Hernquist}}]{2019MNRAS.485.4817D}
{Donnari}, M., {Pillepich}, A., {Nelson}, D., {et~al.} 2019, \mnras, 485, 4817

\bibitem[{Elek {et~al.}(2022)Elek, Burchett, Prochaska, \& Forbes}]{PMID:34905603}
Elek, O., Burchett, J.~N., Prochaska, J.~X., \& Forbes, A.~G. 2022, Artificial life, 28, 22—57

\bibitem[{{Ferreira} {et~al.}(2022){Ferreira}, {Adams}, {Conselice}, {Sazonova}, {Austin}, {Caruana}, {Ferrari}, {Verma}, {Trussler}, {Broadhurst}, {Diego}, {Frye}, {Pascale}, {Wilkins}, {Windhorst}, \& {Zitrin}}]{2022ApJ...938L...2F}
{Ferreira}, L., {Adams}, N., {Conselice}, C.~J., {et~al.} 2022, \apjl, 938, L2

\bibitem[{{Ferreira} {et~al.}(2013){Ferreira}, {Jain}, \& {Sloth}}]{2013JCAP...10..004F}
{Ferreira}, R. J.~Z., {Jain}, R.~K., \& {Sloth}, M.~S. 2013, \jcap, 2013, 004

\bibitem[{{Gao} {et~al.}(2014){Gao}, {Li}, \& {Zhang}}]{2014ApJ...788..189G}
{Gao}, H., {Li}, Z., \& {Zhang}, B. 2014, \apj, 788, 189

\bibitem[{{Gordon} {et~al.}(2023){Gordon}, {Fong}, {Kilpatrick}, {Eftekhari}, {Leja}, {Prochaska}, {Nugent}, {Bhandari}, {Blanchard}, {Caleb}, {Day}, {Deller}, {Dong}, {Glowacki}, {Gourdji}, {Mannings}, {Mahoney}, {Marnoch}, {Miller}, {Paterson}, {Rastinejad}, {Ryder}, {Sadler}, {Scott}, {Sears}, {Shannon}, {Simha}, {Stappers}, \& {Tejos}}]{2023ApJ...954...80G}
{Gordon}, A.~C., {Fong}, W.-f., {Kilpatrick}, C.~D., {et~al.} 2023, \apj, 954, 80

\bibitem[{{Grand} {et~al.}(2017){Grand}, {G{\'o}mez}, {Marinacci}, {Pakmor}, {Springel}, {Campbell}, {Frenk}, {Jenkins}, \& {White}}]{2017MNRAS.467..179G}
{Grand}, R. J.~J., {G{\'o}mez}, F.~A., {Marinacci}, F., {et~al.} 2017, \mnras, 467, 179

\bibitem[{{Hackstein} {et~al.}(2019){Hackstein}, {Br{\"u}ggen}, {Vazza}, {Gaensler}, \& {Heesen}}]{2019MNRAS.488.4220H}
{Hackstein}, S., {Br{\"u}ggen}, M., {Vazza}, F., {Gaensler}, B.~M., \& {Heesen}, V. 2019, \mnras, 488, 4220

\bibitem[{{Hackstein} {et~al.}(2020){Hackstein}, {Br{\"u}ggen}, {Vazza}, \& {Rodrigues}}]{2020MNRAS.498.4811H}
{Hackstein}, S., {Br{\"u}ggen}, M., {Vazza}, F., \& {Rodrigues}, L.~F.~S. 2020, \mnras, 498, 4811

\bibitem[{{Hackstein} {et~al.}(2016){Hackstein}, {Vazza}, {Br{\"u}ggen}, {Sigl}, \& {Dundovic}}]{2016MNRAS.462.3660H}
{Hackstein}, S., {Vazza}, F., {Br{\"u}ggen}, M., {Sigl}, G., \& {Dundovic}, A. 2016, \mnras, 462, 3660

\bibitem[{{Hashimoto} {et~al.}(2020){Hashimoto}, {Goto}, {On}, {Lu}, {Santos}, {Ho}, {Wang}, {Kim}, \& {Hsiao}}]{2020MNRAS.497.4107H}
{Hashimoto}, T., {Goto}, T., {On}, A. Y.~L., {et~al.} 2020, \mnras, 497, 4107

\bibitem[{{Haverkorn} {et~al.}(2008){Haverkorn}, {Brown}, {Gaensler}, \& {McClure-Griffiths}}]{2008ApJ...680..362H}
{Haverkorn}, M., {Brown}, J.~C., {Gaensler}, B.~M., \& {McClure-Griffiths}, N.~M. 2008, \apj, 680, 362

\bibitem[{{Heintz} {et~al.}(2020){Heintz}, {Prochaska}, {Simha}, {Platts}, {Fong}, {Tejos}, {Ryder}, {Aggerwal}, {Bhandari}, {Day}, {Deller}, {Kilpatrick}, {Law}, {Macquart}, {Mannings}, {Marnoch}, {Sadler}, \& {Shannon}}]{2020ApJ...903..152H}
{Heintz}, K.~E., {Prochaska}, J.~X., {Simha}, S., {et~al.} 2020, \apj, 903, 152

\bibitem[{{Hutschenreuter} {et~al.}(2022){Hutschenreuter}, {Anderson}, {Betti}, {Bower}, {Brown}, {Br{\"u}ggen}, {Carretti}, {Clarke}, {Clegg}, {Costa}, {Croft}, {Van Eck}, {Gaensler}, {de Gasperin}, {Haverkorn}, {Heald}, {Hull}, {Inoue}, {Johnston-Hollitt}, {Kaczmarek}, {Law}, {Ma}, {MacMahon}, {Mao}, {Riseley}, {Roy}, {Shanahan}, {Shimwell}, {Stil}, {Sobey}, {O'Sullivan}, {Tasse}, {Vacca}, {Vernstrom}, {Williams}, {Wright}, \& {En{\ss}lin}}]{2022A&A...657A..43H}
{Hutschenreuter}, S., {Anderson}, C.~S., {Betti}, S., {et~al.} 2022, \aap, 657, A43

\bibitem[{{James} {et~al.}(2022){James}, {Ghosh}, {Prochaska}, {Bannister}, {Bhandari}, {Day}, {Deller}, {Glowacki}, {Gordon}, {Heintz}, {Marnoch}, {Ryder}, {Scott}, {Shannon}, \& {Tejos}}]{2022MNRAS.516.4862J}
{James}, C.~W., {Ghosh}, E.~M., {Prochaska}, J.~X., {et~al.} 2022, \mnras, 516, 4862

\bibitem[{{Jansson} \& {Farrar}(2012)}]{2012ApJ...757...14J}
{Jansson}, R. \& {Farrar}, G.~R. 2012, \apj, 757, 14

\bibitem[{{Jaroszy{\'n}ski}(2020)}]{2020AcA....70...87J}
{Jaroszy{\'n}ski}, M. 2020, \actaa, 70, 87

\bibitem[{{Kaasinen} {et~al.}(2017){Kaasinen}, {Bian}, {Groves}, {Kewley}, \& {Gupta}}]{2017MNRAS.465.3220K}
{Kaasinen}, M., {Bian}, F., {Groves}, B., {Kewley}, L.~J., \& {Gupta}, A. 2017, \mnras, 465, 3220

\bibitem[{{Kahniashvili} {et~al.}(2012){Kahniashvili}, {Brandenburg}, {Campanelli}, {Ratra}, \& {Tevzadze}}]{2012PhRvD..86j3005K}
{Kahniashvili}, T., {Brandenburg}, A., {Campanelli}, L., {Ratra}, B., \& {Tevzadze}, A.~G. 2012, \prd, 86, 103005

\bibitem[{{Kahniashvili} {et~al.}(2013){Kahniashvili}, {Tevzadze}, {Brandenburg}, \& {Neronov}}]{2013PhRvD..87h3007K}
{Kahniashvili}, T., {Tevzadze}, A.~G., {Brandenburg}, A., \& {Neronov}, A. 2013, \prd, 87, 083007

\bibitem[{{Keating} \& {Pen}(2020)}]{2020MNRAS.496L.106K}
{Keating}, L.~C. \& {Pen}, U.-L. 2020, \mnras, 496, L106

\bibitem[{{Krause} {et~al.}(2018){Krause}, {Irwin}, {Wiegert}, {Miskolczi}, {Damas-Segovia}, {Beck}, {Li}, {Heald}, {M{\"u}ller}, {Stein}, {Rand}, {Heesen}, {Walterbos}, {Dettmar}, {Vargas}, {English}, \& {Murphy}}]{2018A&A...611A..72K}
{Krause}, M., {Irwin}, J., {Wiegert}, T., {et~al.} 2018, \aap, 611, A72

\bibitem[{{Labb{\'e}} {et~al.}(2023){Labb{\'e}}, {van Dokkum}, {Nelson}, {Bezanson}, {Suess}, {Leja}, {Brammer}, {Whitaker}, {Mathews}, {Stefanon}, \& {Wang}}]{2023Natur.616..266L}
{Labb{\'e}}, I., {van Dokkum}, P., {Nelson}, E., {et~al.} 2023, \nat, 616, 266

\bibitem[{{Lacey} {et~al.}(2016){Lacey}, {Baugh}, {Frenk}, {Benson}, {Bower}, {Cole}, {Gonzalez-Perez}, {Helly}, {Lagos}, \& {Mitchell}}]{2016MNRAS.462.3854L}
{Lacey}, C.~G., {Baugh}, C.~M., {Frenk}, C.~S., {et~al.} 2016, \mnras, 462, 3854

\bibitem[{{Lyutikov}(2022)}]{2022ApJ...933L...6L}
{Lyutikov}, M. 2022, \apjl, 933, L6

\bibitem[{{Macquart} {et~al.}(2015){Macquart}, {Keane}, {Grainge}, {McQuinn}, {Fender}, {Hessels}, {Deller}, {Bhat}, {Breton}, {Chatterjee}, {Law}, {Lorimer}, {Ofek}, {Pietka}, {Spitler}, {Stappers}, \& {Trott}}]{2015aska.confE..55M}
{Macquart}, J.~P., {Keane}, E., {Grainge}, K., {et~al.} 2015, in {Advancing Astrophysics with the Square Kilometre Array (AASKA14)}, 55

\bibitem[{{Macquart} {et~al.}(2020){Macquart}, {Prochaska}, {McQuinn}, {Bannister}, {Bhandari}, {Day}, {Deller}, {Ekers}, {James}, {Marnoch}, {Os{\l}owski}, {Phillips}, {Ryder}, {Scott}, {Shannon}, \& {Tejos}}]{2020Natur.581..391M}
{Macquart}, J.~P., {Prochaska}, J.~X., {McQuinn}, M., {et~al.} 2020, \nat, 581, 391

\bibitem[{{Mannings} {et~al.}(2021){Mannings}, {Fong}, {Simha}, {Prochaska}, {Rafelski}, {Kilpatrick}, {Tejos}, {Heintz}, {Bannister}, {Bhandari}, {Day}, {Deller}, {Ryder}, {Shannon}, \& {Tendulkar}}]{2021ApJ...917...75M}
{Mannings}, A.~G., {Fong}, W.-f., {Simha}, S., {et~al.} 2021, \apj, 917, 75

\bibitem[{{Mannings} {et~al.}(2023){Mannings}, {Pakmor}, {Prochaska}, {van de Voort}, {Simha}, {Shannon}, {Tejos}, {Deller}, \& {Rafelski}}]{2023ApJ...954..179M}
{Mannings}, A.~G., {Pakmor}, R., {Prochaska}, J.~X., {et~al.} 2023, \apj, 954, 179

\bibitem[{{Marcote} {et~al.}(2022){Marcote}, {Kirsten}, {Hessels}, {Nimmo}, {Paragi}, \& {Project}}]{2022evlb.confE..35M}
{Marcote}, B., {Kirsten}, F., {Hessels}, J., {et~al.} 2022, in {European VLBI Network Mini-Symposium and Users' Meeting 2021}, Vol. 2021, 35

\bibitem[{{Marinacci} {et~al.}(2018){Marinacci}, {Vogelsberger}, {Pakmor}, {Torrey}, {Springel}, {Hernquist}, {Nelson}, {Weinberger}, {Pillepich}, {Naiman}, \& {Genel}}]{2018MNRAS.480.5113M}
{Marinacci}, F., {Vogelsberger}, M., {Pakmor}, R., {et~al.} 2018, \mnras, 480, 5113

\bibitem[{{Masui} {et~al.}(2015){Masui}, {Lin}, {Sievers}, {Anderson}, {Chang}, {Chen}, {Ganguly}, {Jarvis}, {Kuo}, {Li}, {Liao}, {McLaughlin}, {Pen}, {Peterson}, {Roman}, {Timbie}, {Voytek}, \& {Yadav}}]{2015Natur.528..523M}
{Masui}, K., {Lin}, H.-H., {Sievers}, J., {et~al.} 2015, \nat, 528, 523

\bibitem[{{McQuinn}(2014)}]{2014ApJ...780L..33M}
{McQuinn}, M. 2014, \apjl, 780, L33

\bibitem[{{Michilli} {et~al.}(2018){Michilli}, {Seymour}, {Hessels}, {Spitler}, {Gajjar}, {Archibald}, {Bower}, {Chatterjee}, {Cordes}, {Gourdji}, {Heald}, {Kaspi}, {Law}, {Sobey}, {Adams}, {Bassa}, {Bogdanov}, {Brinkman}, {Demorest}, {Fernandez}, {Hellbourg}, {Lazio}, {Lynch}, {Maddox}, {Marcote}, {McLaughlin}, {Paragi}, {Ransom}, {Scholz}, {Siemion}, {Tendulkar}, {van Rooy}, {Wharton}, \& {Whitlow}}]{2018Natur.553..182M}
{Michilli}, D., {Seymour}, A., {Hessels}, J.~W.~T., {et~al.} 2018, \nat, 553, 182

\bibitem[{{Mitchell} {et~al.}(2018){Mitchell}, {Lacey}, {Lagos}, {Frenk}, {Bower}, {Cole}, {Helly}, {Schaller}, {Gonzalez-Perez}, \& {Theuns}}]{2018MNRAS.474..492M}
{Mitchell}, P.~D., {Lacey}, C.~G., {Lagos}, C. D.~P., {et~al.} 2018, \mnras, 474, 492

\bibitem[{{Mo} {et~al.}(2023){Mo}, {Zhu}, {Wang}, {Tang}, \& {Feng}}]{2023MNRAS.518..539M}
{Mo}, J.-F., {Zhu}, W., {Wang}, Y., {Tang}, L., \& {Feng}, L.-L. 2023, \mnras, 518, 539

\bibitem[{{Moss} {et~al.}(2000){Moss}, {Shukurov}, \& {Sokoloff}}]{2000A&A...358.1142M}
{Moss}, D., {Shukurov}, A., \& {Sokoloff}, D. 2000, \aap, 358, 1142

\bibitem[{{Nelson} {et~al.}(2019{\natexlab{a}}){Nelson}, {Pillepich}, {Springel}, {Pakmor}, {Weinberger}, {Genel}, {Torrey}, {Vogelsberger}, {Marinacci}, \& {Hernquist}}]{2019MNRAS.490.3234N}
{Nelson}, D., {Pillepich}, A., {Springel}, V., {et~al.} 2019{\natexlab{a}}, \mnras, 490, 3234

\bibitem[{{Nelson} {et~al.}(2019{\natexlab{b}}){Nelson}, {Springel}, {Pillepich}, {Rodriguez-Gomez}, {Torrey}, {Genel}, {Vogelsberger}, {Pakmor}, {Marinacci}, {Weinberger}, {Kelley}, {Lovell}, {Diemer}, \& {Hernquist}}]{2019ComAC...6....2N}
{Nelson}, D., {Springel}, V., {Pillepich}, A., {et~al.} 2019{\natexlab{b}}, Computational Astrophysics and Cosmology, 6, 2

\bibitem[{{Noeske} {et~al.}(2007){Noeske}, {Weiner}, {Faber}, {Papovich}, {Koo}, {Somerville}, {Bundy}, {Conselice}, {Newman}, {Schiminovich}, {Le Floc'h}, {Coil}, {Rieke}, {Lotz}, {Primack}, {Barmby}, {Cooper}, {Davis}, {Ellis}, {Fazio}, {Guhathakurta}, {Huang}, {Kassin}, {Martin}, {Phillips}, {Rich}, {Small}, {Willmer}, \& {Wilson}}]{2007ApJ...660L..43N}
{Noeske}, K.~G., {Weiner}, B.~J., {Faber}, S.~M., {et~al.} 2007, \apjl, 660, L43

\bibitem[{{Oppermann} {et~al.}(2012){Oppermann}, {Junklewitz}, {Robbers}, {Bell}, {En{\ss}lin}, {Bonafede}, {Braun}, {Brown}, {Clarke}, {Feain}, {Gaensler}, {Hammond}, {Harvey-Smith}, {Heald}, {Johnston-Hollitt}, {Klein}, {Kronberg}, {Mao}, {McClure-Griffiths}, {O'Sullivan}, {Pratley}, {Robishaw}, {Roy}, {Schnitzeler}, {Sotomayor-Beltran}, {Stevens}, {Stil}, {Sunstrum}, {Tanna}, {Taylor}, \& {Van Eck}}]{2012A&A...542A..93O}
{Oppermann}, N., {Junklewitz}, H., {Robbers}, G., {et~al.} 2012, \aap, 542, A93

\bibitem[{{O'Sullivan} {et~al.}(2020){O'Sullivan}, {Br{\"u}ggen}, {Vazza}, {Carretti}, {Locatelli}, {Stuardi}, {Vacca}, {Vernstrom}, {Heald}, {Horellou}, {Shimwell}, {Hardcastle}, {Tasse}, \& {R{\"o}ttgering}}]{2020MNRAS.495.2607O}
{O'Sullivan}, S.~P., {Br{\"u}ggen}, M., {Vazza}, F., {et~al.} 2020, \mnras, 495, 2607

\bibitem[{{O'Sullivan} {et~al.}(2019){O'Sullivan}, {Machalski}, {Van Eck}, {Heald}, {Br{\"u}ggen}, {Fynbo}, {Heintz}, {Lara-Lopez}, {Vacca}, {Hardcastle}, {Shimwell}, {Tasse}, {Vazza}, {Andernach}, {Birkinshaw}, {Haverkorn}, {Horellou}, {Williams}, {Harwood}, {Brunetti}, {Anderson}, {Mao}, {Nikiel-Wroczy{\'n}ski}, {Takahashi}, {Carretti}, {Vernstrom}, {van Weeren}, {Orr{\'u}}, {Morabito}, \& {Callingham}}]{2019A&A...622A..16O}
{O'Sullivan}, S.~P., {Machalski}, J., {Van Eck}, C.~L., {et~al.} 2019, \aap, 622, A16

\bibitem[{{Pakmor} {et~al.}(2017){Pakmor}, {G{\'o}mez}, {Grand}, {Marinacci}, {Simpson}, {Springel}, {Campbell}, {Frenk}, {Guillet}, {Pfrommer}, \& {White}}]{2017MNRAS.469.3185P}
{Pakmor}, R., {G{\'o}mez}, F.~A., {Grand}, R. J.~J., {et~al.} 2017, \mnras, 469, 3185

\bibitem[{{Pakmor} {et~al.}(2018){Pakmor}, {Guillet}, {Pfrommer}, {G{\'o}mez}, {Grand}, {Marinacci}, {Simpson}, \& {Springel}}]{2018MNRAS.481.4410P}
{Pakmor}, R., {Guillet}, T., {Pfrommer}, C., {et~al.} 2018, \mnras, 481, 4410

\bibitem[{{Pellegrini} {et~al.}(2020){Pellegrini}, {Reissl}, {Rahner}, {Klessen}, {Glover}, {Pakmor}, {Herrera-Camus}, \& {Grand}}]{2020MNRAS.498.3193P}
{Pellegrini}, E.~W., {Reissl}, S., {Rahner}, D., {et~al.} 2020, \mnras, 498, 3193

\bibitem[{{Petroff} {et~al.}(2016){Petroff}, {Barr}, {Jameson}, {Keane}, {Bailes}, {Kramer}, {Morello}, {Tabbara}, \& {van Straten}}]{2016PASA...33...45P}
{Petroff}, E., {Barr}, E.~D., {Jameson}, A., {et~al.} 2016, \pasa, 33, e045

\bibitem[{{Petroff} {et~al.}(2019){Petroff}, {Hessels}, \& {Lorimer}}]{2019A&ARv..27....4P}
{Petroff}, E., {Hessels}, J.~W.~T., \& {Lorimer}, D.~R. 2019, \aapr, 27, 4

\bibitem[{{Pillepich} {et~al.}(2019){Pillepich}, {Nelson}, {Springel}, {Pakmor}, {Torrey}, {Weinberger}, {Vogelsberger}, {Marinacci}, {Genel}, {van der Wel}, \& {Hernquist}}]{2019MNRAS.490.3196P}
{Pillepich}, A., {Nelson}, D., {Springel}, V., {et~al.} 2019, \mnras, 490, 3196

\bibitem[{{Pillepich} {et~al.}(2018){Pillepich}, {Springel}, {Nelson}, {Genel}, {Naiman}, {Pakmor}, {Hernquist}, {Torrey}, {Vogelsberger}, {Weinberger}, \& {Marinacci}}]{2018MNRAS.473.4077P}
{Pillepich}, A., {Springel}, V., {Nelson}, D., {et~al.} 2018, \mnras, 473, 4077

\bibitem[{{Piro} \& {Gaensler}(2018)}]{2018ApJ...861..150P}
{Piro}, A.~L. \& {Gaensler}, B.~M. 2018, \apj, 861, 150

\bibitem[{{Planck Collaboration} {et~al.}(2016{\natexlab{a}}){Planck Collaboration}, {Ade}, {Aghanim}, {Arnaud}, {Arroja}, {Ashdown}, {Aumont}, {Baccigalupi}, {Ballardini}, {Banday}, {Barreiro}, {Bartolo}, {Battaner}, {Benabed}, {Beno\^{\i}t}, {Benoit-L{\'e}vy}, {Bernard}, {Bersanelli}, {Bielewicz}, {Bock}, {Bonaldi}, {Bonavera}, {Bond}, {Borrill}, {Bouchet}, {Bucher}, {Burigana}, {Butler}, {Calabrese}, {Cardoso}, {Catalano}, {Chamballu}, {Chiang}, {Chluba}, {Christensen}, {Church}, {Clements}, {Colombi}, {Colombo}, {Combet}, {Couchot}, {Coulais}, {Crill}, {Curto}, {Cuttaia}, {Danese}, {Davies}, {Davis}, {de Bernardis}, {de Rosa}, {de Zotti}, {Delabrouille}, {D{\'e}sert}, {Diego}, {Dolag}, {Dole}, {Donzelli}, {Dor{\'e}}, {Douspis}, {Ducout}, {Dupac}, {Efstathiou}, {Elsner}, {En{\ss}lin}, {Eriksen}, {Fergusson}, {Finelli}, {Florido}, {Forni}, {Frailis}, {Fraisse}, {Franceschi}, {Frejsel}, {Galeotta}, {Galli}, {Ganga}, {Giard}, {Giraud-H{\'e}raud}, {Gjerl{\o}w}, {Gonz{\'a}lez-Nuevo}, {G{\'o}rski}, {Gratton},
  {Gregorio}, {Gruppuso}, {Gudmundsson}, {Hansen}, {Hanson}, {Harrison}, {Helou}, {Henrot-Versill{\'e}}, {Hern{\'a}ndez-Monteagudo}, {Herranz}, {Hildebrandt}, {Hivon}, {Hobson}, {Holmes}, {Hornstrup}, {Hovest}, {Huffenberger}, {Hurier}, {Jaffe}, {Jaffe}, {Jones}, {Juvela}, {Keih{\"a}nen}, {Keskitalo}, {Kim}, {Kisner}, {Knoche}, {Kunz}, {Kurki-Suonio}, {Lagache}, {L{\"a}hteenm{\"a}ki}, {Lamarre}, {Lasenby}, {Lattanzi}, {Lawrence}, {Leahy}, {Leonardi}, {Lesgourgues}, {Levrier}, {Liguori}, {Lilje}, {Linden-V{\o}rnle}, {L{\'o}pez-Caniego}, {Lubin}, {Mac\'{\i}as-P{\'e}rez}, {Maggio}, {Maino}, {Mandolesi}, {Mangilli}, {Maris}, {Martin}, {Mart\'{\i}nez-Gonz{\'a}lez}, {Masi}, {Matarrese}, {McGehee}, {Meinhold}, {Melchiorri}, {Mendes}, {Mennella}, {Migliaccio}, {Mitra}, {Miville-Desch{\^e}nes}, {Molinari}, {Moneti}, {Montier}, {Morgante}, {Mortlock}, {Moss}, {Munshi}, {Murphy}, {Naselsky}, {Nati}, {Natoli}, {Netterfield}, {N{\o}rgaard-Nielsen}, {Noviello}, {Novikov}, {Novikov}, {Oppermann}, {Oxborrow}, {Paci},
  {Pagano}, {Pajot}, {Paoletti}, {Pasian}, {Patanchon}, {Perdereau}, {Perotto}, {Perrotta}, {Pettorino}, {Piacentini}, {Piat}, {Pierpaoli}, {Pietrobon}, {Plaszczynski}, {Pointecouteau}, {Polenta}, {Popa}, {Pratt}, {Pr{\'e}zeau}, {Prunet}, {Puget}, {Rachen}, {Rebolo}, {Reinecke}, {Remazeilles}, {Renault}, {Renzi}, {Ristorcelli}, {Rocha}, {Rosset}, {Rossetti}, {Roudier}, {Rubi{\~n}o-Mart\'{\i}n}, {Ruiz-Granados}, {Rusholme}, {Sandri}, {Santos}, {Savelainen}, {Savini}, {Scott}, {Seiffert}, {Shellard}, {Shiraishi}, {Spencer}, {Stolyarov}, {Stompor}, {Sudiwala}, {Sunyaev}, {Sutton}, {Suur-Uski}, {Sygnet}, {Tauber}, {Terenzi}, {Toffolatti}, {Tomasi}, {Tristram}, {Tucci}, {Tuovinen}, {Umana}, {Valenziano}, {Valiviita}, {Van Tent}, {Vielva}, {Villa}, {Wade}, {Wandelt}, {Wehus}, {Yvon}, {Zacchei}, \& {Zonca}}]{2016A&A...594A..19P}
{Planck Collaboration}, {Ade}, P.~A.~R., {Aghanim}, N., {et~al.} 2016{\natexlab{a}}, \aap, 594, A19

\bibitem[{{Planck Collaboration} {et~al.}(2016{\natexlab{b}}){Planck Collaboration}, {Ade}, {Aghanim}, {Arnaud}, {Ashdown}, {Aumont}, {Baccigalupi}, {Banday}, {Barreiro}, {Bartlett}, {Bartolo}, {Battaner}, {Battye}, {Benabed}, {Beno{\^\i}t}, {Benoit-L{\'e}vy}, {Bernard}, {Bersanelli}, {Bielewicz}, {Bock}, {Bonaldi}, {Bonavera}, {Bond}, {Borrill}, {Bouchet}, {Boulanger}, {Bucher}, {Burigana}, {Butler}, {Calabrese}, {Cardoso}, {Catalano}, {Challinor}, {Chamballu}, {Chary}, {Chiang}, {Chluba}, {Christensen}, {Church}, {Clements}, {Colombi}, {Colombo}, {Combet}, {Coulais}, {Crill}, {Curto}, {Cuttaia}, {Danese}, {Davies}, {Davis}, {de Bernardis}, {de Rosa}, {de Zotti}, {Delabrouille}, {D{\'e}sert}, {Di Valentino}, {Dickinson}, {Diego}, {Dolag}, {Dole}, {Donzelli}, {Dor{\'e}}, {Douspis}, {Ducout}, {Dunkley}, {Dupac}, {Efstathiou}, {Elsner}, {En{\ss}lin}, {Eriksen}, {Farhang}, {Fergusson}, {Finelli}, {Forni}, {Frailis}, {Fraisse}, {Franceschi}, {Frejsel}, {Galeotta}, {Galli}, {Ganga}, {Gauthier}, {Gerbino}, {Ghosh},
  {Giard}, {Giraud-H{\'e}raud}, {Giusarma}, {Gjerl{\o}w}, {Gonz{\'a}lez-Nuevo}, {G{\'o}rski}, {Gratton}, {Gregorio}, {Gruppuso}, {Gudmundsson}, {Hamann}, {Hansen}, {Hanson}, {Harrison}, {Helou}, {Henrot-Versill{\'e}}, {Hern{\'a}ndez-Monteagudo}, {Herranz}, {Hildebrandt}, {Hivon}, {Hobson}, {Holmes}, {Hornstrup}, {Hovest}, {Huang}, {Huffenberger}, {Hurier}, {Jaffe}, {Jaffe}, {Jones}, {Juvela}, {Keih{\"a}nen}, {Keskitalo}, {Kisner}, {Kneissl}, {Knoche}, {Knox}, {Kunz}, {Kurki-Suonio}, {Lagache}, {L{\"a}hteenm{\"a}ki}, {Lamarre}, {Lasenby}, {Lattanzi}, {Lawrence}, {Leahy}, {Leonardi}, {Lesgourgues}, {Levrier}, {Lewis}, {Liguori}, {Lilje}, {Linden-V{\o}rnle}, {L{\'o}pez-Caniego}, {Lubin}, {Mac{\'\i}as-P{\'e}rez}, {Maggio}, {Maino}, {Mandolesi}, {Mangilli}, {Marchini}, {Maris}, {Martin}, {Martinelli}, {Mart{\'\i}nez-Gonz{\'a}lez}, {Masi}, {Matarrese}, {McGehee}, {Meinhold}, {Melchiorri}, {Melin}, {Mendes}, {Mennella}, {Migliaccio}, {Millea}, {Mitra}, {Miville-Desch{\^e}nes}, {Moneti}, {Montier}, {Morgante},
  {Mortlock}, {Moss}, {Munshi}, {Murphy}, {Naselsky}, {Nati}, {Natoli}, {Netterfield}, {N{\o}rgaard-Nielsen}, {Noviello}, {Novikov}, {Novikov}, {Oxborrow}, {Paci}, {Pagano}, {Pajot}, {Paladini}, {Paoletti}, {Partridge}, {Pasian}, {Patanchon}, {Pearson}, {Perdereau}, {Perotto}, {Perrotta}, {Pettorino}, {Piacentini}, {Piat}, {Pierpaoli}, {Pietrobon}, {Plaszczynski}, {Pointecouteau}, {Polenta}, {Popa}, {Pratt}, {Pr{\'e}zeau}, {Prunet}, {Puget}, {Rachen}, {Reach}, {Rebolo}, {Reinecke}, {Remazeilles}, {Renault}, {Renzi}, {Ristorcelli}, {Rocha}, {Rosset}, {Rossetti}, {Roudier}, {Rouill{\'e} d'Orfeuil}, {Rowan-Robinson}, {Rubi{\~n}o-Mart{\'\i}n}, {Rusholme}, {Said}, {Salvatelli}, {Salvati}, {Sandri}, {Santos}, {Savelainen}, {Savini}, {Scott}, {Seiffert}, {Serra}, {Shellard}, {Spencer}, {Spinelli}, {Stolyarov}, {Stompor}, {Sudiwala}, {Sunyaev}, {Sutton}, {Suur-Uski}, {Sygnet}, {Tauber}, {Terenzi}, {Toffolatti}, {Tomasi}, {Tristram}, {Trombetti}, {Tucci}, {Tuovinen}, {T{\"u}rler}, {Umana}, {Valenziano}, {Valiviita},
  {Van Tent}, {Vielva}, {Villa}, {Wade}, {Wandelt}, {Wehus}, {White}, {White}, {Wilkinson}, {Yvon}, {Zacchei}, \& {Zonca}}]{2016A&A...594A..13P}
{Planck Collaboration}, {Ade}, P.~A.~R., {Aghanim}, N., {et~al.} 2016{\natexlab{b}}, \aap, 594, A13

\bibitem[{{Platts} {et~al.}(2019){Platts}, {Weltman}, {Walters}, {Tendulkar}, {Gordin}, \& {Kandhai}}]{2019PhR...821....1P}
{Platts}, E., {Weltman}, A., {Walters}, A., {et~al.} 2019, \physrep, 821, 1

\bibitem[{{Prochaska} \& {Zheng}(2019)}]{2019MNRAS.485..648P}
{Prochaska}, J.~X. \& {Zheng}, Y. 2019, \mnras, 485, 648

\bibitem[{{Rodrigues} \& {Chamandy}(2020)}]{2020ascl.soft08011R}
{Rodrigues}, L. F.~S. \& {Chamandy}, L. 2020, {Magnetizer: Computing magnetic fields of evolving galaxies}, Astrophysics Source Code Library, record ascl:2008.011

\bibitem[{{Rodrigues} {et~al.}(2019){Rodrigues}, {Chamandy}, {Shukurov}, {Baugh}, \& {Taylor}}]{2019MNRAS.483.2424R}
{Rodrigues}, L.~F.~S., {Chamandy}, L., {Shukurov}, A., {Baugh}, C.~M., \& {Taylor}, A.~R. 2019, \mnras, 483, 2424

\bibitem[{{Ryder} {et~al.}(2023){Ryder}, {Bannister}, {Bhandari}, {Deller}, {Ekers}, {Glowacki}, {Gordon}, {Gourdji}, {James}, {Kilpatrick}, {Lu}, {Marnoch}, {Moss}, {Prochaska}, {Qiu}, {Sadler}, {Simha}, {Sammons}, {Scott}, {Tejos}, \& {Shannon}}]{2023Sci...382..294R}
{Ryder}, S.~D., {Bannister}, K.~W., {Bhandari}, S., {et~al.} 2023, Science, 382, 294

\bibitem[{{Shimakawa} {et~al.}(2015){Shimakawa}, {Kodama}, {Steidel}, {Tadaki}, {Tanaka}, {Strom}, {Hayashi}, {Koyama}, {Suzuki}, \& {Yamamoto}}]{2015MNRAS.451.1284S}
{Shimakawa}, R., {Kodama}, T., {Steidel}, C.~C., {et~al.} 2015, \mnras, 451, 1284

\bibitem[{Shukurov \& Subramanian(2021)}]{shukurov_subramanian_2021}
Shukurov, A. \& Subramanian, K. 2021, Astrophysical Magnetic Fields: From Galaxies to the Early Universe, Cambridge Astrophysics (Cambridge University Press)

\bibitem[{{Sigl} {et~al.}(2003){Sigl}, {Miniati}, \& {Ensslin}}]{2003PhRvD..68d3002S}
{Sigl}, G., {Miniati}, F., \& {Ensslin}, T.~A. 2003, \prd, 68, 043002

\bibitem[{{Simha} {et~al.}(2020){Simha}, {Burchett}, {Prochaska}, {Chittidi}, {Elek}, {Tejos}, {Jorgenson}, {Bannister}, {Bhandari}, {Day}, {Deller}, {Forbes}, {Macquart}, {Ryder}, \& {Shannon}}]{2020ApJ...901..134S}
{Simha}, S., {Burchett}, J.~N., {Prochaska}, J.~X., {et~al.} 2020, \apj, 901, 134

\bibitem[{{Springel} \& {Hernquist}(2003)}]{2003MNRAS.339..289S}
{Springel}, V. \& {Hernquist}, L. 2003, \mnras, 339, 289

\bibitem[{{Tombesi} {et~al.}(2010){Tombesi}, {Cappi}, {Reeves}, {Palumbo}, {Yaqoob}, {Braito}, \& {Dadina}}]{2010A&A...521A..57T}
{Tombesi}, F., {Cappi}, M., {Reeves}, J.~N., {et~al.} 2010, \aap, 521, A57

\bibitem[{{Tímea Kovács}(2024)}]{phdthesis}
{Tímea Kovács}. 2024, PhD thesis, Rheinische Friedrich-Wilhelms-Universität Bonn, \url{https://hdl.handle.net/20.500.11811/11620}

\bibitem[{{Vazza} {et~al.}(2021){Vazza}, {Locatelli}, {Rajpurohit}, {Banfi}, {Dom\'{\i}nguez-Fern{\'a}ndez}, {Wittor}, {Angelinelli}, {Inchingolo}, {Brienza}, {Hackstein}, {Dallacasa}, {Gheller}, {Br{\"u}ggen}, {Brunetti}, {Bonafede}, {Ettori}, {Stuardi}, {Paoletti}, \& {Finelli}}]{2021Galax...9..109V}
{Vazza}, F., {Locatelli}, N., {Rajpurohit}, K., {et~al.} 2021, Galaxies, 9, 109

\bibitem[{{Walker} {et~al.}(2020){Walker}, {Ma}, \& {Breton}}]{2020A&A...638A..37W}
{Walker}, C. R.~H., {Ma}, Y.-Z., \& {Breton}, R.~P. 2020, \aap, 638, A37

\bibitem[{{Walker} {et~al.}(2024){Walker}, {Spitler}, {Ma}, {Cheng}, {Artale}, \& {Hummels}}]{2024A&A...683A..71W}
{Walker}, C. R.~H., {Spitler}, L.~G., {Ma}, Y.-Z., {et~al.} 2024, \aap, 683, A71

\bibitem[{{Weinberger} {et~al.}(2017){Weinberger}, {Springel}, {Hernquist}, {Pillepich}, {Marinacci}, {Pakmor}, {Nelson}, {Genel}, {Vogelsberger}, {Naiman}, \& {Torrey}}]{2017MNRAS.465.3291W}
{Weinberger}, R., {Springel}, V., {Hernquist}, L., {et~al.} 2017, \mnras, 465, 3291

\bibitem[{{Weinberger} {et~al.}(2018){Weinberger}, {Springel}, {Pakmor}, {Nelson}, {Genel}, {Pillepich}, {Vogelsberger}, {Marinacci}, {Naiman}, {Torrey}, \& {Hernquist}}]{2018MNRAS.479.4056W}
{Weinberger}, R., {Springel}, V., {Pakmor}, R., {et~al.} 2018, \mnras, 479, 4056

\bibitem[{{Xu} \& {Han}(2015)}]{2015RAA....15.1629X}
{Xu}, J. \& {Han}, J.~L. 2015, Research in Astronomy and Astrophysics, 15, 1629

\bibitem[{{Yao} {et~al.}(2017){Yao}, {Manchester}, \& {Wang}}]{2017ApJ...835...29Y}
{Yao}, J.~M., {Manchester}, R.~N., \& {Wang}, N. 2017, \apj, 835, 29

\bibitem[{{Zahid} {et~al.}(2012){Zahid}, {Dima}, {Kewley}, {Erb}, \& {Dav{\'e}}}]{2012ApJ...757...54Z}
{Zahid}, H.~J., {Dima}, G.~I., {Kewley}, L.~J., {Erb}, D.~K., \& {Dav{\'e}}, R. 2012, \apj, 757, 54

\bibitem[{{Zhang} {et~al.}(2020){Zhang}, {Yu}, {He}, \& {Wang}}]{2020ApJ...900..170Z}
{Zhang}, G.~Q., {Yu}, H., {He}, J.~H., \& {Wang}, F.~Y. 2020, \apj, 900, 170

\bibitem[{{Zheng} {et~al.}(2014){Zheng}, {Ofek}, {Kulkarni}, {Neill}, \& {Juric}}]{2014ApJ...797...71Z}
{Zheng}, Z., {Ofek}, E.~O., {Kulkarni}, S.~R., {Neill}, J.~D., \& {Juric}, M. 2014, \apj, 797, 71

\bibitem[{{Zhou} {et~al.}(2014){Zhou}, {Li}, {Wang}, {Fan}, \& {Wei}}]{2014PhRvD..89j7303Z}
{Zhou}, B., {Li}, X., {Wang}, T., {Fan}, Y.-Z., \& {Wei}, D.-M. 2014, \prd, 89, 107303

\bibitem[{{Zinger} {et~al.}(2020){Zinger}, {Pillepich}, {Nelson}, {Weinberger}, {Pakmor}, {Springel}, {Hernquist}, {Marinacci}, \& {Vogelsberger}}]{2020MNRAS.499..768Z}
{Zinger}, E., {Pillepich}, A., {Nelson}, D., {et~al.} 2020, \mnras, 499, 768

\end{thebibliography}

\begin{appendix}

\section{Effects of different parameter choices}
\label{appendixA}

\subsection{Integral path length}

We performed calculations using different integral path lengths, to see how the choice affects the DM and RM. We integrated out to 2 x $r_{\rm SF,99}$ from the same FRB positions and same galaxy inclination as the original, using a subset of 200 galaxies at three different redshifts ($z=0$, $z=0.5$, $z=2$). While some sightlines can have a significant difference depending on the integral length, such as the largest 7\,870 pc cm$^{-3}$ for DM and even 260\,000 rad m$^{-2}$ for RM, the general properties of the distributions stay mostly consistent. DM keeps increasing with larger integral limit (probably due to the circumgalactic and intergalactic medium), such as 30\% larger in the case of an integral limit of 3 x $r_{\rm SF,99}$. We have decided to define $r_{\rm SF,99}$ as the edge of the galaxy and our integral limit. However, our results can be scaled up if one wants to assume a different integral length. In the case of DM, the medians increases by $\sim20$\%, and the $w_{\rm DM, rest}$ by $\sim10$\% from 1 x $r_{\rm SF,99}$ to 2 x $r_{\rm SF,99}$. The RM shows a smaller change, with the median being around 0 $\radm$ (between $-1$ and $+1$ $\radm$) in both cases, and $w_{\rm RM, rest}$ increasing by $\sim$2\%.

\subsection{Integral step size}
The average cell size in TNG50 is between 70 and 140 pc, however in dense regions the cell size can be even only a few pcs. Because of this, we explored different integral step sizes: 5, 10, 20, 50 and 200 pc. We calculated the DM and RM for the same starting points in the same galaxies, with the same inclinations. We found that the calculated DM changes  by $\pm$ 14, $\pm$ 6 and $\pm$ 3 pc cm$^{-3}$ for the sightlines if we use 50, 20 and 10 pc step size, respectively. Nevertheless, the overall shape and properties of the DM distributions remain the same. The differences in the min, max, mean, median, 3 sigma width of the 5, 10 and 20 pc runs are less than 0.2\%. In the case of the d$l$ = 50 pc run, the lower range of the 3 sigma width is 2\% lower than for d$l$ = 5 pc (3.55 instead 3.62 pc cm$^{-3}$),  otherwise all parameters have a difference of 0.1\% or less. After dividing the galaxies into different stellar masses, the resulting DM and RM distributions also do not change significantly between the runs with different step sizes. This is also the case for different offsets. We chose d$l$ = 20 pc, as the PDF properties do not change significantly with smaller d$l$.

\section{Redshift evolution fits}
 
\label{appendix:z_evol}
We fitted how the statistical properties and the parameters of the fitted distributions of DM$_{\rm host, rf}$ and RM$_{\rm host, rf}$ change with redshift. For clarity, we provide the forms of our fitting functions below:
\begin{subequations}
\begin{flalign}
&    \text{Power law:}     &
    a \cdot z^{b} + c
             &   \label{eq:linear_EOM_vector_mass}\\
&    \text{Exponential:} &
    A_{\rm DM} \exp(-B_{\rm DM} z) + C_{\rm DM} &   \label{eq:linear_EOM_vector_mom}\\
&    \text{Curved power law:} &
    A_{\rm RM} \cdot z^{D_{\rm RM}+B_{\rm RM} \cdot z} + C_{\rm RM} &   \label{eq:linear_EOM_vector_mom},
\end{flalign}
\end{subequations}
where $z$ is redshift, and $a$, $b$, $c$, $A_{\rm DM}$, $B_{\rm DM}$, $C_{\rm DM}$, $A_{\rm RM}$, $B_{\rm RM}$, $C_{\rm RM}$, $D_{\rm RM}$ are the fit parameters.

The median of DM$_{\rm host,rf}$, and $w_{\rm DM, rest}$ both increase with redshift, following a power law (see in Table \ref{tab:DM_param}). The $\mu$ and $\sigma$ parameters of the fitted DM PDFs also change with redshift, but we find they are better described by an exponential function (see Table \ref{tab:RM_param_fit}). We note that the difference in the best fit function forms (exponential instead of power law) is due to the fact that $\mu$ and $\sigma$ are related to the variable's natural logarithm.

We find $w_{\rm RM, rf}$ can be fitted by a curved power law as a function of redshift (Table \ref{tab:DM_param}), and its uncertainty is estimated with the bootstrapping method. We note that $w_{\rm RM, rf}$ can also be fitted by a broken power law, but the curved power law has the advantage of the fit changing more smoothly, without a sharp peak. The parameters of the fitted RM PDFs ($a_1$, $a_2$, $a_3$, $\gamma$, $\sigma_1$, and $\sigma_2$) also change as curved power laws. Apart from the normalization parameters of the Gaussian components ($a_1$ and $a_2$), all parameters increase towards higher redshift. We list the results of the fits in Table \ref{tab:RM_param_fit}.

\begin{figure}
    \centering
    \includegraphics[width=8cm]{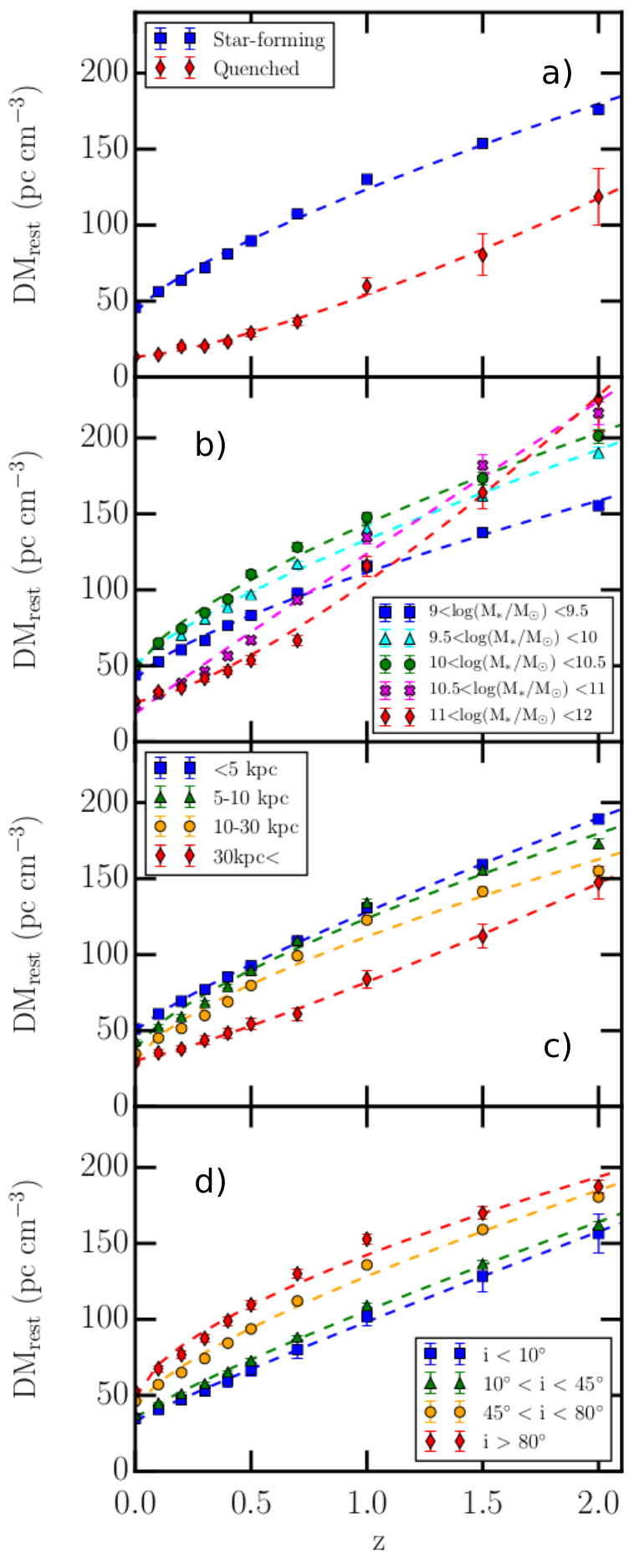}

    \caption{The redshift evolution of median DM$_{\rm host,rf}$ of different subsets of galaxies. A power law can be fitted to each group. The error bars are from bootstrapping. \textbf{a}: Quenched and star-forming galaxies. \textbf{b}: Different stellar mass bins. \textbf{c:} Sightlines with different inclinations. \textbf{d}: Sightlines with different $b_{\rm offset}$. 
}
    \label{fig:DM_subsets}
\end{figure}

\begin{figure}
    \centering
    \includegraphics[width=8cm]{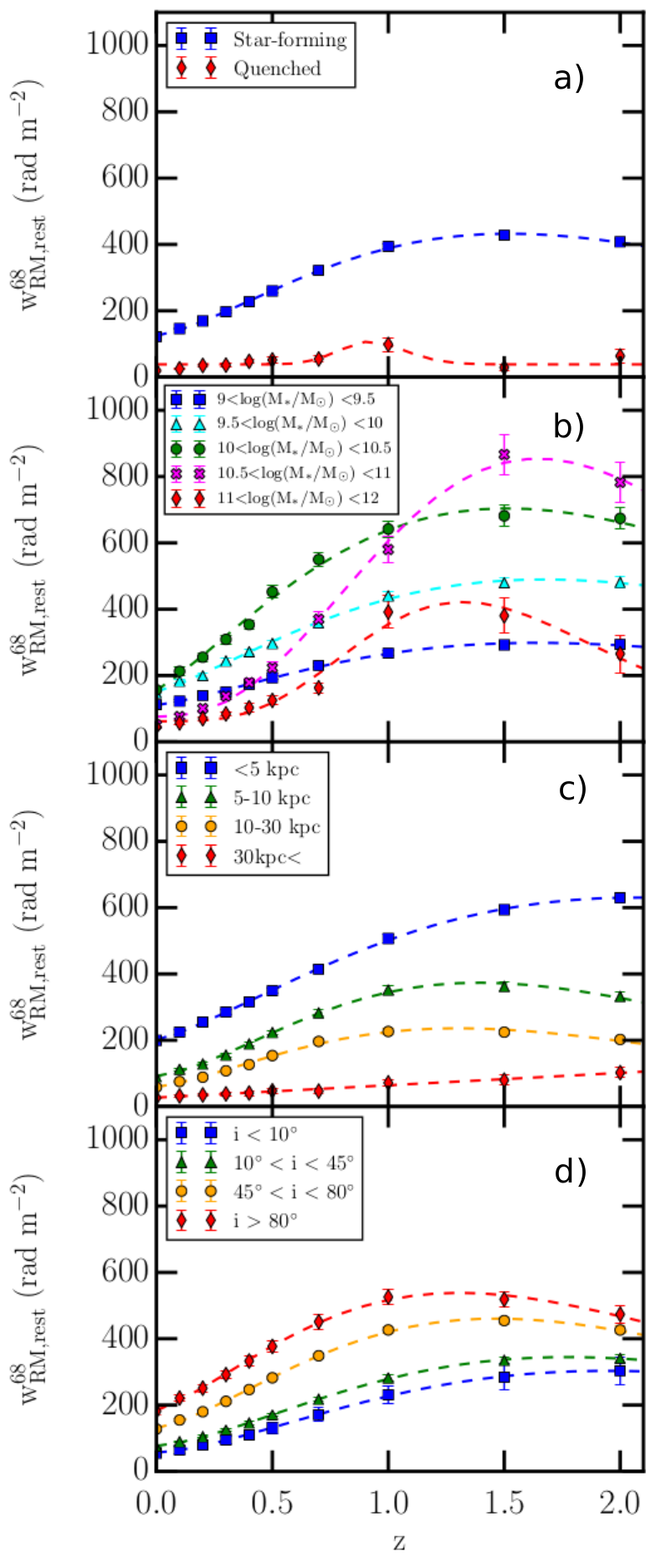}    

    \caption{The redshift evolution of the width of the rest frame RM distribution ($w_{\rm RM, rest}$) of different subsets of galaxies.  A curved power law can be fitted to each group. The error bars are from bootstrapping. \textbf{a}: Quenched and star-forming galaxies. At high $z$ we only have a few (<10) quenched galaxies, which makes it difficult to obtain a good fit. \textbf{b}: Different stellar mass bins. \textbf{c:} Different inclinations. \textbf{d:} Sightlines with different $b_{\rm offset}$.}
    
    \label{fig:RM_subsets}
\end{figure}

\onecolumn

\begin{table}[htb]
\caption{The fitted parameters for the redshift evolution of DM median, $w_{\rm DM,rf}$, and $w_{\rm RM,rf}$.}

\centering
\scalebox{.65}{
    \begin{tabular}{c|| c c c | c c c | c c c c }
    
         \hline
         
         & \multicolumn{3}{c|}{median DM$_{\rm host,rf}$ fit } & \multicolumn{3}{c|}{$w_{\rm DM,rf}$ fit } & \multicolumn{4}{c}{$w_{\rm RM,rf}$ fit }\\
         & $a$ & $b$ & $c$ & $a$ & $b$ & $c$ &  $A_{\rm RM}$ & $D_{\rm RM}$ & $B_{\rm RM}$ & $C_{\rm RM}$\\
         \hline
         \hline 
         All galaxies & 82 $\pm$ 4 & 0.8 $\pm$ 0.1 & 40 $\pm$ 3 & 241 $\pm$ 40 & 0.6 $\pm$ 0.1 & 93 $\pm$ 34 & 267 $\pm$ 4 & 1.24 $\pm$ 0.04 & $-$0.57 $\pm$ 0.03 & 117 $\pm$ 3 \\
         \hline
         Star-forming & 81 $\pm$ 4 & 0.8 $\pm$ 0.1 & 43 $\pm$ 3 & 113 $\pm$ 8 & 0.7 $\pm$ 0.1 & 94 $\pm$ 7 & 264 $\pm$ 4 & 1.24 $\pm$ 0.04 & $-$0.57 $\pm$ 0.03 & 125 $\pm$ 3 \\ 
         Red    & 41 $\pm$ 3 & 1.4 $\pm$ 0.1 & 13 $\pm$ 2 & 82 $\pm$ 7 & 1.3 $\pm$ 0.1 & 33 $\pm$ 4 & 60 $\pm$ 17 & 20.41 $\pm$ 304.02 & $-$24.0 $\pm$ 433.73 & 38 $\pm$ 7 \\  
         
         \hline
	 9 <$\logM$< 9.5 & 70 $\pm$ 3 & 0.74 $\pm$ 0.05 & 41 $\pm$ 3 & 95 $\pm$ 6 & 0.7 $\pm$ 0.1 & 87 $\pm$ 5 & 155 $\pm$ 4 & 1.16 $\pm$ 0.07 & $-$0.47 $\pm$ 0.05 & 111 $\pm$ 3 \\ 
         9.5 <$\logM$< 10 &  83 $\pm$ 4 & 0.78 $\pm$ 0.05 & 50 $\pm$ 3 & 107 $\pm$ 8 & 0.7 $\pm$ 0.1 & 105 $\pm$ 7 & 280 $\pm$ 8 & 1.1 $\pm$ 0.07 & $-$0.44 $\pm$ 0.05 & 151 $\pm$ 6 \\ 
         10 <$\logM$< 10.5 & 97 $\pm$ 5 & 0.7 $\pm$ 0.05 & 47 $\pm$ 4 & 121 $\pm$ 8 & 0.6 $\pm$ 0.1 & 121 $\pm$ 6 & 484 $\pm$ 20 & 1.04 $\pm$ 0.1 & $-$0.49 $\pm$ 0.07 & 155 $\pm$ 17 \\ 
         10.5 <$\logM$< 11& 105 $\pm$ 8 & 0.97 $\pm$ 0.08 & 18 $\pm$ 6 & 188 $\pm$ 15 & 0.7 $\pm$ 0.1 & 56 $\pm$ 12 & 530 $\pm$ 20 & 2.32 $\pm$ 0.19 & $-$0.94 $\pm$ 0.1 & 76 $\pm$ 12 \\ 
        11 < $\logM$ < 12 & 79 $\pm$ 6 & 1.35 $\pm$ 0.09 & 26 $\pm$ 4 & 168 $\pm$ 18 & 1.2 $\pm$ 0.1 & 52 $\pm$ 12 & 293 $\pm$ 27 & 3.4 $\pm$ 0.65 & $-$2.02 $\pm$ 0.41 & 62 $\pm$ 15 \\ 
         \hline
         $i$ < 10$^\circ$ & 65 $\pm$ 2 & 0.93 $\pm$ 0.03 & 33 $\pm$ 1 & 94 $\pm$ 5 & 0.9 $\pm$ 0.1 & 62 $\pm$ 4 & 171 $\pm$ 3 & 1.38 $\pm$ 0.05 & $-$0.43 $\pm$ 0.03 & 56 $\pm$ 2 \\ 
         10$^\circ < i < 45^\circ$ & 70 $\pm$ 2 & 0.87 $\pm$ 0.04 & 35 $\pm$ 2 & 98 $\pm$ 5 & 0.9 $\pm$ 0.1 & 67 $\pm$ 4 & 202 $\pm$ 2 & 1.35 $\pm$ 0.03 & $-$0.48 $\pm$ 0.02 & 77 $\pm$ 2 \\ 
         45$^\circ < i < 80^\circ$ & 86 $\pm$ 5 & 0.73 $\pm$ 0.05 & 42 $\pm$ 4 & 117 $\pm$ 8 & 0.7 $\pm$ 0.1 & 95 $\pm$ 7 & 291 $\pm$ 4 & 1.22 $\pm$ 0.04 & $-$0.6 $\pm$ 0.03 & 130 $\pm$ 4 \\ 
         80$^\circ < i $ & 94 $\pm$ 7 & 0.63 $\pm$ 0.07 & 48 $\pm$ 6 & 124 $\pm$ 15 & 0.4 $\pm$ 0.1 & 142 $\pm$ 14 & 333 $\pm$ 8 & 1.11 $\pm$ 0.07 & $-$0.67 $\pm$ 0.05 & 184 $\pm$ 7 \\ 
        \hline
         $b_{\rm offset}$ < 5 kpc & 78 $\pm$ 2 & 0.84 $\pm$ 0.02 & 50 $\pm$ 1 & 96 $\pm$ 5 & 0.67 $\pm$ 0.05 & 112 $\pm$ 4 & 301 $\pm$ 3 & 1.16 $\pm$ 0.03 & $-$0.33 $\pm$ 0.02 & 201 $\pm$ 2 \\ 
         5 kpc < $b_{\rm offset}$ < 10 kpc    & 86 $\pm$ 7 & 0.72 $\pm$ 0.08 & 38 $\pm$ 6 & 121 $\pm$ 10 & 0.69 $\pm$ 0.08 & 86 $\pm$ 9 & 253 $\pm$ 6 & 1.3 $\pm$ 0.07 & $-$0.7 $\pm$ 0.05 & 92 $\pm$ 5 \\ 
         10 kpc < $b_{\rm offset}$ < 30 kpc &82 $\pm$ 7 & 0.7 $\pm$ 0.08 & 30 $\pm$ 6 & 135 $\pm$ 14 & 0.64 $\pm$ 0.09 & 68 $\pm$ 12 & 166 $\pm$ 6 & 1.18 $\pm$ 0.1 & $-$0.72 $\pm$ 0.07 & 59 $\pm$ 5 \\ 
         30 kpc < $b_{\rm offset}$ & 52 $\pm$ 2 & 1.18 $\pm$ 0.05 & 30 $\pm$ 1 & 136 $\pm$ 13 & 0.88 $\pm$ 0.1 & 55 $\pm$ 10 & 37 $\pm$ 5 & 0.99 $\pm$ 0.32 & 0.0 $\pm$ 0.17 & 27 $\pm$ 4 \\ 
\hline

    \end{tabular}
   }
    \tablefoot{This table shows information for the actual rest frame DM$_{\rm host}$ and RM$_{\rm host}$ distributions (without fitting). The DM median and 1$\sigma$ width ($w_{\rm DM,rf}$) is fitted by a power law, and the RM 1$\sigma$ width ($w_{\rm RM,rf}$) by a curved power law as a function of redshift. The different lines of the table correspond to using every galaxy, galaxies with different star formation activity, galaxies in different stellar mass bins, galaxies with different inclinations and FRBs with different projected offsets from the center of the host galaxy ($b_{\rm offset}$). We also list the fitted parameters of the curved power law relation between the width of the actual RM distribution (without fitting) and redshift for the same subsets of galaxies.
    }
    \label{tab:DM_param}
\end{table}

\begin{scriptsize}

\begin{longtable}{c c c c c c}

    \caption{The curved power law fits of the redshift evolution of the parameters of the RM and DM PDFs.}  
    \\
    %\hhline{======}
    \hline \hline
    Galaxies & Par. & A$_{\rm RM,DM}$ & B$_{\rm RM,DM}$ & C$_{\rm RM,DM}$ & D$_{\rm RM}$ \\
    \hline
    \endfirsthead
    \caption{continued}\\
    \hline
    \hline
    Galaxies & Par. & A$_{\rm RM}$ & B$_{\rm RM}$ & C$_{\rm RM}$ & D$_{\rm RM}$ \\
    \hline
    \endhead
    
    \multirow{8}{*}{ All}& $a_1$ &0.24 $\pm$ 0.01 & $-$0.38 $\pm$ 0.05 & 0.57 $\pm$ 0.01 & 0.68 $\pm$ 0.06 \\ 
    & $a_2$   &$-$0.06 $\pm$ 0.0 & $-$0.7 $\pm$ 0.09 & 0.19 $\pm$ 0.0 & 1.24 $\pm$ 0.14 \\ 
     & $a_3$   &$-$0.18 $\pm$ 0.02 & $-$0.29 $\pm$ 0.06 & 0.25 $\pm$ 0.02 & 0.5 $\pm$ 0.08 \\ 
    & $\gamma$  &76.45 $\pm$ 0.67 & $-$0.59 $\pm$ 0.02 & 72.46 $\pm$ 0.51 & 1.37 $\pm$ 0.03 \\ 
    & $\sigma_1$  &1.11 $\pm$ 0.13 & $-$0.38 $\pm$ 0.17 & 6.84 $\pm$ 0.1 & 0.64 $\pm$ 0.16 \\ 
    & $\sigma_2$  &12.25 $\pm$ 0.33 & $-$0.52 $\pm$ 0.05 & 20.64 $\pm$ 0.26 & 1.13 $\pm$ 0.07 \\ 
    & $\mu$& $-$1.48 $\pm$ 0.02 & 1.24 $\pm$ 0.05 & 5.33 $\pm$ 0.02 \\ 
    & $\sigma$ & 0.49 $\pm$ 0.01 & 2.15 $\pm$ 0.05 & 0.8 $\pm$ 0.003 \\ 
    \hline
    \multirow{8}{*}{ Star-forming}& $a_1$ & 0.23 $\pm$ 0.01 & $-$0.38 $\pm$ 0.04 & 0.59 $\pm$ 0.01 & 0.68 $\pm$ 0.05 \\ 
    & $a_2$ &$-$0.06 $\pm$ 0.0 & $-$0.66 $\pm$ 0.09 & 0.2 $\pm$ 0.0 & 1.2 $\pm$ 0.13 \\ 
    & $a_3$ &$-$0.16 $\pm$ 0.01 & $-$0.31 $\pm$ 0.06 & 0.23 $\pm$ 0.01 & 0.51 $\pm$ 0.07 \\ 
    & $\gamma$&75.44 $\pm$ 0.77 & $-$0.6 $\pm$ 0.02 & 73.81 $\pm$ 0.54 & 1.37 $\pm$ 0.03 \\ 
    & $\sigma_1$ &1.07 $\pm$ 0.12 & $-$0.53 $\pm$ 0.2 & 6.99 $\pm$ 0.09 & 0.65 $\pm$ 0.17 \\ 
    & $\sigma_2$  &12.01 $\pm$ 0.31 & $-$0.55 $\pm$ 0.05 & 20.99 $\pm$ 0.22 & 1.14 $\pm$ 0.07 \\ 
    &$\mu$ & $-$1.43 $\pm$ 0.02 & 1.21 $\pm$ 0.05 & 5.332 $\pm$ 0.021 \\ 
    & $\sigma$ &0.44 $\pm$ 0.01 & 2.22 $\pm$ 0.08 & 0.801 $\pm$ 0.004 \\ 
    \hline
    \multirow{8}{*}{ Quenched}&$a_1$ &  0.16 $\pm$ 0.1 & 0.18 $\pm$ 0.1 & 0.35 $\pm$ 0.1 & 0.2 $\pm$ 0.2 \\ 
    & $a_2$ &0.06 $\pm$ 0.01 & $-$5.55 $\pm$ 5.2 & 0.12 $\pm$ 0.0 & 11.34 $\pm$ 10.41 \\ 
    & $a_3$ &$-$0.26 $\pm$ 0.14 & 0.81 $\pm$ 0.58 & 0.58 $\pm$ 0.14 & 0.27 $\pm$ 0.38 \\ 
    & $\gamma$&$-$1.19 $\pm$ 28.71 & 1.39 $\pm$ 13.45 & 44.35 $\pm$ 28.84 & 0.0 $\pm$ 22.46 \\ 
    & $\sigma_1$ &1.58 $\pm$ 13.15 & $-$0.97 $\pm$ 8.74 & 1.54 $\pm$ 13.06 & 0.0 $\pm$ 1.43 \\ 
    & $\sigma_2$  &7.52 $\pm$ 1.29 & $-$4.69 $\pm$ 4.6 & 14.05 $\pm$ 1.16 & 4.28 $\pm$ 3.1 \\ 
    &$\mu$ & $-$3.05 $\pm$ 0.31 & 0.56 $\pm$ 0.12 & 5.786 $\pm$ 0.355 \\ 
    & $\sigma$ & 0.05 $\pm$ 0.18 & 3.21 $\pm$ 20.14 & 1.02 $\pm$ 0.048 \\ 
    \hline
    \multirow{8}{*}{ $9 <\logM< 9.5$}& a$_1$ & 0.23 $\pm$ 0.02 & $-$0.38 $\pm$ 0.09 & 0.6 $\pm$ 0.02 & 0.66 $\pm$ 0.1 \\ 
    & $a_2$&$-$0.04 $\pm$ 0.01 & $-$0.57 $\pm$ 0.36 & 0.18 $\pm$ 0.01 & 1.05 $\pm$ 0.46 \\ 
    & $a_3$ &$-$0.16 $\pm$ 0.02 & $-$0.29 $\pm$ 0.09 & 0.25 $\pm$ 0.02 & 0.44 $\pm$ 0.12 \\ 
    & $\gamma$&50.1 $\pm$ 2.52 & $-$0.7 $\pm$ 0.14 & 65.98 $\pm$ 1.58 & 1.7 $\pm$ 0.2 \\ 
    & $\sigma_1$ &1.97 $\pm$ 0.11 & $-$0.85 $\pm$ 0.15 & 6.79 $\pm$ 0.08 & 1.63 $\pm$ 0.22 \\ 
    & $\sigma_2$  &13.12 $\pm$ 0.82 & $-$0.96 $\pm$ 0.19 & 20.02 $\pm$ 0.4 & 2.13 $\pm$ 0.28 \\ 
    & $\mu$ & $-$1.29 $\pm$ 0.02 & 1.14 $\pm$ 0.05 & 5.22 $\pm$ 0.02 \\ 
    & $\sigma$ & 0.6 $\pm$ 0.05 & 3.46 $\pm$ 0.36 & 0.82 $\pm$ 0.01 \\
    \hline
    \multirow{8}{*}{ $9.5 <\logM< 10$}& a$_1$ & 0.21 $\pm$ 0.01 & $-$0.3 $\pm$ 0.09 & 0.63 $\pm$ 0.01 & 0.5 $\pm$ 0.06 \\ 
    & $a_2$&$-$0.06 $\pm$ 0.01 & $-$0.17 $\pm$ 0.16 & 0.2 $\pm$ 0.0 & 0.54 $\pm$ 0.11 \\ 
    & $a_3$ &$-$0.1 $\pm$ 0.01 & $-$0.28 $\pm$ 0.12 & 0.19 $\pm$ 0.01 & 0.29 $\pm$ 0.08 \\ 
    & $\gamma$&109.63 $\pm$ 8.32 & $-$0.72 $\pm$ 0.17 & 82.65 $\pm$ 1.92 & 1.57 $\pm$ 0.17 \\ 
    & $\sigma_1$ &4.92 $\pm$ 0.28 & $-$1.16 $\pm$ 0.16 & 7.21 $\pm$ 0.09 & 1.8 $\pm$ 0.17 \\ 
    & $\sigma_2$  &40.45 $\pm$ 3.57 & $-$1.27 $\pm$ 0.22 & 22.28 $\pm$ 0.45 & 2.14 $\pm$ 0.21 \\ 
    & $\mu$ & $-$1.36 $\pm$ 0.04 & 1.12 $\pm$ 0.11 & 5.42 $\pm$ 0.05 \\ 
    & $\sigma$ &0.45 $\pm$ 0.02 & 2.72 $\pm$ 0.23 & 0.77 $\pm$ 0.01 \\ 
    \hline
    \multirow{8}{*}{ $10 <\logM<10.5$}& a$_1$ &0.26 $\pm$ 0.01 & $-$0.4 $\pm$ 0.08 & 0.56 $\pm$ 0.01 & 0.62 $\pm$ 0.06 \\ 
    & $a_2$&$-$0.08 $\pm$ 0.01 & $-$0.28 $\pm$ 0.15 & 0.22 $\pm$ 0.0 & 0.62 $\pm$ 0.13 \\ 
    & $a_3$ &$-$0.13 $\pm$ 0.01 & $-$0.3 $\pm$ 0.11 & 0.21 $\pm$ 0.01 & 0.35 $\pm$ 0.09 \\ 
    & $\gamma$&178.25 $\pm$ 17.63 & $-$0.83 $\pm$ 0.19 & 98.8 $\pm$ 3.3 & 1.59 $\pm$ 0.19 \\ 
    & $\sigma_1$ &6.17 $\pm$ 0.97 & $-$1.66 $\pm$ 0.43 & 7.37 $\pm$ 0.2 & 2.12 $\pm$ 0.44 \\ 
    & $\sigma_2$  &59.5 $\pm$ 8.35 & $-$1.69 $\pm$ 0.31 & 24.78 $\pm$ 0.69 & 2.5 $\pm$ 0.31 \\ 
    & $\mu$ & $-$1.43 $\pm$ 0.04 & 1.53 $\pm$ 0.12 & 5.37 $\pm$ 0.04 \\ 
    & $\sigma$ & 0.48 $\pm$ 0.03 & 1.65 $\pm$ 0.3 & 0.75 $\pm$ 0.03 \\ 
    \hline

    \pagebreak
    
    \multirow{8}{*}{ $10.5 <\logM)<11$} & a$_1$ & 0.31 $\pm$ 0.02 & $-$0.43 $\pm$ 0.08 & 0.41 $\pm$ 0.01 & 0.84 $\pm$ 0.1 \\ 
    & $a_2$&$-$0.04 $\pm$ 0.01 & $-$1.43 $\pm$ 0.82 & 0.22 $\pm$ 0.01 & 3.17 $\pm$ 1.59 \\ 
    & $a_3$ &$-$0.29 $\pm$ 0.02 & $-$0.28 $\pm$ 0.06 & 0.38 $\pm$ 0.02 & 0.48 $\pm$ 0.08 \\ 
    & $\gamma$&167.68 $\pm$ 11.47 & $-$0.74 $\pm$ 0.17 & 50.79 $\pm$ 3.35 & 1.9 $\pm$ 0.23 \\ 
    & $\sigma_1$ &2.05 $\pm$ 0.23 & $-$0.38 $\pm$ 0.22 & 6.71 $\pm$ 0.1 & 1.11 $\pm$ 0.25 \\ 
    & $\sigma_2$  &26.2 $\pm$ 2.91 & $-$1.24 $\pm$ 0.3 & 19.52 $\pm$ 0.78 & 2.39 $\pm$ 0.42 \\ 
    & $\mu$ & $-$2.64 $\pm$ 0.09 & 1.06 $\pm$ 0.08 & 5.79 $\pm$ 0.1 \\ 
    & $\sigma$ & 0.87 $\pm$ 1.12 & 0.19 $\pm$ 0.29 & 0.23 $\pm$ 1.13 \\
    \hline
    \multirow{8}{*}{ 11<$\logM< 12$} & a$_1$ & $-$0.01 $\pm$ 0.06 & 1.43 $\pm$ 2.82 & 0.62 $\pm$ 0.05 & 0.0 $\pm$ 1.84 \\ 
    & $a_2$&0.03 $\pm$ 0.1 & 0.64 $\pm$ 0.39 & 0.14 $\pm$ 0.1 & 0.47 $\pm$ 4.33 \\ 
    & $a_3$ &$-$1.15 $\pm$ 16.54 & $-$0.02 $\pm$ 0.32 & 1.35 $\pm$ 16.54 & 0.04 $\pm$ 0.65 \\ 
    & $\gamma$&104.54 $\pm$ 16.89 & $-$1.07 $\pm$ 0.56 & 34.4 $\pm$ 4.12 & 2.29 $\pm$ 0.71 \\ 
    & $\sigma_1$ &0.75 $\pm$ 0.36 & $-$0.4 $\pm$ 0.74 & 6.09 $\pm$ 0.31 & 1.63 $\pm$ 1.54 \\ 
    & $\sigma_2$  &14.03 $\pm$ 2.79 & $-$1.27 $\pm$ 0.64 & 15.15 $\pm$ 1.8 & 2.41 $\pm$ 0.97 \\ 
    & $\mu$ & $-$2.64 $\pm$ 0.09 & 1.06 $\pm$ 0.08 & 5.79 $\pm$ 0.1 \\ 
    & $\sigma$ & 0.87 $\pm$ 1.12 & 0.19 $\pm$ 0.29 & 0.23 $\pm$ 1.13 \\ 
    \hline
    \multirow{8}{*}{ $i$ < 10$^\circ$} & a$_1$ & 0.19 $\pm$ 0.02 & $-$0.44 $\pm$ 0.17 & 0.64 $\pm$ 0.02 & 0.6 $\pm$ 0.16 \\ 
    & $a_2$&$-$0.05 $\pm$ 0.02 & $-$4.64 $\pm$ 5.1 & 0.15 $\pm$ 0.01 & 3.61 $\pm$ 3.8 \\ 
    & $a_3$ &$-$0.26 $\pm$ 0.07 & $-$0.26 $\pm$ 0.16 & 0.35 $\pm$ 0.07 & 0.39 $\pm$ 0.2 \\ 
    & $\gamma$&65.95 $\pm$ 2.62 & $-$0.33 $\pm$ 0.09 & 27.86 $\pm$ 1.06 & 1.36 $\pm$ 0.11 \\ 
    & $\sigma_1$ &7.25 $\pm$ 1.47 & $-$1.22 $\pm$ 0.58 & 2.57 $\pm$ 0.47 & 1.99 $\pm$ 0.69 \\ 
    & $\sigma_2$  &17.96 $\pm$ 1.93 & $-$0.35 $\pm$ 0.26 & 9.77 $\pm$ 0.85 & 1.47 $\pm$ 0.35 \\ 
    &  $\mu$ &$-$1.79 $\pm$ 0.03 & 0.95 $\pm$ 0.03 & 5.35 $\pm$ 0.03 \\ 
    &$\sigma$ &0.32 $\pm$ 0.01 & 2.55 $\pm$ 0.19 & 0.75 $\pm$ 0.01 \\ 
    \hline
    \multirow{8}{*}{ 10$^\circ < i < 45^\circ$} & a$_1$ & 0.23 $\pm$ 0.01 & $-$0.39 $\pm$ 0.06 & 0.6 $\pm$ 0.01 & 0.57 $\pm$ 0.05 \\ 
    & $a_2$&$-$0.01 $\pm$ 0.0 & $-$4.95 $\pm$ 3.76 & 0.14 $\pm$ 0.0 & 5.79 $\pm$ 4.57 \\ 
    & $a_3$ &$-$0.19 $\pm$ 0.03 & $-$0.25 $\pm$ 0.09 & 0.29 $\pm$ 0.03 & 0.37 $\pm$ 0.1 \\ 
    & $\gamma$&73.39 $\pm$ 1.04 & $-$0.56 $\pm$ 0.04 & 44.3 $\pm$ 0.44 & 1.61 $\pm$ 0.05 \\ 
    & $\sigma_1$ &2.34 $\pm$ 0.18 & $-$0.9 $\pm$ 0.22 & 6.54 $\pm$ 0.11 & 1.88 $\pm$ 0.33 \\ 
    & $\sigma_2$  &18.51 $\pm$ 0.51 & $-$1.07 $\pm$ 0.09 & 16.42 $\pm$ 0.2 & 2.41 $\pm$ 0.13 \\ 
    & $\mu$ & $-$1.68 $\pm$ 0.03 & 1.03 $\pm$ 0.04 & 5.33 $\pm$ 0.03 \\ 
    &$\sigma$ &0.35 $\pm$ 0.01 & 2.47 $\pm$ 0.15 & 0.76 $\pm$ 0.01 \\ 
    \hline
    \multirow{8}{*}{ 45$^\circ < i < 80^\circ$} & a$_1$ & 0.2 $\pm$ 0.01 & $-$0.36 $\pm$ 0.07 & 0.6 $\pm$ 0.01 & 0.6 $\pm$ 0.07 \\ 
    & $a_2$&$-$0.04 $\pm$ 0.01 & $-$0.56 $\pm$ 0.2 & 0.19 $\pm$ 0.0 & 1.01 $\pm$ 0.27 \\ 
    & $a_3$ &$-$0.14 $\pm$ 0.02 & $-$0.26 $\pm$ 0.1 & 0.23 $\pm$ 0.02 & 0.38 $\pm$ 0.1 \\ 
    & $\gamma$&100.65 $\pm$ 2.96 & $-$0.74 $\pm$ 0.07 & 77.61 $\pm$ 1.44 & 1.5 $\pm$ 0.09 \\ 
    & $\sigma_1$ &2.47 $\pm$ 0.15 & $-$0.93 $\pm$ 0.16 & 7.02 $\pm$ 0.09 & 1.73 $\pm$ 0.23 \\ 
    & $\sigma_2$  &24.27 $\pm$ 1.14 & $-$1.04 $\pm$ 0.13 & 21.46 $\pm$ 0.5 & 1.92 $\pm$ 0.18 \\ 
    & $\mu$ &$-$1.42 $\pm$ 0.02 & 1.33 $\pm$ 0.05 & 5.33 $\pm$ 0.02 \\ 
    &$\sigma$ &0.51 $\pm$ 0.0 & 2.25 $\pm$ 0.05 & 0.8 $\pm$ 0.0 \\ 
    \hline
    \multirow{8}{*}{ 80$^\circ < i $} & a$_1$ & 0.18 $\pm$ 0.01 & $-$0.35 $\pm$ 0.08 & 0.59 $\pm$ 0.01 & 0.66 $\pm$ 0.08 \\ 
    & $a_2$&$-$0.06 $\pm$ 0.01 & $-$0.38 $\pm$ 0.13 & 0.21 $\pm$ 0.0 & 0.83 $\pm$ 0.18 \\ 
    & $a_3$ &$-$0.11 $\pm$ 0.02 & $-$0.29 $\pm$ 0.11 & 0.2 $\pm$ 0.02 & 0.41 $\pm$ 0.12 \\ 
    & $\gamma$&107.04 $\pm$ 4.11 & $-$0.86 $\pm$ 0.1 & 107.77 $\pm$ 2.26 & 1.43 $\pm$ 0.12 \\ 
    & $\sigma_1$ &2.42 $\pm$ 0.21 & $-$1.16 $\pm$ 0.27 & 7.13 $\pm$ 0.13 & 1.99 $\pm$ 0.39 \\ 
    & $\sigma_2$  &25.1 $\pm$ 1.59 & $-$0.94 $\pm$ 0.17 & 24.26 $\pm$ 0.82 & 1.56 $\pm$ 0.21 \\ 
    & $\mu$ &$-$1.22 $\pm$ 0.02 & 1.63 $\pm$ 0.08 & 5.31 $\pm$ 0.02 \\ 
    &$\sigma$ &0.8 $\pm$ 0.01 & 1.87 $\pm$ 0.07 & 0.84 $\pm$ 0.01 \\ 
    \hline
    \multirow{8}{*}{ $b_{\rm offset}$ < 5 kpc } & a$_1$ &0.17 $\pm$ 0.01 & $-$0.26 $\pm$ 0.11 & 0.65 $\pm$ 0.01 & 0.6 $\pm$ 0.06 \\ 
    & $a_2$&$-$0.03 $\pm$ 0.01 & 0.16 $\pm$ 0.45 & 0.17 $\pm$ 0.0 & 1.0 $\pm$ 0.5 \\ 
    & $a_3$ & $-$0.09 $\pm$ 0.01 & $-$0.41 $\pm$ 0.16 & 0.18 $\pm$ 0.01 & 0.38 $\pm$ 0.1 \\ 
    & $\gamma$&114.96 $\pm$ 5.06 & $-$0.61 $\pm$ 0.11 & 101.19 $\pm$ 1.35 & 1.58 $\pm$ 0.11 \\ 
    & $\sigma_1$ &11.19 $\pm$ 0.47 & $-$0.86 $\pm$ 0.11 & 7.07 $\pm$ 0.08 & 2.49 $\pm$ 0.14 \\ 
    & $\sigma_2$  &53.76 $\pm$ 3.93 & $-$0.89 $\pm$ 0.21 & 24.94 $\pm$ 0.51 & 2.22 $\pm$ 0.2 \\ 
    & $\mu$ &$-$1.38 $\pm$ 0.03 & 0.8 $\pm$ 0.04 & 5.53 $\pm$ 0.04 \\ 
    &$\sigma$ &0.96 $\pm$ 0.04 & 3.8 $\pm$ 0.33 & 0.76 $\pm$ 0.02 \\
    \hline
    \multirow{8}{*}{ 5 kpc < $b_{\rm offset}$ < 10 kpc} & a$_1$ &0.21 $\pm$ 0.01 & $-$0.35 $\pm$ 0.1 & 0.6 $\pm$ 0.01 & 0.58 $\pm$ 0.07 \\ 
    & $a_2$&$-$0.03 $\pm$ 0.01 & $-$0.54 $\pm$ 0.66 & 0.18 $\pm$ 0.0 & 1.3 $\pm$ 0.72 \\ 
    & $a_3$ &$-$0.17 $\pm$ 0.01 & $-$0.33 $\pm$ 0.1 & 0.25 $\pm$ 0.01 & 0.4 $\pm$ 0.09 \\ 
    & $\gamma$&93.88 $\pm$ 2.61 & $-$0.9 $\pm$ 0.07 & 56.74 $\pm$ 0.48 & 1.7 $\pm$ 0.07 \\ 
    & $\sigma_1$ &3.64 $\pm$ 0.27 & $-$1.06 $\pm$ 0.2 & 6.89 $\pm$ 0.07 & 1.85 $\pm$ 0.22 \\ 
    & $\sigma_2$  &27.13 $\pm$ 1.85 & $-$1.3 $\pm$ 0.19 & 18.81 $\pm$ 0.28 & 2.32 $\pm$ 0.2 \\ 
    & $\mu$ &$-$1.56 $\pm$ 0.03 & 1.36 $\pm$ 0.07 & 5.32 $\pm$ 0.03 \\ 
    &$\sigma$ &0.43 $\pm$ 0.02 & 2.41 $\pm$ 0.25 & 0.83 $\pm$ 0.01 \\ 
    \hline
    \multirow{8}{*}{ 10 kpc < $b_{\rm offset}$ < 30 kpc} & a$_1$ &0.17 $\pm$ 0.04 & $-$0.37 $\pm$ 0.15 & 0.58 $\pm$ 0.04 & 0.49 $\pm$ 0.18 \\ 
    & $a_2$&0.0 $\pm$ 0.01 & 1.09 $\pm$ 11.27 & 0.16 $\pm$ 0.01 & 0.0 $\pm$ 0.69 \\ 
    & $a_3$ &$-$0.18 $\pm$ 0.05 & $-$0.32 $\pm$ 0.14 & 0.3 $\pm$ 0.05 & 0.39 $\pm$ 0.16 \\ 
    & $\gamma$&62.39 $\pm$ 2.49 & $-$0.72 $\pm$ 0.1 & 37.06 $\pm$ 1.83 & 1.3 $\pm$ 0.13 \\ 
    & $\sigma_1$ &0.88 $\pm$ 0.26 & $-$0.28 $\pm$ 0.25 & 6.44 $\pm$ 0.25 & 0.46 $\pm$ 0.27 \\ 
    & $\sigma_2$  &9.36 $\pm$ 0.83 & $-$0.55 $\pm$ 0.17 & 15.2 $\pm$ 0.79 & 1.02 $\pm$ 0.23 \\ 
    & $\mu$ &$-$1.66 $\pm$ 0.04 & 1.49 $\pm$ 0.09 & 5.18 $\pm$ 0.04 \\ 
    &$\sigma$ &0.07 $\pm$ 0.01 & 2.01 $\pm$ 0.8 & 0.94 $\pm$ 0.01 \\
    \hline
    \multirow{8}{*}{ 30 kpc < $b_{\rm offset}$} & a$_1$ &$-$0.17 $\pm$ 0.09 & $-$0.79 $\pm$ 0.59 & 0.57 $\pm$ 0.09 & 0.55 $\pm$ 0.43 \\ 
    & $a_2$&0.06 $\pm$ 0.02 & $-$1.07 $\pm$ 0.48 & 0.11 $\pm$ 0.01 & 2.64 $\pm$ 1.02 \\ 
    & $a_3$ &$-$0.01 $\pm$ 0.28 & 1.57 $\pm$ 9.33 & 0.41 $\pm$ 0.34 & 0.0 $\pm$ 10.36 \\ 
    & $\gamma$&56.4 $\pm$ 2.44 & $-$0.69 $\pm$ 0.11 & 21.59 $\pm$ 1.88 & 1.39 $\pm$ 0.15 \\ 
    & $\sigma_1$ &0.13 $\pm$ 0.15 & $-$2.59 $\pm$ 2.34 & 6.12 $\pm$ 0.08 & 7.25 $\pm$ 5.9 \\ 
    & $\sigma_2$  &5.42 $\pm$ 2.52 & $-$0.97 $\pm$ 0.95 & 16.16 $\pm$ 2.17 & 2.08 $\pm$ 1.79 \\ 
    & $\mu$ &$-$2.3 $\pm$ 0.22 & 0.58 $\pm$ 0.1 & 5.72 $\pm$ 0.24 \\ 
    &$\sigma$ &$-$0.21 $\pm$ 0.05 & 5.29 $\pm$ 2.65 & 1.06 $\pm$ 0.02 \\ 
    \hline
    \multicolumn{6}{l}{\textbf{Notes:} The RM PDF is fitted as the sum of a Lorentzian and two Gaussian functions.} \\
    \multicolumn{6}{l}{$a_1$, $a_2$ and $a_3$ are normalization factors for the RM PDF.}\\
    \multicolumn{6}{l}{$\gamma$ is the parameter of the Lorentzian function, and $\sigma_1$ and $\sigma_2$ of the Gaussian functions.}\\
    \multicolumn{6}{l}{The DM PDF is fitted as a log normal function, with parameters $\mu$ and $\sigma$.}
    
\label{tab:RM_param_fit}

\end{longtable}

\end{scriptsize}
\twocolumn

\section{Application to localized FRBs: FRB190608 as a case study}
\label{FRBexample}

We show one example of how DM$_{\rm host,rf}$ and RM$_{\rm host,rf}$ contribution of a localized FRB with an identified host galaxy (with detailed information about the galaxy properties and the FRB position within the host) can be constrained using TNG50. First, we select analog galaxies with similar properties to the host. Then we calculate DM$_{\rm host,rf}$ and RM$_{\rm host,rf}$ in two ways: (1) by placing FRBs in these galaxies at specific locations, and (2) using our sightlines calculated in Section \ref{Results} (referred to as `database' from here on) that match the host's $i$ and the FRB's $b_{\rm offset}$. We compare these methods to see how well our database of sightlines can be used. We also show that our estimates are consistent with observations.

\begin{figure*}[b]
    \centering
    \includegraphics[width=15cm]{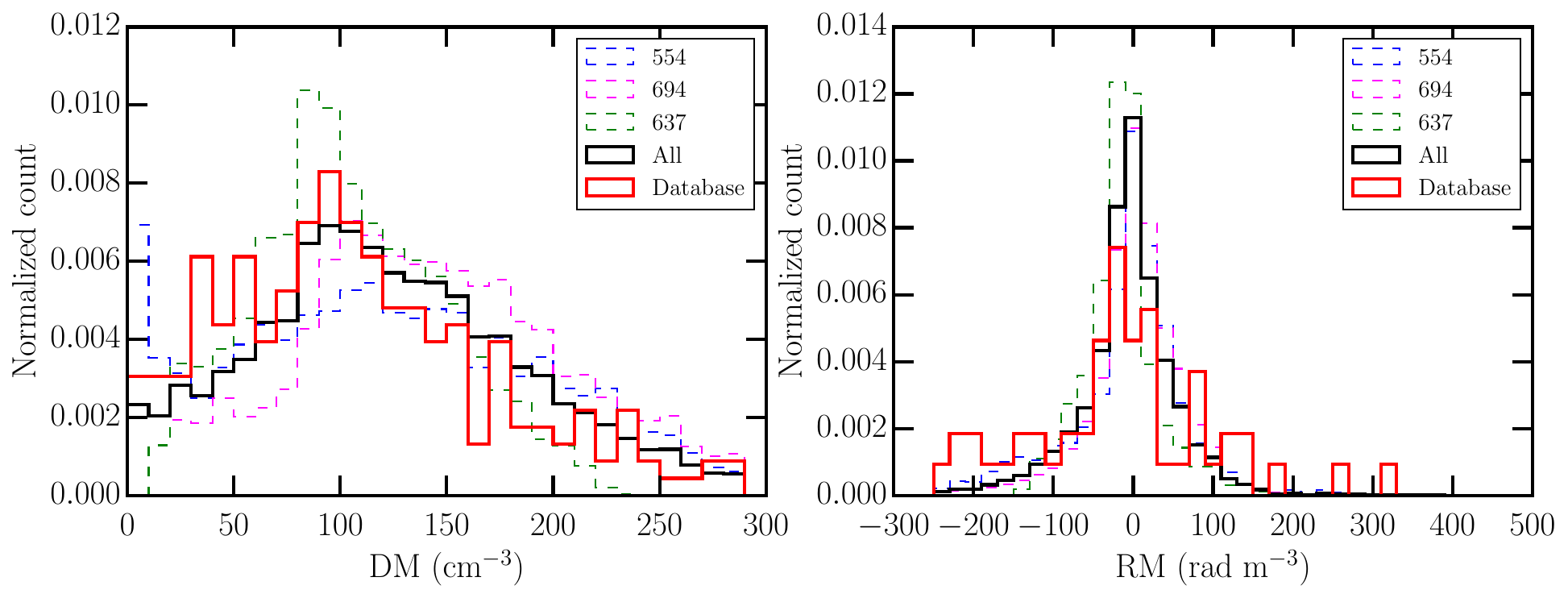}

    \caption{\textbf{Left:} The DM$_{\rm host,rf}$ distribution of FRBs placed in the analogs of the host galaxy of FRB190608. \textbf{Right:} The RM$_{\rm host,rf}$ distribution of FRBs placed in the analogs of the host galaxy of FRB190608.}
    \label{fig:FRB19}
\end{figure*}

\begin{table*}[b]
    \centering
    \caption{The median and 1$\sigma$ range of FRB190806's DM$_{\rm host,rf}$ and RM$_{\rm host,rf}$ from different sightlines. }
    \begin{tabular}{l c|c c c c}
         \hline
         \hline
         Subset of & $N_{\rm sl}$ & median DM$_{\rm host,rf}$ & 1$\sigma$ DM$_{\rm }$ & median RM$_{\rm host,rf}$ & 1$\sigma$ RM$_{\rm }$\\
         sightlines & & ($\dm$) & ($\dm$) & ($\radm$) & $(\radm$) \\ 
         \hline
         Spec. loc. & 3000 & 118 & 60 -- 190 & +7 & $-$74 -- +99 \\
         DB. all & 3000 & 103 & 47~--~207 & +3 & $-$192~--~+231\\
         DB. $b_{\rm offset}$ & 257 & 107 & 49~--~207 & +3 & $-$222~--~+226 \\
         DB. $i$ & 61 & 87 & 50~--~146 & $-17$ & $-$186~--~+89\\
         DB. $b_{\rm offset}$+$i$ & 7 & 93 & 75~--~144 & +24 &$-$143~--~+262 \\
        \hline
        
    \end{tabular}
    
    \tablefoot{The first line lists the results from placing the FRBs at the minor axis of the host galaxy analogs, and the rest of the table shows the results using our `database' of sightlines from Section \ref{Res:DMvsz}. We show the results using all sightlines from the analog galaxies in the database, sightlines selected based on their galaxy inclination, $b_{\rm offset}$, and both. We show how the number of appropriate sightlines decreases when we apply stricter selection rules.}
    
    \label{tab:FRB190608}
\end{table*}

\subsection{Background information of FRB190608 and TNG50 analogs}
The host galaxy of FRB190608 is well known (e.g. \citealt{2020Natur.581..391M}), thus it provides numerous constraints on both galaxy properties and the location of the FRB. The host is a spiral galaxy at $z=0.12$ with $\logM = 10.06^{+0.09}_{-0.12}$ and an SFR of $0.69 \pm 0.21\,M_\odot\,{\rm yr}^{-1}$ \citep{2020ApJ...903..152H}. Its inclination is 43$^{\circ}$ and its position angle (PA) is 54$^{\circ}$. The FRB is in a spiral arm, along the minor axis of the galaxy, at a projected offset of $6.52 \pm 0.82$\,kpc from the center \citep{2021ApJ...917...75M}. The DM and RM of the FRB were measured as well: DM$_{\rm obs} = 338.7 \pm 0.5$\,pc\,cm$^{-3}$ \citep{2020Natur.581..391M} and RM$_{\rm obs} = +353 \pm 2\,\radm$ \citep{2020MNRAS.497.3335D}. These contain all the contributions from the different components along the line of sight, such as that of the MW (DM$_{\rm MW}$~=~38 pc cm$^{-3}$ \citealt{2002astro.ph..7156C}, DM$_{\rm MW, halo}$ = 39 $\dm$ \citealt{2019MNRAS.485..648P}, and RM$_{\rm MW}$ = $-$24 $\pm$ 13 $\radm^{-2}$ \citealt{2022A&A...657A..43H} towards FRB190608).

From the snapshot at $z=0.1$, we found three analogs (IDs 554, 637 and 694) of the host galaxy of FRB190608 from the TNG50 simulation based on their stellar mass (within the error), SFR (same SFR$_{\rm{MS}}$-SFR difference\footnote{We calculate the host's difference from the MS of star-forming galaxies using the relation from \cite{2012ApJ...757...54Z}, then we use the same difference from the simulation derived MS to constrain the SFR of the host galaxy analogs.}), and with an $r_{\rm SF,99}$ larger than the offset of the FRB. All the selected galaxies are the first (most massive) subhalos of their parent halo. 

\subsection{Placing FRBs at specific locations}

We rotate the galaxies to $i = 43^{\circ}$ \citep{2021ApJ...917...75M}, and we place 1000 FRBs along the minor axis of each analog galaxy. We then calculate the DM and RM in the same way as we calculated for the other FRBs in this paper (see Section \ref{Methods}).

The DM distributions of the three host galaxy analogs of FRB190608 are shown in the left panel of Fig. \ref{fig:FRB19}, and the median and $1\sigma$ range are listed in the first row of Table \ref{tab:FRB190608}. As expected, the derived DM are smaller than the observed DM of the FRB, even after accounting for the MW contribution (i.e. DM$_{\rm FRB}-$ DM$_{\rm MW} \sim$ 300 $\dm$). \cite{2021ApJ...922..173C} derived the host DM from the H$\alpha$ emission measure in the observer's frame: the DM of the host ISM is 94 $\pm$ 38 $\dm$ and the DM of the host halo is 55 $\pm$ 25 $\dm$. Including both the ISM and the halo of the host, they derived a range of 100 -- 190 $\dm$, consistent with our results, but suggesting we might have some DM contribution from the halo.

 The RM distributions are shown in the right panel of Fig. \ref{fig:FRB19}, and the median and $1\sigma$ range are listed in the first row of Table \ref{tab:FRB190608}. We note that median RM$_{\rm host,rf}$ might be further from 0 $\radm$ as we have multiple restrictions on the position of the FRB, which can skew the distribution.

\subsection{Using DM and RM from Section \ref{Results}}

We can also provide DM and RM estimates using our database of 16.5 million sightlines calculated in Section \ref{Results}, however, there will be fewer sightlines that fulfill all the placing criteria we used above. In table \ref{tab:FRB190608}, we show the median and 1$\sigma$ ranges of DM and RM of all sightlines that belong to the three analog galaxies regardless of $i$ and $b_{\rm offset}$, of sightlines that have $b_{\rm offset} =$ 6.52 $\pm$ 0.82 kpc, of sightlines with $42.27^{\circ}<i<43.95^{\circ}$, and of sightlines that fulfill both criteria. The DM and RM from different sightline selections are of the same magnitude. We find that constraining the inclination of selected sightlines had a bigger effect on the ranges of DM and RM than constraining $b_{\rm offset}$. This is the case even if we increase the uncertainty of the inclination ($32.27^{\circ}$<i<53.95$^{\circ}$), suggesting it is not caused by the lower relative uncertainty of $i$ compared to $b_{\rm offset}$.

We can see that in the case of DM, even without strict selection criteria, we get a similar range of values from our database and from placing FRBs at a more specified location. However, for RM the possible range is much larger when using our database, thus it might be more sensitive to the exact place in the galaxy. The DM of the $b_{\rm offset}$-selected sightlines, and the RM of the $i$-selected sightlines match the results of our previous test the most. Therefore, we suggest selecting sightlines based on $i$ when one wants to constrain RM$_{\rm host}$ from our database of sightlines.

\subsection{IGM contribution and summary}

We constrain the IGM's DM contribution by subtracting our host contribution estimate and the contribution of the Milky Way from the observed DM. We obtain a residual DM$_{\rm res}$~=~105~--~212 pc cm$^{-3}$. This DM can arise from the IGM and the immediate source environment, making this estimate an upper limit on DM$_{\rm IGM}$. In comparison, \citealt{2020ApJ...901..134S} derived a DM$_{\rm IGM}$ (also including intervening halos) of 98 -- 154 pc cm$^{-3}$ using the Monte Carlo Physarum Machine \citep{PMID:34905603}, similar to our result. We may have some larger DM values because we did not take into account the immediate environment of the source. If we subtract the above contributions from the observed RM, we derive a range of residual RM$_{\rm res}$~=~+283~--~+451 rad~m$^{-2}$, which includes the contributions of the IGM and the immediate source environment. In the case of RM, the contributions along the line of sight can be negative or positive, hence it is difficult to determine if RM$_{\rm IGM}$ is lower or higher than this estimate.

In summary, we showed that IllustrisTNG, in one example, gives reasonable estimates on the DM and RM of the host galaxy of FRB190608, as they are consistent with observations, thus our database of sightlines can be used to constrain the DM and RM contributions of host galaxies. However, we note that the range of values is wide for one single host galaxy contribution. Thus, a large sample of FRBs is needed to statistically constrain DM$_{\rm host,rf}$ and RM$_{\rm host,rf}$. This will be improved by future simulations (with better resolution and updated models), and even better characterization of the host galaxies of FRBs.

\FloatBarrier

\section{Calculation of Magnetic field parameters}
\label{appendixB}
 The magnetic field properties are calculated similarly to \cite{2017MNRAS.469.3185P} and \cite{2018MNRAS.481.4410P}.
\subsection{Magnetic field strength - face-on}

We calculate the projected absolute magnetic field strength in a face-on view for every galaxy, with two examples shown in the second rows of Fig. \ref{fig:B_486}. For this, we calculate the line-of-sight integral of the magnetic energy density ($B^2$) in an r$_{\rm SF,99}$ x  r$_{\rm SF,99}$ x~2$h$ box using Eq. (1) from \cite{2017MNRAS.469.3185P}. We chose $h$ to be 1 kpc (i.e. we integrate from -1 kpc below the disk to +1 kpc above the disk), and the size of a pixel in the map is 80 pc x 80 pc. Using $B^2$ makes the resulting $B$ strength independent of the sign of the field. We calculate the average magnetic field strength in the disk by averaging the magnetic field strength of the pixels.

\begin{figure}
    \centering
    \includegraphics[width=7.8cm]{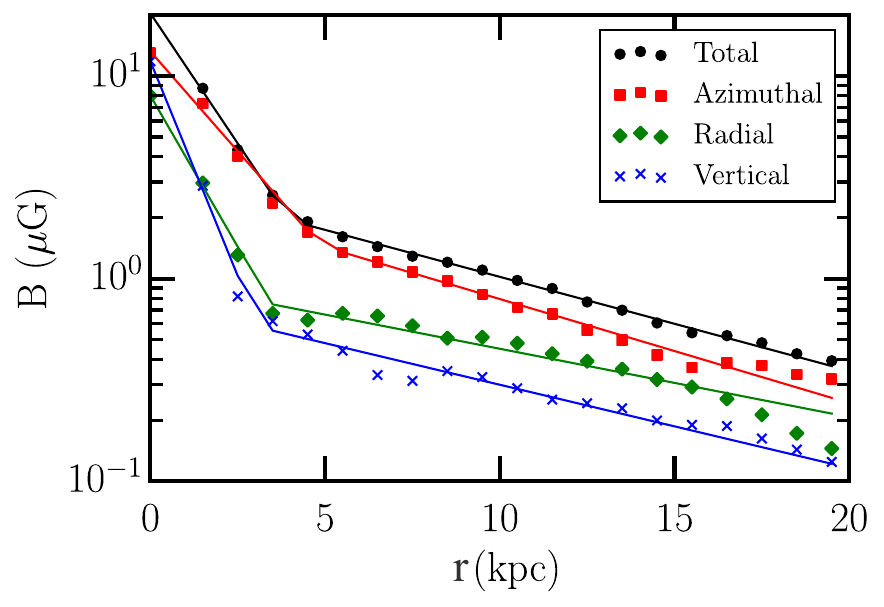}
    \includegraphics[width=7.8cm]{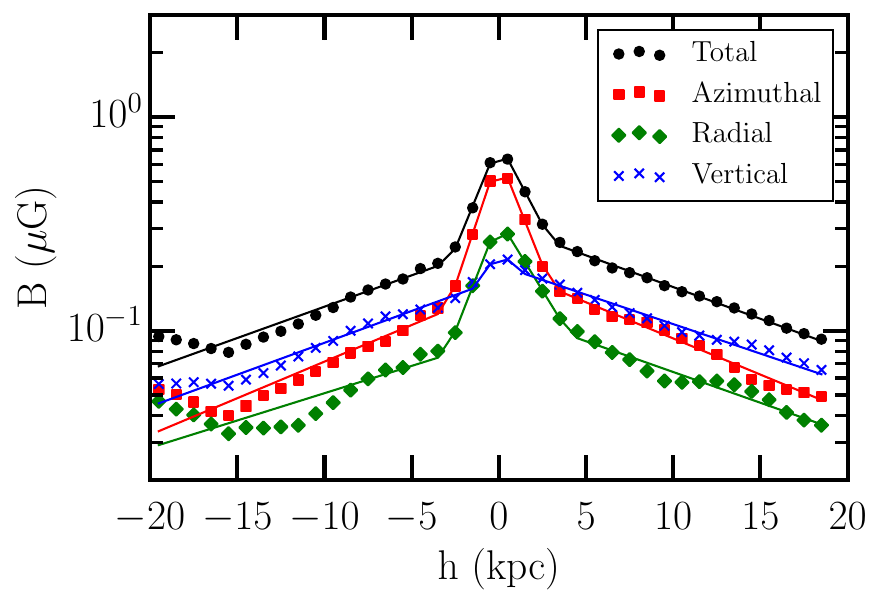}
    
    \caption{The radial (top) and vertical (bottom) profile of the total magnetic field strength and the magnetic field components of an example galaxy (ID487) at $z=0$, with a double exponential function fitted to each, and different fit parameters for the vertical profile below and above the midplane of the galaxy.}
    \label{fig:B_profile_radial}
\end{figure}

We then separate the magnetic field into azimuthal, radial, and vertical components. Then we calculate the radial profiles (an example is shown in \ref{fig:B_profile_radial}) for the total B field and the different components, using Eq. (2) from \cite{2017MNRAS.469.3185P}. We fit a double exponential (following Eq. (3) from \cite{2017MNRAS.469.3185P}) to these radial profiles, and list the median parameters for the total field strength in Table \ref{tab:scale_length} for every redshift. 

\begin{table}[b]
    \centering
    \caption{The medians of the fitted parameters of the radial profiles of total magnetic field strength at each redshift. }
    \bgroup
\def\arraystretch{1.3}
\begin{tabular}{c  c c c c }
 \hline\hline        
  $z$ & $B_{\rm center}$ & $r_0^B$ & $r_{\rm \rm inner}^B$ & $r_{\rm outer}^B$ \\
   & ($\mu$G) & (kpc) & (kpc) & (kpc) \\
     \hline
     0 & $10_{-5}^{+9}$ & $2.8_{-1.4}^{+1.0}$ & $1.6_{-0.4}^{+0.4}$ & $5.3_{-2.3}^{+4.7}$\\
0.1 & $10_{-5}^{+9}$ & $2.8_{-1.4}^{+1.1}$ & $1.6_{-0.4}^{+0.4}$ & $5.1_{-2.1}^{+5.0}$\\
0.2 & $10_{-5}^{+10}$ & $2.7_{-1.5}^{+1.0}$ & $1.6_{-0.4}^{+0.4}$ & $4.9_{-2.0}^{+4.9}$\\
0.3 & $11_{-6}^{+11}$ & $2.5_{-1.2}^{+1.0}$ & $1.5_{-0.4}^{+0.4}$ & $4.7_{-1.9}^{+4.4}$\\
0.4 & $11_{-6}^{+12}$ & $2.4_{-1.3}^{+1.1}$ & $1.5_{-0.3}^{+0.4}$ & $4.7_{-1.9}^{+4.1}$\\
0.5 & $12_{-6}^{+13}$ & $2.4_{-1.2}^{+1.3}$ & $1.5_{-0.3}^{+0.4}$ & $4.6_{-1.9}^{+3.9}$\\
0.7 & $13_{-6}^{+15}$ & $2.3_{-1.1}^{+1.3}$ & $1.5_{-0.4}^{+0.4}$ & $4.2_{-1.6}^{+3.1}$\\
1 & $16_{-7}^{+19}$ & $2.4_{-1.0}^{+1.6}$ & $1.5_{-0.3}^{+0.4}$ & $4.0_{-1.5}^{+2.6}$\\
1.5 & $19_{-9}^{+32}$ & $2.7_{-1.2}^{+1.7}$ & $1.5_{-0.3}^{+0.4}$ & $3.7_{-1.3}^{+2.3}$\\
2 & $22_{-10}^{+42}$ & $2.9_{-1.4}^{+2.1}$ & $1.4_{-0.3}^{+0.4}$ & $3.5_{-1.2}^{+2.5}$\\
     \hline
\end{tabular}
\tablefoot{The error bars indicate the 1$\sigma$ range.}
\egroup
    \label{tab:scale_length}
\end{table}

\begin{table*}[]
    \centering
    \caption{The medians of the fitted parameters of the vertical profiles of total magnetic field strength at each redshift.}
    \bgroup
\def\arraystretch{1.3}

    \begin{tabular}{c c c c c c c c}
     \hline\hline        
  $z$ & $B_{\rm center}$ & $h_{0,1}^B$ & $h_{\rm \rm inner,1}^B$ & $h_{\rm outer,1}^B$ & $h_{0,2}^B$ & $h_{\rm \rm inner,2}^B$ & $h_{\rm outer,2}^B$ \\
   & ($\mu$G) & (kpc) & (kpc) & (kpc) & (kpc) & (kpc) & (kpc) \\
     \hline

0 & $0.5_{-0.2}^{+0.4}$ & $-3.3_{-1.5}^{+1.4}$ & $3.3_{-1.3}^{+2.1}$ & $14.6_{-6.6}^{+17.2}$ & $3.2_{-1.3}^{+1.5}$ & $3.2_{-1.2}^{+2.1}$ & $14.8_{-6.7}^{+18.2}$\\
0.1 & $0.5_{-0.2}^{+0.4}$ & $-3.2_{-1.5}^{+1.3}$ & $3.0_{-1.1}^{+1.9}$ & $14.3_{-6.1}^{+15.7}$ & $3.3_{-1.3}^{+1.4}$ & $3.0_{-1.1}^{+2.0}$ & $14.4_{-6.6}^{+16.6}$\\
0.2 & $0.6_{-0.2}^{+0.4}$ & $-3.2_{-1.5}^{+1.3}$ & $3.0_{-1.1}^{+1.8}$ & $14.7_{-6.5}^{+16.8}$ & $3.3_{-1.3}^{+1.4}$ & $3.0_{-1.1}^{+1.7}$ & $14.9_{-6.9}^{+19.6}$\\
0.3 & $0.6_{-0.2}^{+0.4}$ & $-3.3_{-1.5}^{+1.3}$ & $2.9_{-1.0}^{+1.8}$ & $15.3_{-6.5}^{+19.3}$ & $3.2_{-1.3}^{+1.5}$ & $2.9_{-1.0}^{+1.8}$ & $14.7_{-6.7}^{+19.5}$\\
0.4 & $0.6_{-0.3}^{+0.4}$ & $-3.3_{-1.6}^{+1.3}$ & $2.9_{-1.0}^{+1.8}$ & $15.6_{-7.1}^{+21.8}$ & $3.3_{-1.3}^{+1.6}$ & $2.9_{-1.0}^{+1.8}$ & $15.5_{-7.3}^{+23.5}$\\
0.5 & $0.6_{-0.3}^{+0.5}$ & $-3.3_{-1.7}^{+1.3}$ & $2.9_{-1.0}^{+1.7}$ & $15.1_{-6.9}^{+21.5}$ & $3.2_{-1.3}^{+1.6}$ & $2.9_{-1.0}^{+1.7}$ & $14.6_{-6.7}^{+21.3}$\\
0.7 & $0.7_{-0.3}^{+0.5}$ & $-3.2_{-1.5}^{+1.3}$ & $2.8_{-0.9}^{+1.7}$ & $15.1_{-6.8}^{+18.9}$ & $3.1_{-1.3}^{+1.4}$ & $2.8_{-1.0}^{+1.7}$ & $15.4_{-7.1}^{+20.2}$\\
1 & $0.7_{-0.3}^{+0.6}$ & $-3.0_{-1.5}^{+1.2}$ & $2.8_{-1.0}^{+1.7}$ & $15.2_{-6.5}^{+18.7}$ & $3.0_{-1.3}^{+1.6}$ & $2.7_{-1.0}^{+1.8}$ & $15.5_{-7.0}^{+18.3}$\\
1.5 & $0.7_{-0.3}^{+0.5}$ & $-2.8_{-1.6}^{+1.3}$ & $2.8_{-1.0}^{+1.7}$ & $15.1_{-6.0}^{+19.2}$ & $2.7_{-1.3}^{+1.6}$ & $2.8_{-1.0}^{+1.7}$ & $15.3_{-6.3}^{+15.8}$\\
2 & $0.6_{-0.3}^{+0.4}$ & $-2.5_{-1.5}^{+1.2}$ & $2.8_{-1.1}^{+1.8}$ & $15.8_{-5.8}^{+14.2}$ & $2.4_{-1.2}^{+1.5}$ & $2.8_{-1.1}^{+1.7}$ & $15.5_{-6.0}^{+14.8}$\\\hline
\end{tabular}
\tablefoot{The error bars indicate the 1$\sigma$ range.}
\egroup
    \label{tab:scale_height}
\end{table*}

\subsection{Magnetic field strength - edge-on}

We calculate the magnetic field strength in edge-on view (examples shown in the top four panels of Fig. \ref{fig:B_486} for each galaxy, similarly to face-on view. We integrate through 2$r_{\rm SF,99}$. In Fig. \ref{fig:B_profile_radial} we show the height profile of the $B$ field strength (total and its components) for one example galaxy, calculated from slices with a 1 kpc width. A double exponential can be fitted to both sides of this profile, and the profiles below and above the disk of the galaxy are not symmetric for every galaxy. The medians of the fitted parameters at each redshift are listed in Table \ref{tab:scale_height}.

\subsection{Magnetic field structure}
\begin{figure*}
    \centering
    \includegraphics[width=15.3cm]{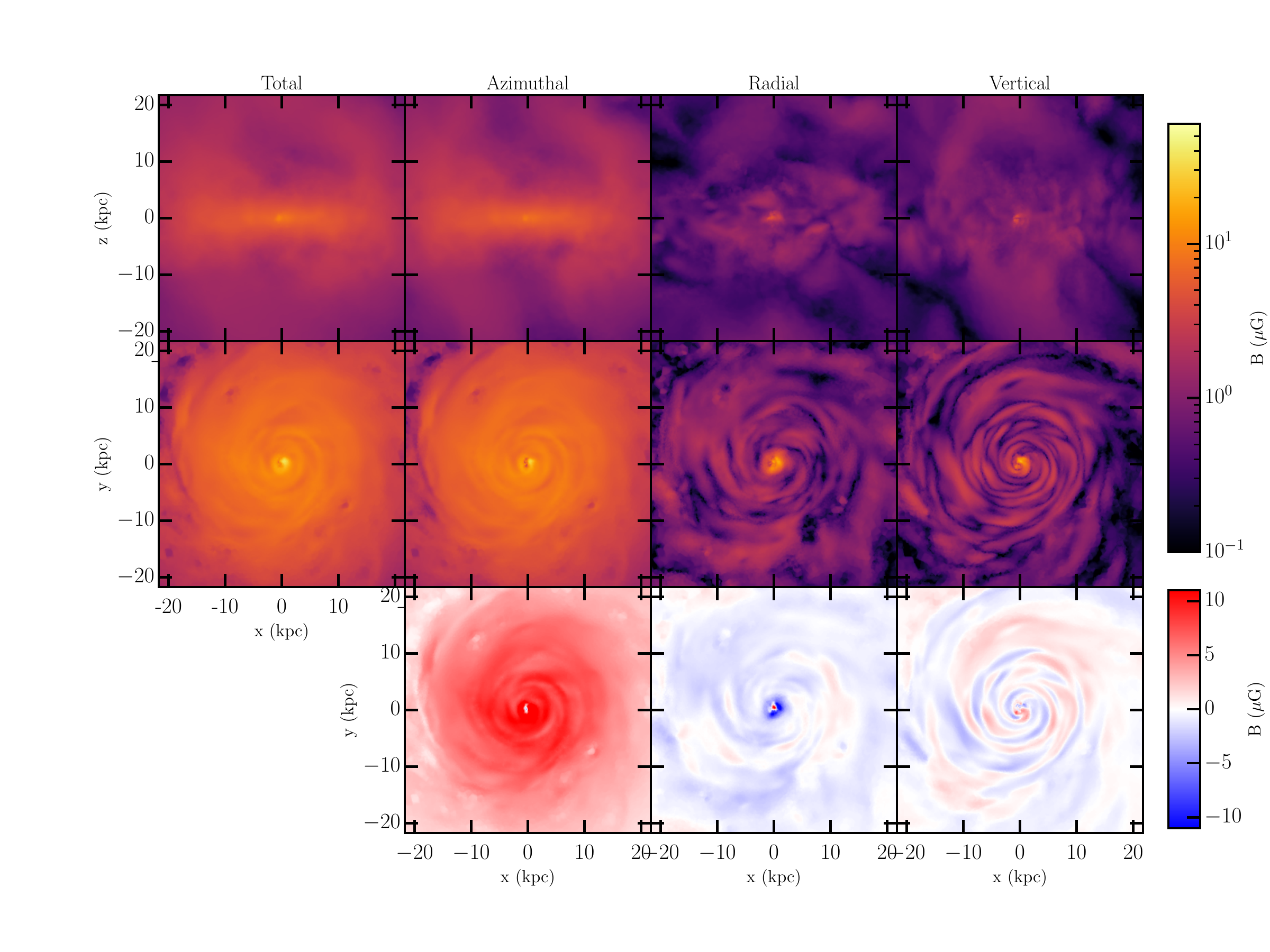}
    \includegraphics[width=15.3cm]{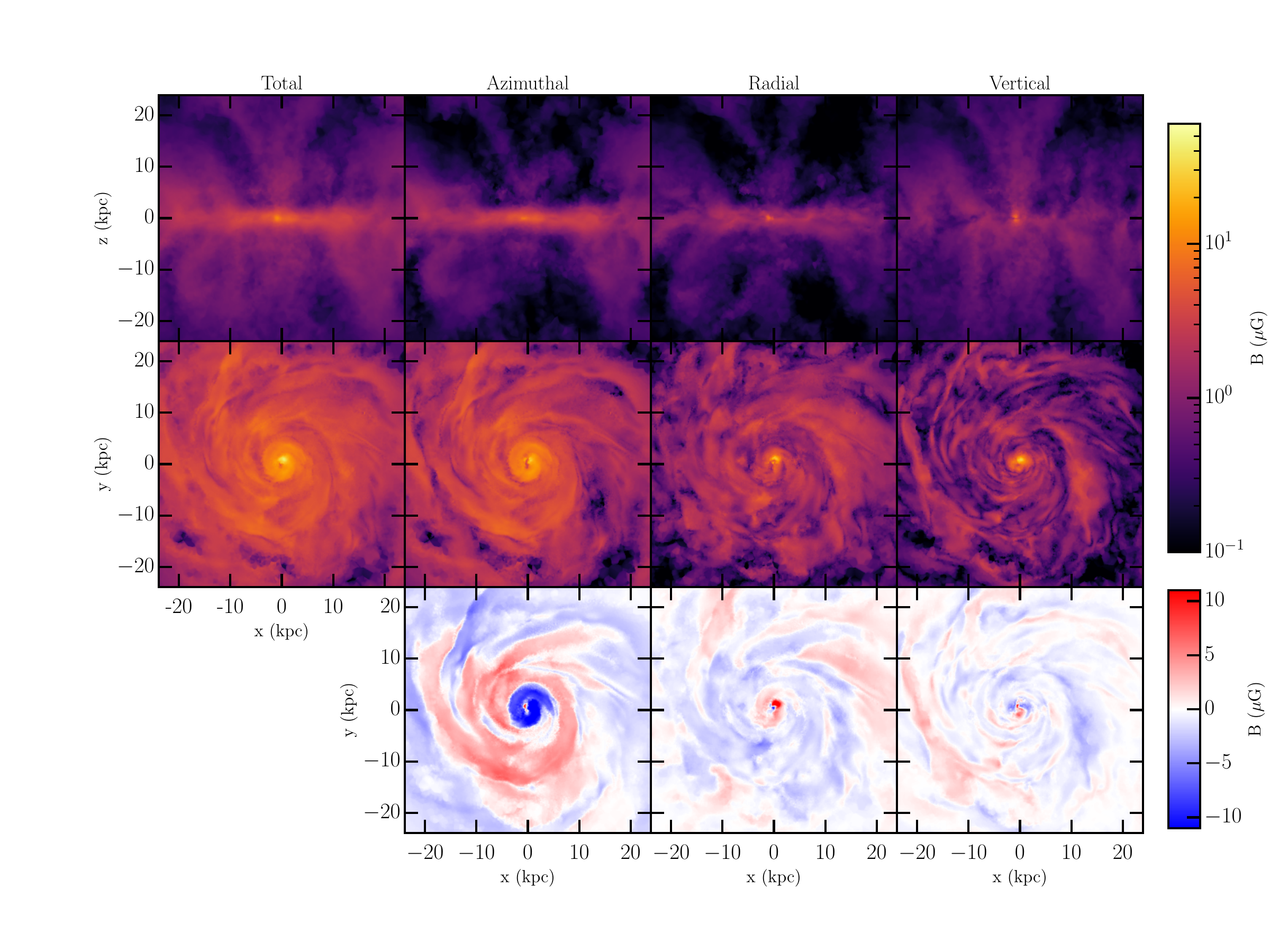}
\caption{We show the magnetic field strength maps (total, and components $-$ azimuthal, radial and vertical) of two example galaxies (ID486 and ID487) at $z=0$. These maps are calculated for every galaxy. The top rows show the magnetic field strength maps in an edge-on view, and the middle rows show them in a face-on view. The bottom rows show the magnetic field strength maps in a face-on view, calculated by keeping the sign of the B field while integrating the magnetic field components.
    For galaxy ID486 the azimuthal field's sign remains the same throughout the disk, but for galaxy ID487 the bottom row reveals reversals in the azimuthal magnetic field of the galaxy.}
    \label{fig:B_486}
\end{figure*}

We investigated the structure of magnetic fields, by integrating the magnetic field strength of each component ($B_{\rm component}$, which can be azimuthal, radial, or vertical) separately through a thin projection of 2 kpc depth of the galaxy (the resulting maps are shown in the bottom rows of Fig. \ref{fig:B_486}):
\begin{equation}
  B\left(x,y\right) = \frac{1}{2h} \int_{-h}^{+h} B_{\rm component}\left(x,y,z\right) \rm{d}z.
  \label{eq:B4}
\end{equation}
This preserves their signs, and provides us information about whether the large-scale fields or random fields dominate in a galaxy, and whether there are any field reversals. We take the average of the magnetic field strength of each map derived this way. If this is not $\sim$0 $\mu$G, it suggests that the B field is large-scale and not random. However, if it is $\sim$0 $\mu$G, the B field is either random, or it is large-scale but has one or more field reversals in one of the field components: large regions with similar B field strengths but different signs.

\end{appendix}

\end{document}